\documentclass[reprint,aps,twocolumn,superscriptaddress,amsmath,amssymb,floatfix]{revtex4-2}

\usepackage{graphicx}% Include figure files
\usepackage[dvipsnames]{xcolor}% color
\usepackage{dcolumn}% Align table columns on decimal point
\usepackage{bm}% bold math
\usepackage[colorlinks=true,citecolor=blue,filecolor=blue,linkcolor=blue,urlcolor=blue,pdftex]{hyperref}
\usepackage{tikz}
\usepackage{orcidlink} % For ORCID links in author list
\usepackage{currfile}
\usetikzlibrary{shapes,arrows,positioning}
\def\equationautorefname~#1\null{Equation~(#1)\null}
\newcommand{\isotope}[2]{\(^{#2}\mathrm{#1}\)}
\newcommand{\unit}[1]{\, \mathrm{#1}}
\newcommand{\bologna}{\affiliation{Department of Physics and Astronomy, University of Bologna and INFN-Bologna, 40126 Bologna, Italy}}
\newcommand{\chicago}{\affiliation{Department of Physics \& Kavli Institute for Cosmological Physics, University of Chicago, Chicago, IL 60637, USA}}
\newcommand{\coimbra}{\affiliation{LIBPhys, Department of Physics, University of Coimbra, 3004-516 Coimbra, Portugal}}
\newcommand{\columbia}{\affiliation{Physics Department, Columbia University, New York, NY 10027, USA}}
\newcommand{\lngs}{\affiliation{INFN-Laboratori Nazionali del Gran Sasso and Gran Sasso Science Institute, 67100 L'Aquila, Italy}}
\newcommand{\mainz}{\affiliation{Institut f\"ur Physik \& Exzellenzcluster PRISMA$^{+}$, Johannes Gutenberg-Universit\"at Mainz, 55099 Mainz, Germany}}
\newcommand{\mpik}{\affiliation{Max-Planck-Institut f\"ur Kernphysik, 69117 Heidelberg, Germany}}
\newcommand{\munster}{\affiliation{Institut f\"ur Kernphysik, Westf\"alische Wilhelms-Universit\"at M\"unster, 48149 M\"unster, Germany}}
\newcommand{\nikhef}{\affiliation{Nikhef and the University of Amsterdam, Science Park, 1098XG Amsterdam, Netherlands}}
\newcommand{\nyuad}{\affiliation{New York University Abu Dhabi - Center for Astro, Particle and Planetary Physics, Abu Dhabi, United Arab Emirates}}
\newcommand{\purdue}{\affiliation{Department of Physics and Astronomy, Purdue University, West Lafayette, IN 47907, USA}}
\newcommand{\rice}{\affiliation{Department of Physics and Astronomy, Rice University, Houston, TX 77005, USA}}
\newcommand{\stockholm}{\affiliation{Oskar Klein Centre, Department of Physics, Stockholm University, AlbaNova, Stockholm SE-10691, Sweden}}
\newcommand{\subatech}{\affiliation{SUBATECH, IMT Atlantique, CNRS/IN2P3, Universit\'e de Nantes, Nantes 44307, France}}
\newcommand{\torino}{\affiliation{INAF-Astrophysical Observatory of Torino, Department of Physics, University  of  Torino and  INFN-Torino,  10125  Torino,  Italy}}
\newcommand{\ucsd}{\affiliation{Department of Physics, University of California San Diego, La Jolla, CA 92093, USA}}
\newcommand{\wis}{\affiliation{Department of Particle Physics and Astrophysics, Weizmann Institute of Science, Rehovot 7610001, Israel}}
\newcommand{\zurich}{\affiliation{Physik-Institut, University of Z\"urich, 8057  Z\"urich, Switzerland}}
\newcommand{\paris}{\affiliation{LPNHE, Sorbonne Universit\'{e}, CNRS/IN2P3, 75005 Paris, France}}
\newcommand{\freiburg}{\affiliation{Physikalisches Institut, Universit\"at Freiburg, 79104 Freiburg, Germany}}
\newcommand{\napels}{\affiliation{Department of Physics ``Ettore Pancini'', University of Napoli and INFN-Napoli, 80126 Napoli, Italy}}
\newcommand{\nagoya}{\affiliation{Kobayashi-Maskawa Institute for the Origin of Particles and the Universe, and Institute for Space-Earth Environmental Research, Nagoya University, Furo-cho, Chikusa-ku, Nagoya, Aichi 464-8602, Japan}}
\newcommand{\laquila}{\affiliation{Department of Physics and Chemistry, University of L'Aquila, 67100 L'Aquila, Italy}}
\newcommand{\tokyo}{\affiliation{Kamioka Observatory, Institute for Cosmic Ray Research, and Kavli Institute for the Physics and Mathematics of the Universe (WPI), University of Tokyo, Higashi-Mozumi, Kamioka, Hida, Gifu 506-1205, Japan}}
\newcommand{\kobe}{\affiliation{Department of Physics, Kobe University, Kobe, Hyogo 657-8501, Japan}}
\newcommand{\kit}{\affiliation{Institute for Astroparticle Physics, Karlsruhe Institute of Technology, 76021 Karlsruhe, Germany}}
\newcommand{\tsinghua}{\affiliation{Department of Physics \& Center for High Energy Physics, Tsinghua University, Beijing 100084, P.R. China}}
\newcommand{\ferrara}{\affiliation{INFN-Ferrara and Dip. di Fisica e Scienze della Terra, Universit\`a di Ferrara, 44122 Ferrara, Italy}}
\newcommand{\groningen}{\affiliation{Nikhef and the University of Groningen, Van Swinderen Institute, 9747AG Groningen, Netherlands}}
\newcommand{\westlake}{\affiliation{Department of Physics, School of Science, Westlake University, Hangzhou 310030, P.R. China}}
\newcommand{\shenzhen}{\affiliation{School of Science and Engineering, The Chinese University of Hong Kong, Shenzhen, Guangdong, 518172, P.R. China}}

\begin{document}
\author{E.~Aprile\,\orcidlink{0000-0001-6595-7098}}\columbia
\author{J.~Aalbers\,\orcidlink{0000-0003-0030-0030}}\groningen
\author{K.~Abe\,\orcidlink{0009-0000-9620-788X}}\tokyo
\author{S.~Ahmed Maouloud\,\orcidlink{0000-0002-0844-4576}}\paris
\author{L.~Althueser\,\orcidlink{0000-0002-5468-4298}}\munster
\author{B.~Andrieu\,\orcidlink{0009-0002-6485-4163}}\paris
\author{E.~Angelino\,\orcidlink{0000-0002-6695-4355}}\torino
\author{J.~R.~Angevaare\,\orcidlink{0000-0003-3392-8123}}\nikhef
\author{D.~Ant\'on Martin}\chicago
\author{F.~Arneodo\,\orcidlink{0000-0002-1061-0510}}\nyuad
% ...
\collaboration{XENON Collaboration}\email[]{xenon@lngs.infn.it}\noaffiliation

\noaffiliation
% \begin{document}

% \preprint{APS/123-QED}
\date{\today}
\title{Offline tagging of radon-induced backgrounds in XENON1T and applicability to other liquid xenon time projection chambers}% Force line breaks with \\

\begin{abstract}
This paper details the first application of a software tagging algorithm to reduce radon-induced backgrounds in liquid noble element time projection chambers, such as XENON1T and XENONnT. The convection velocity field in XENON1T was mapped out using \isotope{Rn}{222} and \isotope{Po}{218} events, and the root-mean-square convection speed was measured to be \(0.30 \pm 0.01 \unit{cm/s}\). Given this velocity field, \isotope{Pb}{214} background events can be tagged when they are followed by \isotope{Bi}{214} and \isotope{Po}{214} decays, or preceded by \isotope{Po}{218} decays. This was achieved by evolving a point cloud in the direction of a measured convection velocity field, and searching for \isotope{Bi}{214} and \isotope{Po}{214} decays or \isotope{Po}{218} decays within a volume defined by the point cloud. In XENON1T, this tagging system achieved a \isotope{Pb}{214} background reduction of \(6.2^{+0.4}_{-0.9}\%\) with an exposure loss of \(1.8\pm 0.2 \%\), despite the timescales of convection being smaller than the relevant decay times. 
%The tagging algorithm was also used to produce a population of tagged events with a large enhancement in the \isotope{Pb}{214} fraction. 
We show that the performance can be improved in XENONnT, 
%where an exposure loss of only \(4.3\%\) would be required to produce an estimated \isotope{Pb}{214} background reduction of \(25\%\) for high convection speeds of \(0.8 \unit{cm/s}\), 
and that the performance of such a software-tagging approach can be expected to be further improved in a diffusion-limited scenario. Finally, a similar method might be useful to tag the cosmogenic \isotope{Xe}{137} background, which is relevant to the search for neutrinoless double-beta decay.
\end{abstract}

\maketitle

%\tableofcontents

\section{Introduction}

Liquid xenon time projection chambers (TPCs) such as XENON1T~\cite{XENON:2017lvq}, XENONnT~\cite{XENON:2020kmp}, and LZ~\cite{LZ:2019sgr} are constructed with the primary goal of searching for dark matter in the form of weakly-interacting massive particles (WIMPs)~\cite{XENON:2017lvq, XENON:2018voc}. These TPCs as well as the dedicated EXO-200 TPC also search for neutrinoless double-beta decay (\(0\nu\beta\beta\))~\cite{XENON:2022evz,EXO-200:2019rkq,LZ:2021blo}. Other physics channels include measurements of double electron capture in \isotope{Xe}{124}~\cite{XENON:2019dti}, solar axions, non-standard neutrino interactions, and bosonic dark matter~\cite{XENON:2020rca, XENON:2022ltv}.

Achieving low levels of radioactive backgrounds is critical to the aforementioned physics channels because the sensitivity scales as $\mathrm{signal}/\sqrt{\mathrm{background}}$ (see \autoref{ssec:optimisation}. The decay chain of \isotope{Rn}{222} includes \isotope{Pb}{214}, an isotope that undergoes beta decay. At low energies, this is a major source of backgrounds in the electronic-recoil (ER) channel of xenon-based dark matter experiments, and is also important to the nuclear-recoil (NR) channel due to imperfect ER/NR discrimination~\cite{DARWIN:2016hyl, LZ:2019sgr, PandaX:2018wtu, XENON:2017lvq, XENON:2020kmp}. The reason \isotope{Rn}{222} contamination is a major source of backgrounds is because \isotope{Rn}{222} is produced from the emanation of \isotope{Ra}{226}, which is present at low levels in almost all materials~\cite{XENON:2021mrg}. In addition, \isotope{Rn}{222} is miscible with xenon, and the half-life of \(t_{1/2} = 3.8 \unit{days}\)~\cite{Singh:2019qyr} allows it to move throughout the detector. The isotope in the \isotope{Rn}{222} decay chain that decays to produce the relevant low-energy background is \isotope{Pb}{214}, see \autoref{fig:rn222_decay_chain}. Because of this, substantial efforts have been made to reduce the radon level using dedicated hardware solutions~\cite{Abe2012RadonRF, Arthurs:2020ans, Chen:2023llu,LUX:2016wel, LZ:2020fty, Murra:2022mlr, Pushkin:2018wdl, XENON:2017fdb, XENON:2021mrg}. The software-based approach introduced in this paper performs better at lower background levels, thus complementing these hardware-based methods to further reduce radon-induced backgrounds using offline analysis. Such a method can also be used to suppress radon chain backgrounds in liquid argon TPCs, where hardware-based approaches for the mitigation of radon-chain backgrounds are similarly being pursued~\cite{Avasthi:2022tjr}. The key challenge in a software-based approach is that the convection timescale is $\sim\frac{100\unit{cm}}{0.3\unit{cm/s}} = 300\unit{s}$, which is significantly shorter than the decay times of both \isotope{Pb}{214} and \isotope{Bi}{214}; these are $27 \unit{min}$ and $20 \unit{min}$, respectively. This implies that over one decay halflife, the radioactive isotopes move substantially in the TPC, ruling out naive approaches that simply veto spherical volumes around \isotope{Po}{218} events.

As such, efforts to tag \isotope{Pb}{214} events in XENON1T based on other events in the same decay chain require a measurement of the convection velocity field. 
% This is in contrast to solid-state detectors, such as DAMIC~\cite{DAMIC:2013bio}, where decays from the same decay chain are expected to be spatially coincident even when events are separated by tens of days~\cite{DAMIC:2020wkw}. 
The measurement of convection in XENON1T and properties of the velocity field are detailed in \autoref{sec:convection}. The algorithm to track isotopes along the measured velocity field and thus veto \isotope{Pb}{214} events is detailed in \autoref{sec:software_veto_algo}. Results, including demonstrations of the technique on XENON1T data, projections to XENONnT and future liquid xenon detectors, and application to cosmogenic \isotope{Xe}{137} which is a background for the search for neutrinoless double-beta decay (\(0\nu\beta\beta\)), are discussed in \autoref{sec:results}. Finally, the conclusion is presented in \autoref{sec:conclusion}. 

\subsection{XENON1T and XENONnT}\label{ssec:1T_and_nT}
The XENON1T experiment used a cylindrical dual-phase TPC, with a diameter of \(96 \unit{cm}\) and a \(2.0 \unit{tonnes}\) active liquid xenon target~\cite{XENON:2017lvq}. Two arrays of photomultiplier tubes (PMTs) were installed in the top and bottom of the TPC, respectively. The vertical sides of the TPC are constructed out of UV-reflective Polytetrafluoroethlyene to increase the light collection efficiency. A cathode at the bottom of the TPC and a gate electrode \(97 \unit{cm}\) above the cathode produced a drift field of \(81 \unit{V/cm}\). The anode was \(5 \unit{mm}\) above the gate and \(\sim 2.5\unit{mm}\) above the liquid-gas interface, and with the gate produced an extraction field of \(8.1 \unit{kV/cm}\). The cryostat containing the TPC was positioned inside a \(740 \unit{m^3}\) water Cherenkov muon veto which allowed for the active detection and veto of muons and muon-induced backgrounds.

XENONnT is an in-place upgrade of XENON1T that uses much of the existing infrastructure. It features an enlarged TPC, a novel neutron veto system, and various improvements in the xenon handling system, which allow for an improved xenon purity. The TPC diameter is \(1.3 \unit{m}\), and the separation between gate and cathode extended to \(1.5 \unit{m}\). This results in an active liquid xenon target mass of \(5.9 \unit{tonnes}\)~\cite{XENON:2020kmp}. The muon veto uses the same design as in XENON1T. However, there is now a neutron veto surrounding the main cryostat containing the TPC, which is optically separated from the muon veto. The neutron veto aims to reduce the radiogenic neutron background by detecting neutrons which scatter in the TPC volume and are then captured in the neutron veto \cite{XENON:2020kmp}.

Events in the TPC are detected via two signals. First, a prompt scintillation signal (S1) comprised of $175\unit{nm}$ photons~\cite{FUJII2015293} is produced at the site of a particle interaction due to the decay of excited atoms~\cite{Szydagis:2022ikv}. Additionally, ionization electrons are produced; some of these recombine with xenon ions, contributing to the S1 as well~\cite{Szydagis:2022ikv}. These photons are detected by the PMTs. Ionization electrons that do not recombine are drifted towards the liquid-gas interface by the drift field, and extracted into the xenon gas by the extraction field, where a secondary scintillation signal (S2) is produced. The size of these S1 and S2 signals are measured in units of photoelectrons (PE).

The 3D position of the interaction can be reconstructed using the hit pattern of the S2 signal on the top PMT array $(x,y)$ and the drift time between the S1 and S2 signals $(z)$. 
% In XENON1T, the $(x,y)$ position is primarily reconstructed using an artificial neural network trained on simulated data~\cite{XENON:2018voc, XENON:2019ykp}. Non-uniformity of the drift field is corrected for when reconstructing the interaction position~\cite{XENON:2019ykp}. 
Details regarding position reconstruction methods in XENON1T can be found in~\cite{XENON:2019ykp}.
The reconstructed position also allows for the computation of corrected S1 and S2 signal sizes, based on position-dependent signal efficiencies~\cite{XENON:2019ykp}.

\subsection{The \texorpdfstring{\isotope{Rn}{222}}{Rn-222} decay chain}
The decay chain of \isotope{Rn}{222} is shown in \autoref{fig:rn222_decay_chain}. \isotope{Pb}{214} is responsible for the low-energy ER background. This is because the beta spectrum of \isotope{Pb}{214} extends to low energies and is flat to the percent-level below \(50 \unit{keV}\)~\cite{XENON:2020rca}. Alpha decays, on the other hand, are mono-energetic and have a different S1/S2 ratio from ER or NR events~\cite{Aprile:2006kx, Jorg:2021hzu}, and hence are easy to select. This differing S1/S2 ratio is related to recombination of electrons and ions, as described earlier in~\autoref{ssec:1T_and_nT}.
\tikzstyle{block} = [rectangle, draw, fill=white, 
    text width=2.4cm, text centered, rounded corners, minimum height=1cm]
\tikzstyle{arrow} = [thick,->,>=stealth]

\begin{figure}[htp]
    \centering
    \begin{tikzpicture}[scale=1, node distance =1.5cm and 1.5cm, auto]
        \node at (0,0) [block] (Ra226) {\isotope{Ra}{226}\\\(1600 \pm 7 \unit{yr}\)};
        \node at (0,-2.6cm) [block] (Rn222) {\isotope{Rn}{222}\\\(3.8222 \pm 0.0002 \unit{d}\)};
        \node at (0,-5.2cm) [block, text=blue, draw=blue] (Po218) {\isotope{Po}{218}\\\(3.097 \pm 0.012 \unit{min}\)};
        \node at (0,-7.8cm) [block, text=Maroon, draw=Maroon] (Pb214) {\isotope{Pb}{214}\\\(27.06 \pm 0.07 \unit{min}\)};
        \node at (2.8,-6.5cm) [block, text=blue, draw=blue] (Bi214) {\isotope{Bi}{214} \\\(19.71 \pm 0.02 \unit{min}\)};
        \node at (5.6,-5.2cm) [block, text=blue, draw=blue] (Po214) {\isotope{Po}{214}\\\(163.6 \pm 0.3 \unit{\mu s}\)};
        \node at (5.6,-7.8cm) [block] (Pb210) {\isotope{Pb}{210}\\\(22.20 \pm 0.22 \unit{yr}\)};

        \draw [arrow] (Ra226) -- node [align=right, anchor=north west]{\(\alpha\)} (Rn222);
        \draw [arrow] (Rn222) -- node [align=right, anchor=north west]{\(\alpha\)} (Po218);
        \draw [arrow, draw=blue] (Po218) -- node [align=right, anchor=north west, text=blue]{\(\alpha\)} (Pb214);
        \draw [arrow, draw=Maroon] (Pb214) -- node [align=right, anchor=north west, text=Maroon]{\(\beta\)} (Bi214);
        \draw [arrow, draw=blue] (Bi214) -- node [align=right, anchor=north west, text=blue]{\(\beta\)} (Po214);
        \draw [arrow, draw=blue] (Po214) -- node [align=right, anchor=north west, text=blue]{\(\alpha\)} (Pb210);
    \end{tikzpicture}
    \caption{Decay chain of \isotope{Rn}{222}, part of the uranium series. Only branches with branching fraction above \(99.5 \%\) are shown. Data retrieved using the NNDC ENSDF, with the following Nuclear Data Sheets citations: \cite{Singh:2011yau, Singh:2019qyr, Zhu:2021qss, ShamsuzzohaBasunia:2014yyr, Kondev:2008roc}. The isotope that decays to produce the background events being tagged in this work, \isotope{Pb}{214}, is coloured red, whereas the isotopes with decays that are used for the tagging of \isotope{Pb}{214} are coloured blue.}
    \label{fig:rn222_decay_chain}
\end{figure}
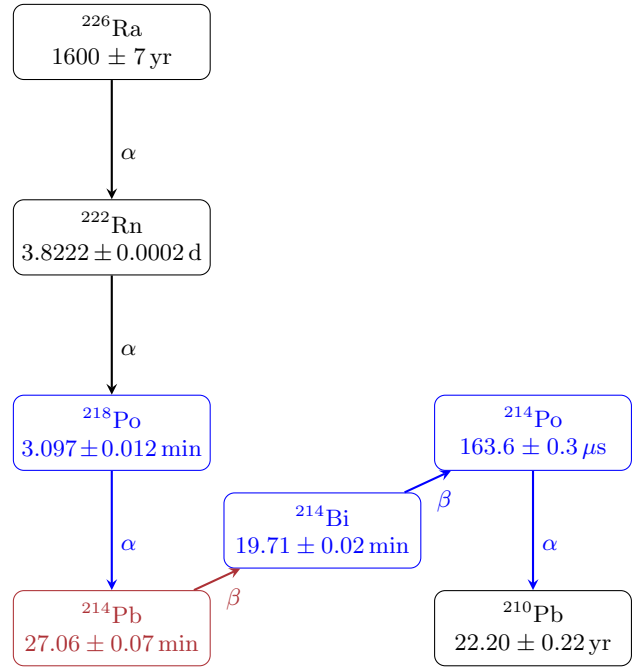

\isotope{Bi}{214} decay does not represent an important background because it is quickly followed by \isotope{Po}{214}, an isotope that undergoes alpha decay with a half-life of \(164 \unit{\mu s}\)~\cite{ShamsuzzohaBasunia:2014yyr}. Because this is much shorter than the drift time in XENON1T and XENONnT, the \isotope{Bi}{214} and \isotope{Po}{214} events are combined into a single event with two S1s, and two or more S2s. This is a unique event topology that is easy to select. Such events are termed BiPo events in this paper.

\section{Measuring convection in the XENON1T detector}\label{sec:convection}

\subsection{Mapping the convection velocity field} \label{ssec:convection_measurement}

Convection has been observed in earlier dual phase liquid xenon TPCs, such as XENON100 and LUX~\cite{XENON:2016rze, Malling:2014oxk}. While the exact boundary conditions driving the convection are not known, the convective flow is likely driven by the thermal flux into the TPC, possibly from both recirculation flows and from the cryostat. The relevant temperature gradient might be either horizontal or vertical.

To measure the convective flow in the XENON1T detector, \isotope{Rn}{222} and \isotope{Po}{218} events were used. \isotope{Rn}{222} and \isotope{Po}{218} undergo alpha decay. 
% As mentioned earlier, these events are monoenergetic and have a different S1/S2 ratio from ER or NR events~\cite{Aprile:2006kx, Jorg:2021hzu}; they can thus be easily identified. 
The $3 \unit{min}$ half-life of \isotope{Po}{218} is short enough that the \isotope{Rn}{222} and \isotope{Po}{218} events can be paired up, but long enough that there can be significant displacement between pairs of events at $\sim 0.1\unit{cm/s}$ speeds. As such, the decays of \isotope{Rn}{222} and \isotope{Po}{218} are particularly suited to the measurement of convection.

These events were selected using Gaussian Mixture clustering~\cite{pedregosa:JMLR:v12:pedregosa11a} using the position-corrected S1, position-corrected S2, width of the S2 peak in nanoseconds, radial coordinate, and z-coordinate of each event~\cite{Masson:2018pte}.

%\(\mathrm{cS1} \in (4.4\times10^4, 8.5\times10^4) \unit{PE}\) and \(\mathrm{cS2} \in (7\times10^4, 3\times10^5)\unit{PE}\)

\begin{figure}[htbp]
 \centering
    \includegraphics[width=\columnwidth]{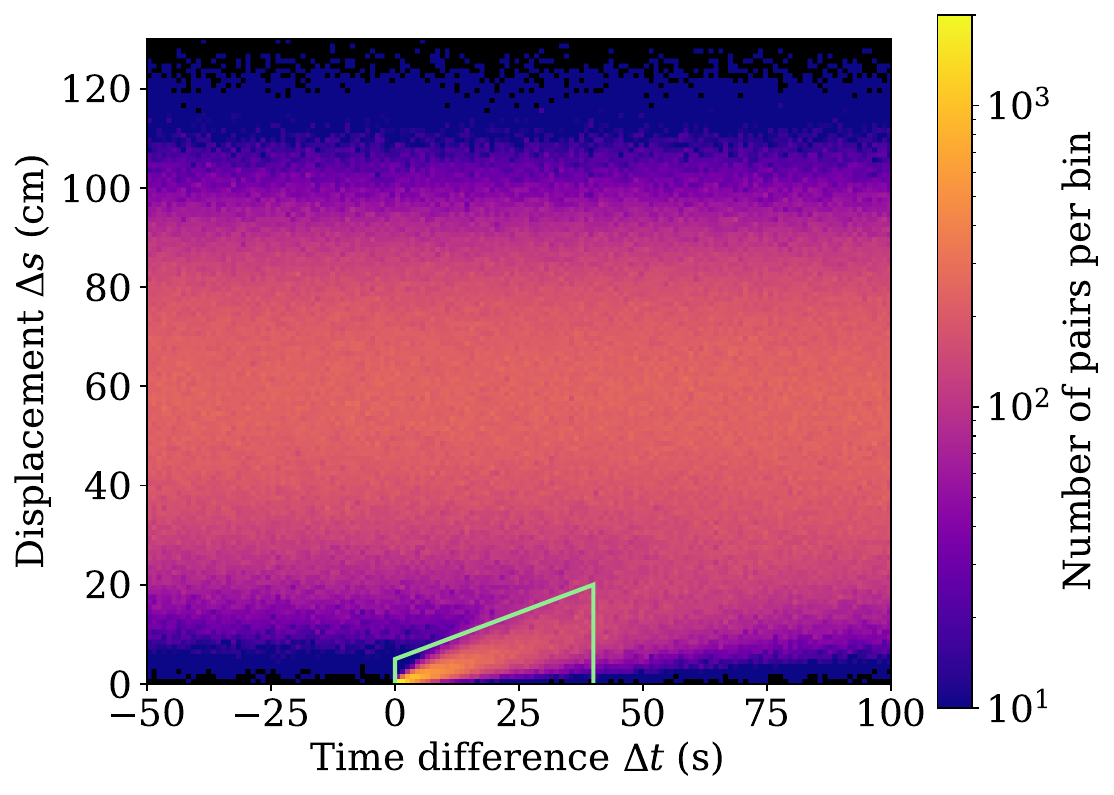}%
    \caption{2D histogram of the time difference \((\Delta t)\) and displacement \((\Delta s)\) of all permutations of \isotope{Rn}{222} and \isotope{Po}{218} pairs. The excess of pairs where \(\Delta s < 20\unit{cm}\) and \(0 \unit{s} < \Delta t < 40 \unit{s}\), is from correctly-paired events, and is highlighted with the light green box. At negative times, the pairs are unphysical and can be used to profile the distribution of incorrect pairs.}\label{fig:dxdt_histogram}
\end{figure}

After event selection, \isotope{Rn}{222} and \isotope{Po}{218} events have to be paired to construct velocity vectors corresponding to the convective flow. However, the rate of \isotope{Rn}{222} decays exceeds \(10 \unit{\mu Bq/kg}\)~\cite{XENON:2020fbs}, corresponding to approximately two \isotope{Rn}{222} events every \(3 \unit{min}\) in a \(1 \unit{tonne}\) fiducial mass. As the half-life of \isotope{Po}{218} is \(\sim 3 \unit{min}\), the pairing of \isotope{Rn}{222} and \isotope{Po}{218} events cannot be done in a naive manner where every \isotope{Po}{218} event is considered to be the daughter of the preceding \isotope{Rn}{222} event. Instead, for each pair of \isotope{Rn}{222} and \isotope{Po}{218} events, the time difference \((\Delta t)\) and displacement \((\Delta s)\) were plotted on a histogram, see~\autoref{fig:dxdt_histogram}. An excess of pairs where \(\Delta s < 20\unit{cm}\) and \(0 \unit{s} < \Delta t < 40 \unit{s}\) becomes apparent. This excess is due to correctly-paired events. In addition, the distribution of \isotope{Rn}{222} and \isotope{Po}{218} events that are not correctly paired is independent of \(\Delta t\), and can be determined using pairs where \(\Delta t < 0\). The observed distribution of these incorrect pairs is largely due to the TPC geometry, and the maxima at $\sim 60\unit{cm}$ is what one would expect from the pair-wise distances of two uniform random distributions in the TPC. One can then compute the purity of each histogram bin as \(f_{\mathrm{pure}} = 1-{N_{bg}}/{N_{bin}}\), where \(N_{bg}\) is the number of incorrect pairs in a bin at the given \(\Delta t\) estimated using negative-time bins, and \(N_{bin}\) is the total number pairs in a bin. 
% With this, it is already possible to obtain a sample of pairs that can be used to measure the convection velocity simply by selecting all histogram bins with purities above a given threshold.

All permutations of \isotope{Rn}{222} and \isotope{Po}{218} pairs with \(0 \unit{s} < \Delta t < 100 \unit{s}\) were used. We iteratively selected the bin with highest \(f_{\mathrm{pure}}\), and then removed all pairs which contain one of these \isotope{Rn}{222} or \isotope{Po}{218} events. This is done because as one iteratively removes events that belong to selected pairs, the total number of events remaining in the pool decreases, thus decreasing the number of incorrectly-matched pairs remaining. This is run for 1000 iterations.
% The average purity of \isotope{Rn}{222}-\isotope{Po}{218} pairs as a function of the total number of pairs selected is shown in~\autoref{fig:iterative_purity} for both the naive and the iterative methods. It can be seen that the iterative method indeed provides a modest improvement. The maximum $\Delta t$ of $100 \unit{s}$ was chosen such that the bins selected by both the naive and the iterative methods remain within this limit; at larger values of $\Delta t$ the purity of histogram bins are too low to be selected by these methods. 

% \begin{figure}[htbp]
%  \centering
%     \includegraphics[width=\columnwidth]{plots/iterative_purity.pdf}%
%     \caption{Plot of purity, \(f_{\mathrm{pure}}\) versus number of pairs for the iterative method, and for a naive method where one simply selects the histogram bins from~\autoref{fig:dxdt_histogram}.}\label{fig:iterative_purity}
% \end{figure}

\begin{figure}[htbp]
 \centering
    \includegraphics[width=\columnwidth]{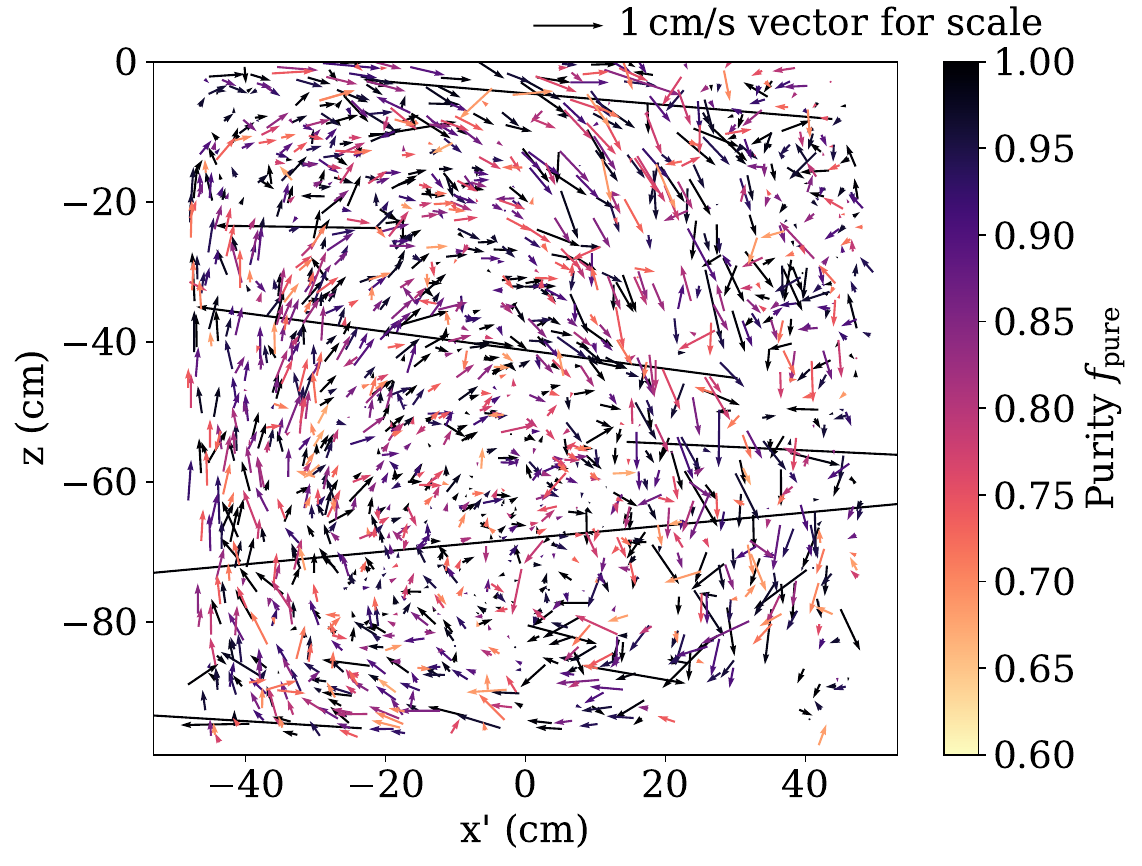}%
    \caption{Slice of velocity field with 107804 vectors showing the convection cell. The \(x'\) coordinate is perpendicular to the angular momentum vector. A \(1\,\unit{cm/s}\) velocity vector is shown in the top right for scale. The purity of a vector, $f_\mathrm{pure}$, is indicated by the colour.}\label{fig:unfiltered_velocity_field}
\end{figure}

Vectors were constructed from selected pairs of events by computing the velocity from the \(\Delta s\) and \(\Delta t\) values of the pair. The velocity field obtained using the iterative method is shown in \autoref{fig:unfiltered_velocity_field}. It can be seen that this velocity field is still noisy, and contains outliers that likely correspond to incorrectly-paired events.

It is notable that there appears to be a single large convection cell, which is expected for convection cells in a cylinder with an aspect ratio of close to 1~\cite{Guenter:PhysRevLett.128.084501}.
% It is notable that there appears to be a single large convection cell. While one might expect a toroidal convection flow due to cylindrical symmetry of the TPC, the observed asymmetric convection flow is expected for convection cells in a cylinder with an aspect ratio of close to 1~\cite{Guenter:PhysRevLett.128.084501}. Attempts to simulate this convection flow have been made, but have been unsuccessful due to lack of knowledge of the boundary conditions driving this flow, and resultant inability to validate these simulations.

\subsection{Filtering of the velocity field}\label{ssec:v_field_filtering}

\begin{figure*}[htp]
 \centering
    \includegraphics[width=0.9\columnwidth]{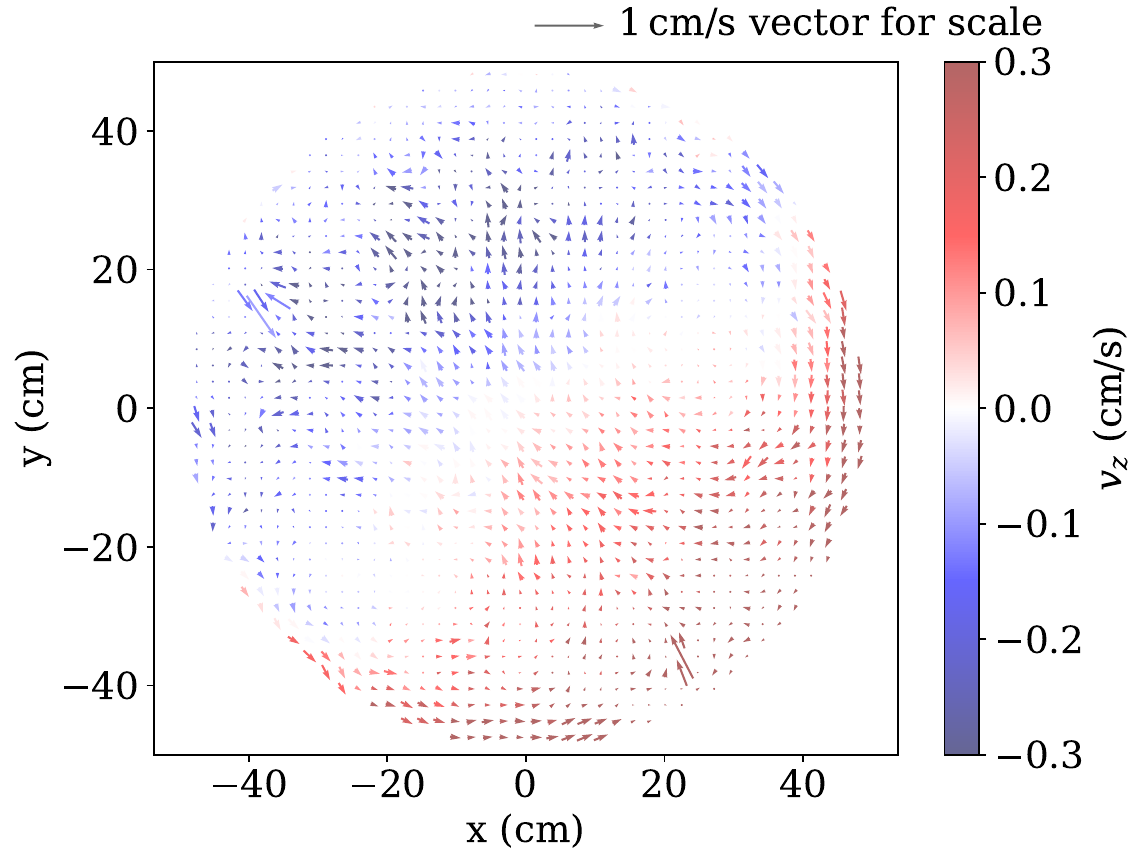}%
    \qquad
    \includegraphics[width=0.93\columnwidth]{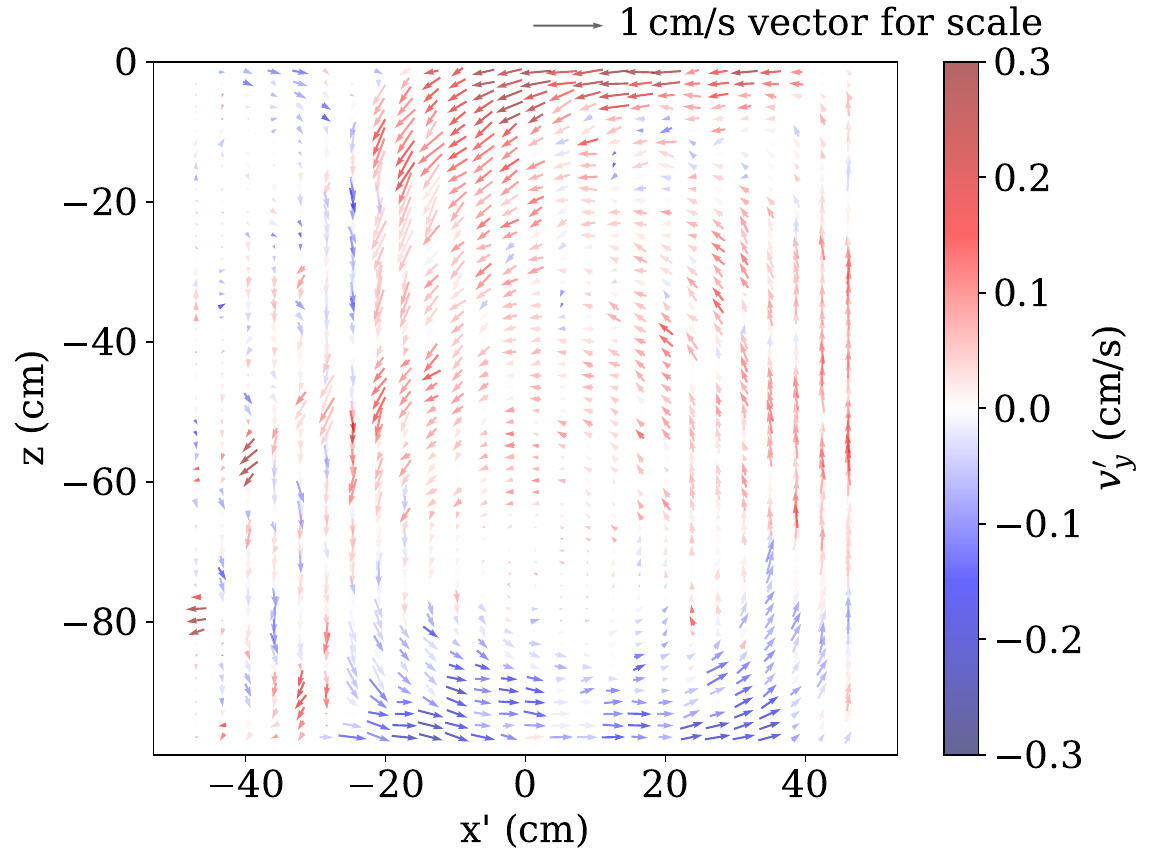}%
    \caption{Slice of the velocity field after it was filtered and put discretized onto a grid, shown from the top view (left) and the side view (right). In the left plot, the \(x'\) coordinate is defined as in \autoref{fig:unfiltered_velocity_field}; in the right plot, a slice at $z=-50\unit{cm}$ is taken. For clarity, only every 6th vector is displayed. A \(1\,\unit{cm/s}\) velocity vector is shown in the top right for scale.}\label{fig:grid_velocity_field}
\end{figure*}

The convection field, of which a slice is shown in \autoref{fig:unfiltered_velocity_field}, was then filtered and discretized onto a grid. The purpose of this was to reduce noise and to speed up computation, as finding the nearest velocity vector to a given position is much faster with data on a regular grid. First, every vector of purity \(f_{\mathrm{pure}}\) was oversampled $25\times f_{\mathrm{pure}}$ times, rounded to the nearest integer. The value of $25$ was chosen to avoid significant computational cost. The x-y position reconstruction uncertainty of \(\alpha\) events was estimated to be \(\sigma_x = 0.3 \, \mathrm{cm}, \sigma_y = 0.3 \, \mathrm{cm}\), based on the spread observed in \isotope{Pb}{210} decays on the TPC surface~\cite{Ye:2020PhDT.........9Y}, and the z-position uncertainty was estimated to be \(\sigma_z = 0.17 \, \mathrm{cm}\) from the displacement of the two decays in BiPo events. Diffusion is not considered as it is much smaller than the position resolution for the relevant timescales of \isotope{Po}{218} decay; this can be seen from~\ref{ssec:performance_xent}. During oversampling, each vector was perturbed randomly based on the position reconstruction uncertainty. The oversampled population of vectors were then put onto a grid with a grid spacing of \(1/3 \unit{cm}\), by computing the geometric median~\cite{HALDANE:10.1093/biomet/35.3-4.414} of the nearest 175 vectors at every grid point, as defined by the midpoint of the vectors. The geometric median has been shown to be particularly robust for noisy datasets~\cite{Lopuhaa:10.1214/aos/1176347978}. The result of this procedure is shown in \autoref{fig:grid_velocity_field}.

\subsection{Root-mean-square convection speed}\label{ssec:v_rms}

The convection vectors obtained in \autoref{ssec:convection_measurement} allow a measurement of the bulk convection properties. To avoid biases due to uneven event densities, the detector was divided into 11 bins in \(r^2 \in [0,47.9^2] \mathrm{cm}^2\), 10 bins in azimuth \(\phi \in [-\pi, \pi]\), and 9 bins in \(z \in [-96.9, 0] \mathrm{cm}\). Every vector was then assigned to a bin, and given a weight \(w_i\) equal to the reciprocal of the number of vectors in that bin. This procedure allows for the computation of a volume-averaged root-mean-square speed:
\begin{equation}\label{eq:v_rms}
    v_{\mathrm{rms}} = \sqrt{\frac{\sum^N_{i=1} v_i^2 w_i}{\sum^N_{i=1} w_i}}
\end{equation}

% \begin{figure}[htp]
%  \centering
%     \includegraphics[width=\columnwidth]{plots/convection_speed_vs_target.pdf}%
%     \caption{Convection speed versus \((\text{target mass})^{1/3}\) for XENON100~\cite{XENON:2016rze}, LUX~\cite{Malling:2014oxk}, and XENON1T (this work), with a linear fit shown in red. It can be seen that convection speed decreases linearly. The target masses corresponding to EXO-200 and LZ are shown by the black dotted lines, as convection was not observed in EXO-200~\cite{EXO-200:2015ura}, and was found to be subdominant to the drift of charged ions in LZ~\cite{LZ:2022oxb}. The target mass of XENONnT is indicated in grey as convection in XENONnT has not yet been analysed in detail.}\label{fig:convection_speed_vs_target}
% \end{figure}

\begin{figure}[htp]
 \centering
    \includegraphics[width=\columnwidth]{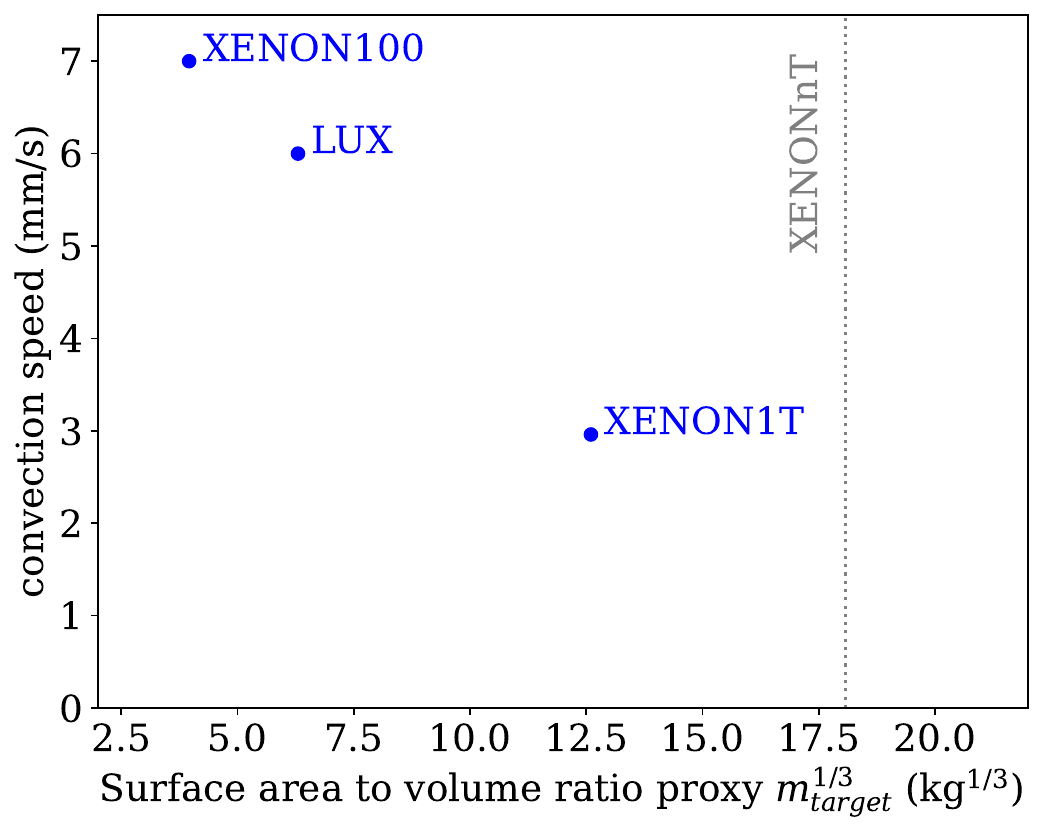}%
    \caption{Convection speed versus \((\text{target mass})^{1/3}\) for XENON100~\cite{XENON:2016rze}, LUX~\cite{Malling:2014oxk}, and XENON1T (this work). \((\text{target mass})^{1/3}\) is used as a proxy for the surface area to volume ratio. It can be seen that convection speed decreases linearly. EXO-200 and LZ are excluded from this plot as convection was not observed in EXO-200~\cite{EXO-200:2015ura}, and was found to be subdominant to the drift of charged ions in LZ~\cite{LZ:2022oxb}. The target mass of XENONnT is indicated in grey as convection in XENONnT has not yet been analysed in detail.}\label{fig:convection_speed_vs_target}
\end{figure}

The uncertainty on each velocity vector can be estimated using the position reconstruction uncertainty as \(\sigma_{i} = \sqrt{2\frac{\sigma_x^2 + \sigma_y^2 + \sigma_z^2}{\Delta t_i^2}}\). The total uncertainty is then given by

\begin{equation}\label{eq:v_rms_err}
    \sigma_{\mathrm{rms}} = \sqrt{\frac{\sum^N_{i=1} v_i^2 \sigma_{i}^2 w_i}{\sum^N_{i=1} w_i}}.
\end{equation}

The root-mean-square speed was thus found to be \(0.30\pm0.01\unit{cm/s}\). This is significantly slower than what was observed in XENON100 and LUX~\cite{XENON:2016rze, Malling:2014oxk}; however, it is significantly higher than EXO-200 and LZ where convection is sub-dominant to the mobility of charged ions~\cite{EXO-200:2015ura, LZ:2022oxb}. This is shown in \autoref{fig:convection_speed_vs_target}, plotted against the cube-root of the target mass, which is a proxy for linear dimension.

The heat flux into a TPC is likely proportional to the surface area, which is the square of linear dimension; when distributed over the entire target mass, which is related to the cube of linear dimension, one expects convection speed to vary linearly with linear dimension. It can be seen that convection speed decreases linearly with larger detectors. 
% , with the exception of EXO-200~\cite{EXO-200:2015ura}. EXO-200 is a single-phase liquid xenon TPC with a unique cooling solution where the liquid xenon vessel is immersed in a cryostat filled with a thermal-transfer fluid~\cite{Auger:2012gs}, and hence the heat flux into the TPC is much lower than in the other liquid xenon experiments which use vacuum-isolated cryostats.

It is notable that the convection velocity implies a convection timescale of $\sim\frac{100\unit{cm}}{0.3\unit{cm/s}} = 300\unit{s}$, which is significantly smaller than the decay time of both \isotope{Pb}{214} and \isotope{Bi}{214}, underscoring the difficulty of a software veto.

\section{\texorpdfstring{\isotope{Pb}{214}}{Pb-214} veto algorithm}\label{sec:software_veto_algo}
The veto algorithm first starts from an ER event that is in the energy range to be due to \isotope{Pb}{214} decay, termed the \isotope{Pb}{214} candidate. A normally-distributed set of points centered around the location of this event is generated, and propagated forwards or backwards along the velocity field just described in~\ref{sec:convection} to search for BiPo or \isotope{Pb}{218} events respectively. If a BiPo or \isotope{Pb}{218} event is found within the point clouds, then the \isotope{Pb}{218} candidate event is labelled as the \isotope{Pb}{218} background and can be vetoed. In the rest of this section section, we detail the algorithm used to select \isotope{Pb}{214} events. 
\subsection{Generation of noise fields}\label{ssec:noise_field}
There is likely uncertainty in the velocity field due to both limited statistics of \isotope{Rn}{220}-\isotope{Po}{218} pairs and the position reconstruction uncertainty; this needs to be properly accounted for. We address this by adding noise fields to the velocity field to induce fluctuations. This is done because one cannot directly use the uncertainties from the position reconstruction uncertainty when integrating the trajectories of the propagated points; furthermore, there is additional uncertainty introduced due to the purity of the selected vector population (see~\ref{sec:convection}). These are constrained by two conditions; first, the noise fields have to be divergence-free to avoid introducing sources and sinks, and second, the fields have to behave smoothly at the TPC boundaries. The constraints are expected to be sufficiently strong to make the arbitrary noise field a realistic proxy for the real conditions. Generation of the noise field started with smoothed Gaussian noise where a \(\sigma = 0.8\unit{cm}\) Gaussian kernel was used for smoothing, chosen to be significantly smaller than the length scale of the convection field as visible in~\ref{fig:grid_velocity_field}. After this, the curl was taken to ensure the noise is divergence free. 16 noise fields were generated, and then permuted by mirroring and rotating the fields, resulting in a total of 256 noise fields.

\begin{figure}[htp]
 \centering
    \includegraphics[width=\columnwidth]{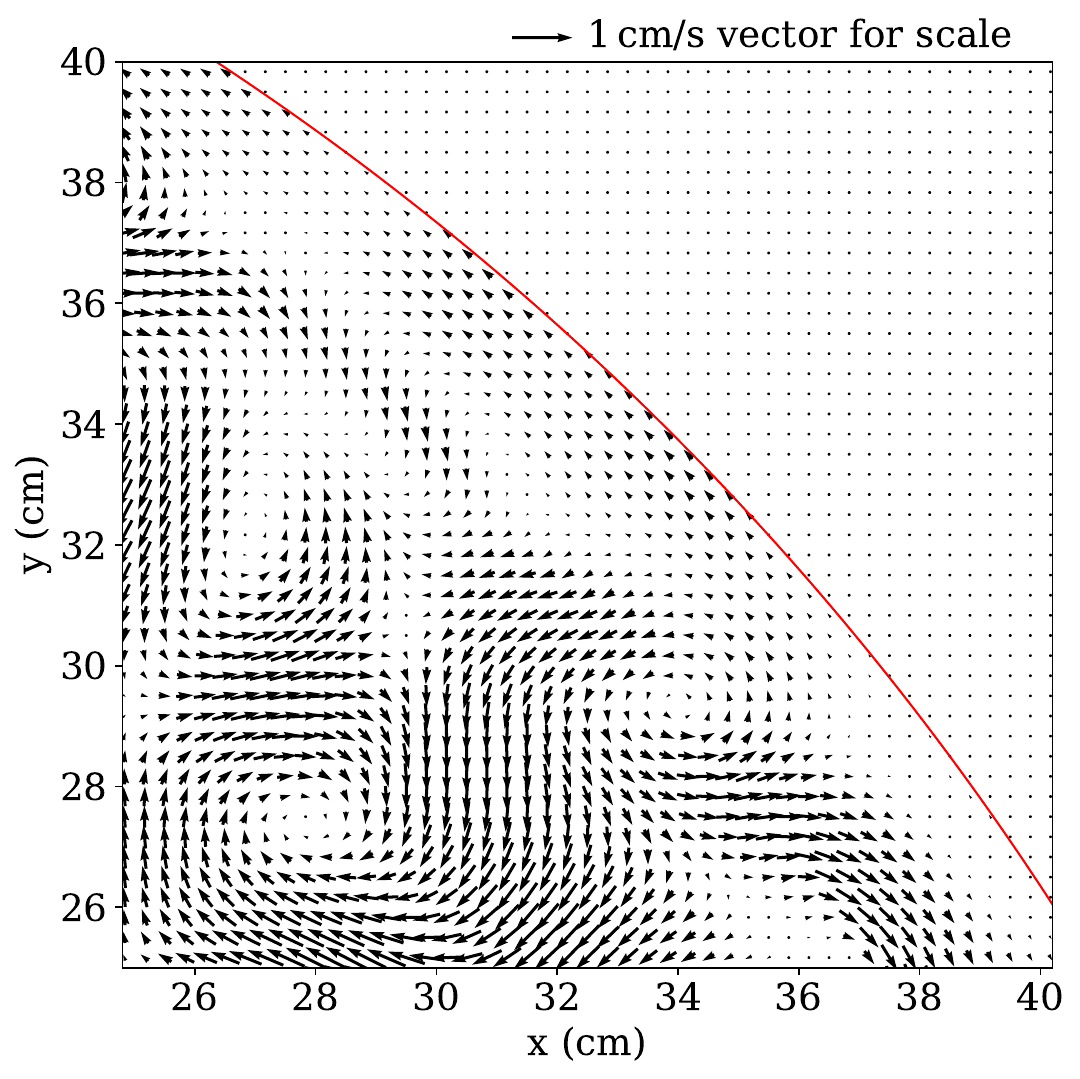}%
    \caption{Zoomed-in sample of the noise field. The edge of the detector is shown in red. It can be seen that the boundaries are handled smoothly, and that the field has no sinks that can trap propagating points.}\label{fig:noise_field_zoomin}
\end{figure}

TPC surfaces must also be handled smoothly, and the velocity component perpendicular to the surfaces must approach zero at the surfaces. This was ensured by smoothly scaling the perpendicular component of the field to zero, starting \(3\unit{cm}\) away from surfaces. This method of using the curl to generate divergence-free noise and handling boundaries is described in~\cite{Bridson:10.1145/1276377.1276435}. A section of the resultant noise field can be seen in \autoref{fig:noise_field_zoomin}.

\subsection{Point cloud propagation}\label{ssec:point_cloud_propagation}
\begin{figure}[htp]
 \centering
    \includegraphics[width=\columnwidth]{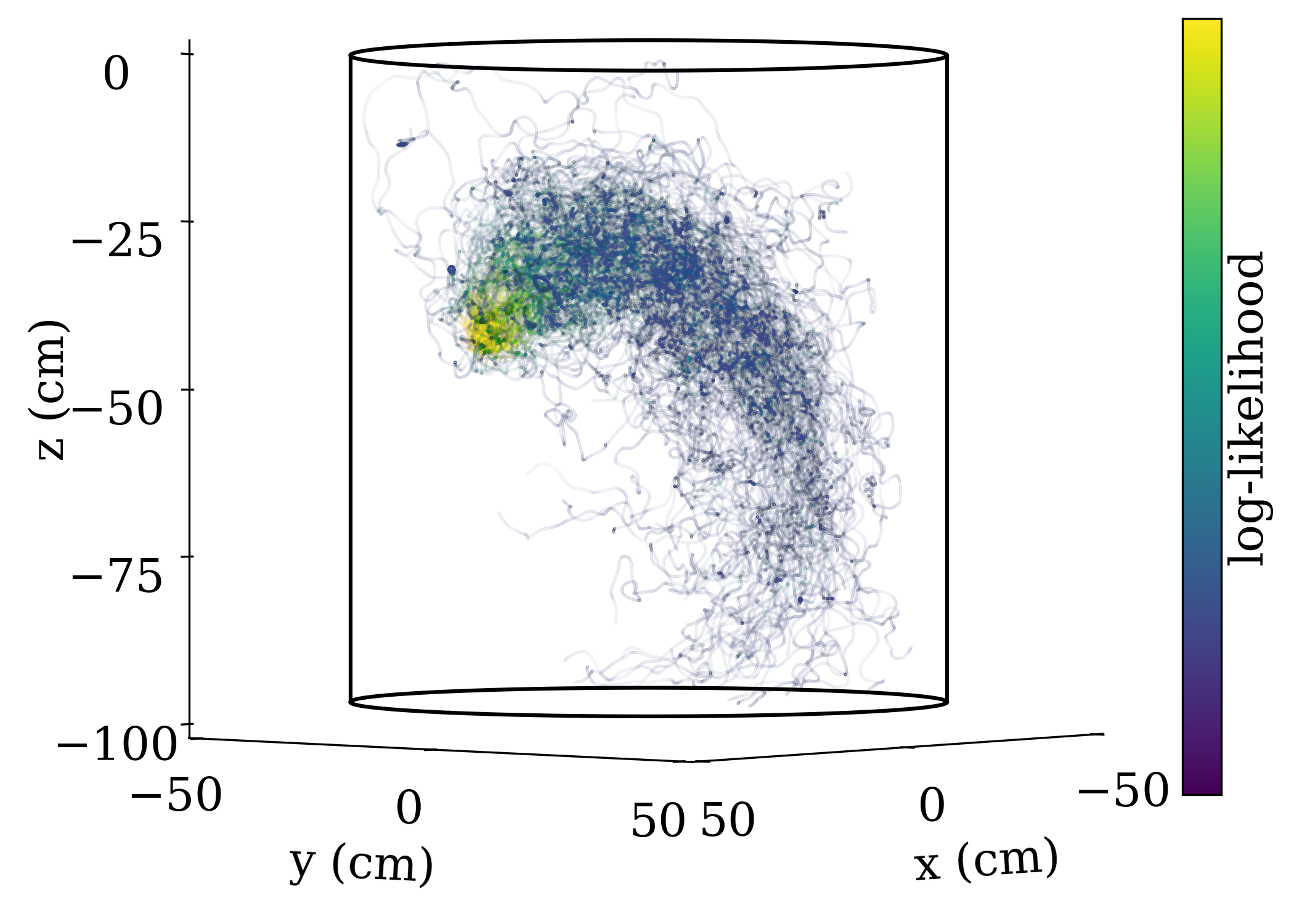}%
    \caption{Plot of a point cloud and the associated log-likelihood at each point. For illustration, the likelihood threshold is relaxed to a value of 8 show the point cloud propagating along the convection velocity field; elsewhere, higher values are used as indicated in the text, varying depending on the use-case. There are 192 points in the initial point cloud, and the timestep size is \(0.05 \unit{s}\)}\label{fig:point_cloud_tight_scatter}
\end{figure}

A veto volume within which one looks for predecessor or daughter events was constructed using a point cloud. The predecessor and daughter events for \isotope{Pb}{214} are \isotope{Po}{218} and BiPo events (see~\autoref{fig:rn222_decay_chain}). BiPo events were selected by choosing events that have two interactions, corresponding to the \isotope{Bi}{214} and \isotope{Po}{214} decays. These interactions were required to be less than \(5\unit{cm}\) apart in each of the x, y, and z directions, and the \(\alpha\)-event is further required to have appropriate position-corrected S1 and S2 values. First, a random event was picked from the set of electronic recoils in XENON1T as the \isotope{Pb}{214} candidate event. A point cloud was then generated around this \isotope{Pb}{214} candidate event and then propagated using the convection and noise velocity fields, with every point in the point cloud exposed to a different randomly-assigned noise field (see~\autoref{ssec:noise_field}). This noise field is introduced to account for the uncertainty of the velocity field; hence, exposing each point to a different noise field can be understood conceptually as exposing each point to a different version of the velocity field, allowing the uncertainty to be sampled. As one only needs to consider \isotope{Pb}{214} candidate events within the energy region of interest to a given analysis, propagating a point cloud from every \isotope{Po}{218} and BiPo event is more computationally expensive than from the smaller number of low-energy \isotope{Pb}{214} events. Point clouds from each \isotope{Pb}{214} candidate event are thus propagated in the forward and backward directions to look for BiPo and \isotope{Po}{218} events, respectively. These search directions are termed the BiPo and \isotope{Po}{218} channels for the rest of this paper. The point clouds are then culled in likelihood-space, based on the log-likelihood of each point. An illustration of a point cloud propagated along the velocity field can be seen in \autoref{fig:point_cloud_tight_scatter}.

There are four main steps involved in the generation and propagation of this point cloud:
\begin{enumerate}
    \item A \isotope{Pb}{214} candidate event is identified. 
    \item A point cloud is generated around the event, representing the position reconstruction uncertainty. The radial position uncertainty is \(\sigma_{R} = 5.2\unit{cm} - (1.61\unit{cm}) \log_{10} (S2/PE) + (0.019\unit{cm})\sqrt{S2/PE}\)~\cite{Ye:2020PhDT.........9Y}, whereas the z-position uncertainty is estimated to be \(\sigma_z = 0.17 \, \mathrm{cm}\) from the displacement of the two decays in BiPo events.
    \item Every 600 timesteps (\(30 \unit{s}\)), a probability density function is built out of the point cloud produced in the past 600 timesteps. This was done using kernel density estimation (KDE) in 4-dimensions. To this end, a uniform kernel of radius 3 cm, and \(0.3 \unit{sec}\) in the time-axis is used. Points that fall below a threshold of log-likelihood \((\xi)\) are culled to speed up computation. This log likelihood threshold is a free parameter.
    \item DBScan clustering~\cite{Ester:10.5555/3001460.3001507} is used to remove outlier points.
    \item The algorithm repeats from step 3, until all points have been removed.
\end{enumerate}

% A time-slice of a point cloud at every 594th (600-6, to avoid edge-effects from KDE) timestep is shown in \autoref{fig:point_cloud_timesteps}, with a log-likelihood constraint of \(9.1\), same as that used in the final analysis in the BiPo channel.

\begin{figure*}[htbp]
 \centering
    \includegraphics[width=\textwidth]{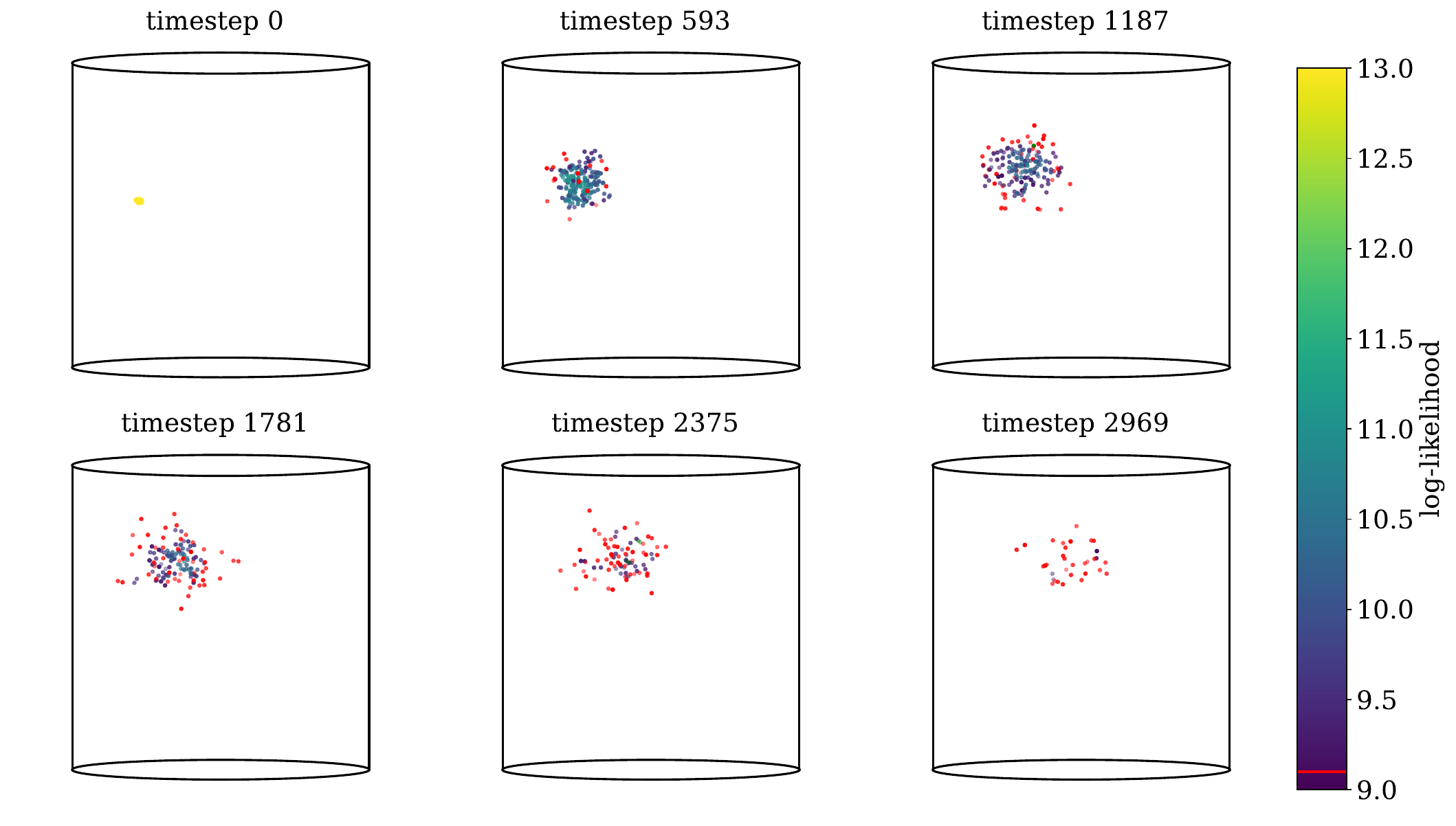}%
    \caption{Plot of point cloud with associated likelihoods at 6 different timesteps (\(0\unit{s}, 29.65\unit{s}, 89.05\unit{s}, 118.75\unit{s},\) and \(148.45\unit{s}\)). Red points are culled by the log-likelihood limit and not propagated further to speed up computation.}\label{fig:point_cloud_timesteps}
\end{figure*}

A time-slice of a point cloud at the end of every iteration is shown in \autoref{fig:point_cloud_timesteps}, with a log-likelihood constraint of \(9.1\), same as that used in the final analysis in the BiPo channel.

\subsection{Optimisation of veto volume}\label{ssec:optimisation}

In this section we will describe the optimisation of parameters governing the operation of the \isotope{Pb}{214} veto algorithm. As the algorithm looks forward in time for BiPo events, and backwards in time for \isotope{Po}{218} events, there are two free parameters representing the log-likelihood thresholds \((\xi_{BiPo}, \xi_{Po})\) that had to be optimised. To this end, electronic recoil data from science run 0 of XENON1T between \(30 \unit{keV}\) and \(70\unit{keV}\) was used~\cite{XENON:2020rca}.

\tikzstyle{block} = [circle, draw, fill=white, 
    text width=3.2em, text centered, rounded corners, minimum size=0.1cm]
\tikzstyle{result} = [draw, fill=white, 
    text width=3.2em, text centered, rounded corners, minimum size=0.1cm]
\tikzstyle{arrow} = [thick,->,>=stealth]

\begin{figure}
    \centering
    \begin{tikzpicture}[scale=1, node distance = 2.5cm, auto]
        \node at (0,0) [block] (o) {};
        \node at (2cm, -2cm) [block] (Pb) {\isotope{Pb}{214}};
        \node at (-2cm, -2cm) [block] (nPb) {$\neg$\isotope{Pb}{214}};
        \node at (1cm, -4cm) [result] (rejPb) {veto};
        \node at (3cm, -4cm) [result] (accPb) {accept};

        \node at (-3cm, -4cm) [result] (nrejPb) {veto};
        \node at (-1cm, -4cm) [result] (naccPb) {accept};

        \draw [arrow] (o) -- node [align=center, anchor=south west]{\(p_{Pb}\)} (Pb);
        \draw [arrow] (o) -- node [align=center, anchor=south east]{\(1-p_{Pb}\)} (nPb);
        \draw [arrow] (Pb) -- node [align=center, anchor=south east]{\(p_{\mathrm{true}}\)} (rejPb);
        \draw [arrow] (Pb) -- node [align=center, anchor=south west]{\(1-p_{\mathrm{true}}\)} (accPb);
        \draw [arrow] (nPb) -- node [align=center, anchor=south east]{\(p_{\mathrm{coinc}}\)} (nrejPb);
        \draw [arrow] (nPb) -- node [align=center, anchor=south west]{\(1-p_{\mathrm{coinc}}\)} (naccPb);
    \end{tikzpicture}
    \caption{Tree diagram describing the probabilities involved in whether an event is vetoed for or not for the aligned direction runs. This tree diagram describes the likelihoods shown in \autoref{eq:vetoed_likelihood} and \autoref{eq:unvetoed_likelihood}. Each event has an energy-dependent probability of being a \isotope{Pb}{214} event ($p_{Pb}$); the probability of vetoing an event then depends on whether said event is a \isotope{Pb}{214} event or not. A higher probability of vetoing a \isotope{Pb}{214} event $(p_{\mathrm{true}})$ and a lower probability of vetoing events that are not \isotope{Pb}{214} $(p_{\mathrm{coinc}})$ indicate better performance.}
    \label{fig:forward_likelihood}
\end{figure}
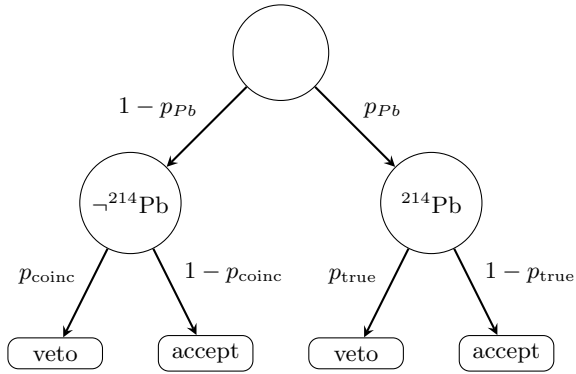

To find the optimal values for these two parameters, the software radon veto was run on the entire dataset twice. For one of the two runs, the velocity field and time directions were reversed, so that due to causality, the \isotope{Pb}{214} candidate cannot be related to the \isotope{Po}{218} or BiPo events. This creates a sample of events that were vetoed purely due to coincidence, allowing for the probability of vetoing an event purely due to coincidence \((p_{\text{coinc}})\) to be profiled. For the aligned-direction run where one searches for BiPo and \isotope{Po}{218} events in the correct directions, the fraction of events that was \isotope{Pb}{214} \((p_{\mathrm{Pb}})\) was determined from a spectral fit from the XENON1T electronic recoil analysis~\cite{XENON:2020rca}. Whether an event gets vetoed depends on both the probability of vetoing a \isotope{Pb}{214} event \((p_{\text{true}})\) and the probability of vetoing an event purely due to coincidence \((p_{\text{coinc}})\), as illustrated in a tree diagram by~\autoref{fig:forward_likelihood}. A likelihood function was thus used to fit \(p_{\text{true}}\) and \(p_{\text{coinc}}\). \(p_{\text{true}}\) and \(p_{\text{coinc}}\) can also be interpreted as the \isotope{Pb}{214} background reduction and the exposure loss, respectively, as the \isotope{Pb}{214} background reduction can be given by the probability of vetoing \isotope{Pb}{214} events, and the exposure loss can be given by the probability of vetoing events that are not \isotope{Pb}{214}, as defined above. The likelihood function for a vetoed event is:

\begin{equation}\label{eq:vetoed_likelihood}
\begin{split}
 \ell_{i, \mathrm{veto}} \left(p_{\text{true}}, p_{\text{coinc}}\right) =& p_{\mathrm{Pb}}(E_i) \times p_{\text{true}} + \\
 &(1-p_{\mathrm{Pb}}(E_i)) \times p_{\text{coinc}}   
 \end{split}
\end{equation}
where \(\ell_{i, \mathrm{veto}} \left(p_{\text{true}}, p_{\text{coinc}}\right)\) is the likelihood for the \(i^\text{th}\) event to be vetoed, \(E_i\) is the energy of the event, and \(p_{\mathrm{Pb}}(E_i)\) is the fraction of events resulting from the decay of \isotope{Pb}{214}, as determined from the XENON1T electronic recoil analysis spectral fit. This can be interpreted as the probability that a given event is \isotope{Pb}{214}, multiplied by the probability of vetoing \isotope{Pb}{214} events, summed with the probability that a given event is not \isotope{Pb}{214}, multiplied by the probability of vetoing events that are not \isotope{Pb}{214}. 

The likelihood function for a candidate event that is not vetoed for the aligned-direction runs is:
\begin{equation}\label{eq:unvetoed_likelihood}
\begin{split}
     \ell_{j, \mathrm{nveto}} \left(p_{\text{true}}, p_{\text{coinc}}\right) =& p_{\mathrm{Pb}}(E_j) \times (1 - p_{\text{true}}) + \\
     & (1-p_{\mathrm{Pb}}(E_i)) \times (1 - p_{\text{coinc}}) \\
     = & 1 - p_{\mathrm{Pb}}(E_j) \times p_{\text{true}} - \\
     &(1-p_{\mathrm{Pb}}(E_j)) \times p_{\text{coinc}}
 \end{split}
\end{equation}
where \(\ell_{j, \mathrm{nveto}} \left(p_{\text{true}}, p_{\text{coinc}}\right)\) is the likelihood for the \(j^\text{th}\) event not vetoed. This can be interpreted as the probability that a given event is \isotope{Pb}{214}, multiplied by the probability of not vetoing \isotope{Pb}{214} events, summed with the probability that a given event is not \isotope{Pb}{214}, multiplied by the probability of not vetoing events that are not \isotope{Pb}{214}. This likelihood can similarly be derived from~\autoref{fig:forward_likelihood}.

\tikzstyle{block} = [circle, draw, fill=white, 
    text width=3.2em, text centered, rounded corners, minimum size=0.1cm]
\tikzstyle{result} = [draw, fill=white, 
    text width=5em, text centered, rounded corners, minimum size=0.1cm]
\tikzstyle{arrow} = [thick,->,>=stealth]

\begin{figure*}[htbp]
    \centering
    \begin{tikzpicture}[scale=1, node distance = 2.5cm, auto]
        \node at (0,0) [block] (o) {\isotope{Pb}{214}};
        \node at (2cm, -2cm) [result] (nrejPo) {not vetoed by \isotope{Po}{218}};
        \node at (-2cm, -2cm) [result] (rejPo) {vetoed by \isotope{Po}{218}};
        \node at (1cm, -4cm) [result] (rejBiPo) {vetoed by BiPo};
        \node at (3cm, -4cm) [result] (nrejBiPo) {not vetoed by BiPo};

        \draw [arrow] (o) -- node [align=center, anchor=south west]{\(1-p_{\mathrm{true, Po}}\)} (nrejPo);
        \draw [arrow] (o) -- node [align=center, anchor=south east]{\(p_{\mathrm{true, Po}}\)} (rejPo);
        \draw [arrow] (nrejPo) -- node [align=center, anchor=south east]{\(p_{\mathrm{true, BiPo}}\)} (rejBiPo);
        \draw [arrow] (nrejPo) -- node [align=center, anchor=south west]{\(1-p_{\mathrm{true, BiPo}}\)} (nrejBiPo);
    \end{tikzpicture}
    \begin{tikzpicture}[scale=1, node distance = 2.5cm, auto]
        \node at (0,0) [block] (o) {$\neg$\isotope{Pb}{214}};
        \node at (2cm, -2cm) [result] (nrejPo) {not vetoed by \isotope{Po}{218}};
        \node at (-2cm, -2cm) [result] (rejPo) {vetoed by \isotope{Po}{218}};
        \node at (1cm, -4cm) [result] (rejBiPo) {vetoed by BiPo};
        \node at (3cm, -4cm) [result] (nrejBiPo) {not vetoed by BiPo};

        \draw [arrow] (o) -- node [align=center, anchor=south west]{\(1-p_{\mathrm{coinc, Po}}\)} (nrejPo);
        \draw [arrow] (o) -- node [align=center, anchor=south east]{\(p_{\mathrm{coinc, Po}}\)} (rejPo);
        \draw [arrow] (nrejPo) -- node [align=center, anchor=south east]{\(p_{\mathrm{coinc, BiPo}}\)} (rejBiPo);
        \draw [arrow] (nrejPo) -- node [align=center, anchor=south west]{\(1-p_{\mathrm{coinc, BiPo}}\)} (nrejBiPo);
    \end{tikzpicture}
    \caption{Tree diagrams describing the probabilities involved in whether \isotope{Pb}{214} events (left) and other events (right) are vetoed, given both \isotope{Po}{218} and BiPo channels. This tree diagram describes the derivation of~\autoref{eq:normalised_sensitivity}. An event is vetoed if the veto is triggered by either the \isotope{Po}{218} or BiPo channels.}
    \label{fig:ptrue_pcoinc}
\end{figure*}
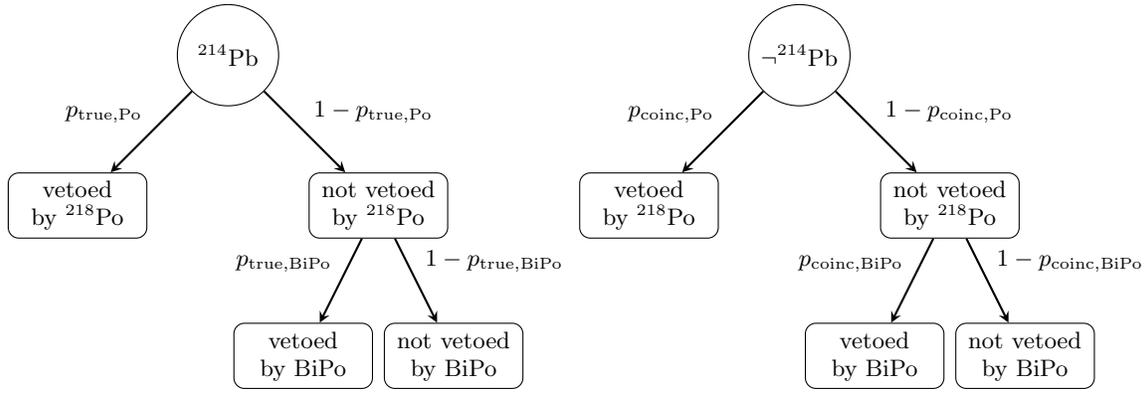

For the reversed runs, the likelihoods were changed to:
\begin{equation}\label{eq:vetoed_likelihood_reversed}
\begin{split}
 \ell_{i, \mathrm{veto}} \left(p_{\text{true}}, p_{\text{coinc}}\right) =& p_{\text{coinc}} \\
 \ell_{j, \mathrm{nveto}} \left(p_{\text{true}}, p_{\text{coinc}}\right) =& 1 - p_{\text{coinc}} \\
 \end{split}
\end{equation}

This was done because for these runs the candidate event cannot be related to any \isotope{Po}{218} or BiPo events found. The likelihood function that was used to fit the probabilities \(p_{\text{true}}\) and \(p_{\text{coinc}}\) was then the sum of the log-likelihoods from each individual candidate event. This was done separately for the \isotope{Po}{218} and BiPo channels, to obtain four probabilities: the probability of vetoing an event due to coincidence via the BiPo channel \((p_{\mathrm{coinc, BiPo}})\), the probability of vetoing an event due to coincidence via the \isotope{Po}{218} channel \((p_{\mathrm{coinc, Po}})\), the probability of vetoing a \isotope{Pb}{214} event via the BiPo channel \((p_{\mathrm{true, BiPo}})\), and the probability of vetoing a \isotope{Pb}{214} event via the \isotope{Po}{218} channel \((p_{\mathrm{true, Po}})\).

If a signal is much smaller than the background, the median asymptotic discovery significance of a counting experiment scales as \({\text{signal}}/{\sqrt{\text{background}}}\)~\cite{Cowan:2010js}. The reduction in signal can be computed as the probability for events that are not \isotope{Pb}{214} to be vetoed. The background reduction can be computed by the reduction in \isotope{Pb}{214} background multiplied by the fraction of the background represented by \isotope{Pb}{214}, summed with the probability for events that are not \isotope{Pb}{214} to be vetoed multiplied by the fraction of background events that are not \isotope{Pb}{214}. An event is vetoed is vetoed by either the \isotope{Po}{218} or the BiPo channel, thus the probabilities for an event to survive each of the two channels can be multiplied; a graphical depiction of these probabilities can be found in \autoref{fig:ptrue_pcoinc}. One can thus compute a normalised sensitivity for a dark matter search:
\begin{equation}\label{eq:normalised_sensitivity}
Z = \frac{\mathrm{signal}}{\sqrt{\mathrm{background}}}    
\end{equation}
where
\begin{equation*}
    \begin{split}
        \mathrm{signal} =& \tilde{p}_{\mathrm{coinc, BiPo}}\times\tilde{p}_{\mathrm{coinc, Po}}, \\
        \mathrm{background} =& 1-(1-\alpha)\times\\
        &\left[1-\tilde{p}_{\mathrm{coinc, BiPo}}\times\tilde{p}_{\mathrm{coinc, Po}}\right]\\
        &- \alpha \left[1-\left(\tilde{p}_{\mathrm{true, BiPo}}\times\tilde{p}_{\mathrm{true, Po}}\right)\right]\\
        \tilde{p}_{\mathrm{coinc, BiPo}} =& 1-p_{\mathrm{coinc, BiPo}}(\xi_{BiPo})\\
        \tilde{p}_{\mathrm{coinc, Po}} =& 1-p_{\mathrm{coinc, Po}}(\xi_{Po})\\
        \tilde{p}_{\mathrm{true, BiPo}} =& 1-p_{\mathrm{true, BiPo}}(\xi_{BiPo})\\
        \tilde{p}_{\mathrm{true, Po}} =& 1-p_{\mathrm{true, Po}}(\xi_{Po})
    \end{split}
\end{equation*}
\(\xi_{Po}\) and \(\xi_{BiPo}\) refer to the likelihood threshold parameters being optimised for the \isotope{Po}{218} and BiPo channels, respectively, and \(\alpha\) refers to the fraction of the background that can be attributed to \isotope{Pb}{214}. This is energy dependent in principle, but is approximated to be a constant \(\alpha=0.8\) for the purposes of this optimisation, as given by the average between \(0 \unit{keV}\) and \(30 \unit{keV}\). A tilde above a parameter, such as $\tilde{p}$ denotes the best fit value of said parameter. This normalised sensitivity is used as a proxy for optimisation of algorithm parameters and evaluation of performance.

Finally, this process was repeated for multiple values of the threshold parameters that govern the veto volume size in the \isotope{Po}{218} and BiPo channels.
% The result of this can be seen in \autoref{fig:data_driven_optimisation}. 
This procedure gives an optimal \isotope{Pb}{214} background reduction of $6.3\%$, and a exposure loss of $1.8\%$, as defined by the signal and background components in~\autoref{eq:normalised_sensitivity}, with optimal thresholds of 9.7 and 9.0 for the \isotope{Po}{218} and BiPo channels respectively. From this, the sensitivity improvement estimated via this procedure is a modest $1.4\%$ in XENON1T. However, as this is an analysis technique, it can still be a cost-effective addition to hardware radon-mitigation efforts, such as the cryogenic distillation system in XENONnT~\cite{Murra:2022mlr}. In addition, as will be shown later in sections \ref{ssec:performance_xent}, much higher performance is possible in systems with lower background radon levels and which have slower convective flows.
% \begin{figure}[htp]
%  \centering
%     \includegraphics[width=0.95\columnwidth]{plots/data_driven_optimisation.pdf}%
%     \caption{Plot of the improvement in sensitivity of XENON1T for a NR dark matter signal at various values of the BiPo and \isotope{Po}{218} point cloud threshold parameters. The colour plot in the background shows the linearly interpolated sensitivity between data points, and the actual data points are indicated with green circles. 1 on the colour scale corresponds to the same sensitivity as what one would achieve without the software radon veto.}\label{fig:data_driven_optimisation}
% \end{figure}

\section{Results and discussion}\label{sec:results}

\subsection{Demonstration of software radon veto}\label{ssec:low_er_demo}

The ER analysis dataset from XENON1T~\cite{XENON:2020rca} was used to demonstrate how this software radon veto would work in practice. For this section, the same data as introduced in \autoref{ssec:optimisation} was used, but with the energy range \([0\unit{keV}, 70\unit{keV}]\) instead. Similarly to \autoref{ssec:optimisation}, both \isotope{Po}{218} and BiPo channels were used to tag events as \isotope{Pb}{214}. The thresholds used are \(\xi_{Po}=9.7\) and \(\xi_{BiPo}=9.0\).

\begin{figure}[htp]
 \centering
    \includegraphics[width=\columnwidth]{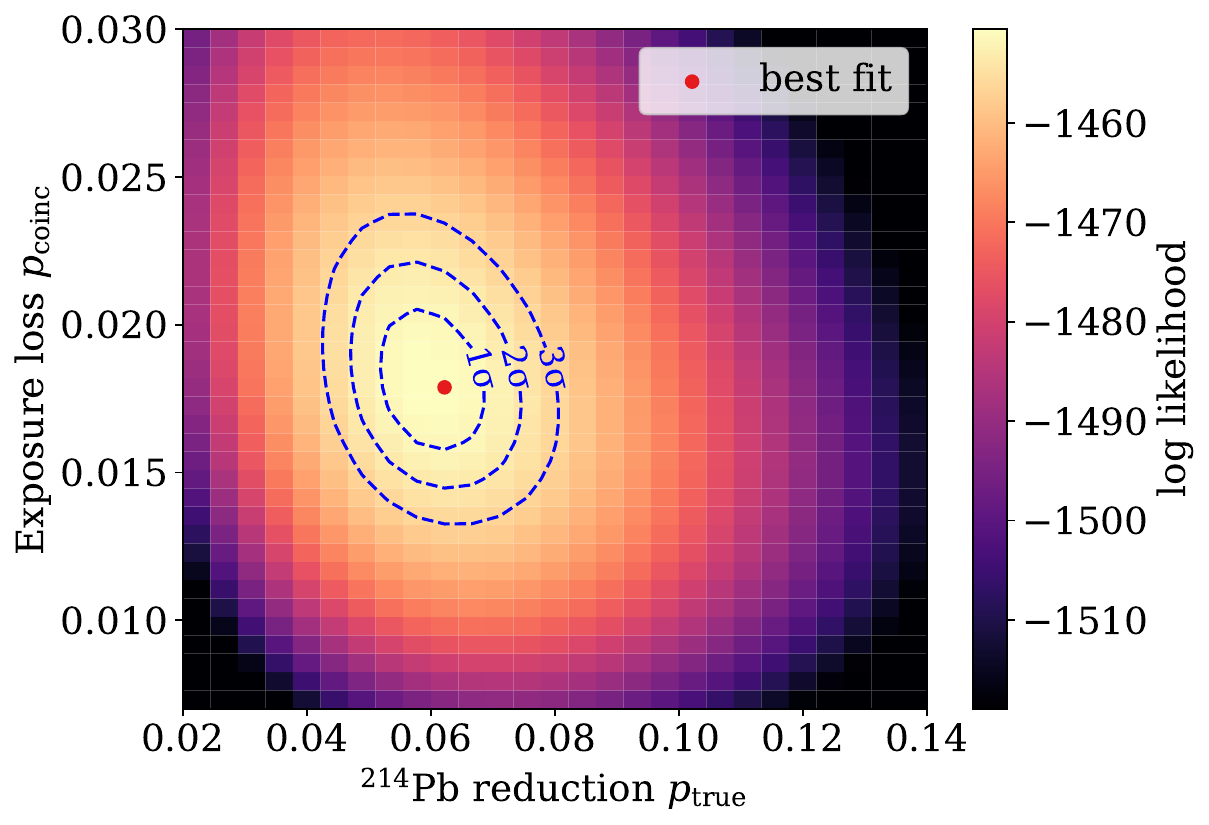}%
    \caption{Likelihood fit using events from the ER analysis dataset and the likelihoods discussed in \autoref{ssec:optimisation}. In this plot, the best fit values of the \isotope{Pb}{214} background reduction \((p_{\text{true}})\) and the exposure loss \((p_{\text{coinc}})\) are marked in red, and the error ellipses are shown in blue.}\label{fig:lowER_likelihood}
\end{figure}

Following this, the likelihoods shown in \autoref{eq:vetoed_likelihood} and \autoref{eq:unvetoed_likelihood} were used to fit \(p_{\text{true}}\) and \(p_{\text{coinc}}\). This is shown in \autoref{fig:lowER_likelihood}. The fit corresponds to an exposure loss of \(1.8\pm 0.2 \%\), and a \(6.2^{+0.4}_{-0.9}\%\) reduction in the \isotope{Pb}{214} background. We can see that, as expected, the final fit of the \isotope{Pb}{214} background reduction \((p_{\text{true}})\) and the exposure loss \((p_{\text{coinc}})\) agrees with the best fit values obtained \autoref{ssec:optimisation}.

\begin{figure}[htp]
 \centering
    \includegraphics[width=\columnwidth]{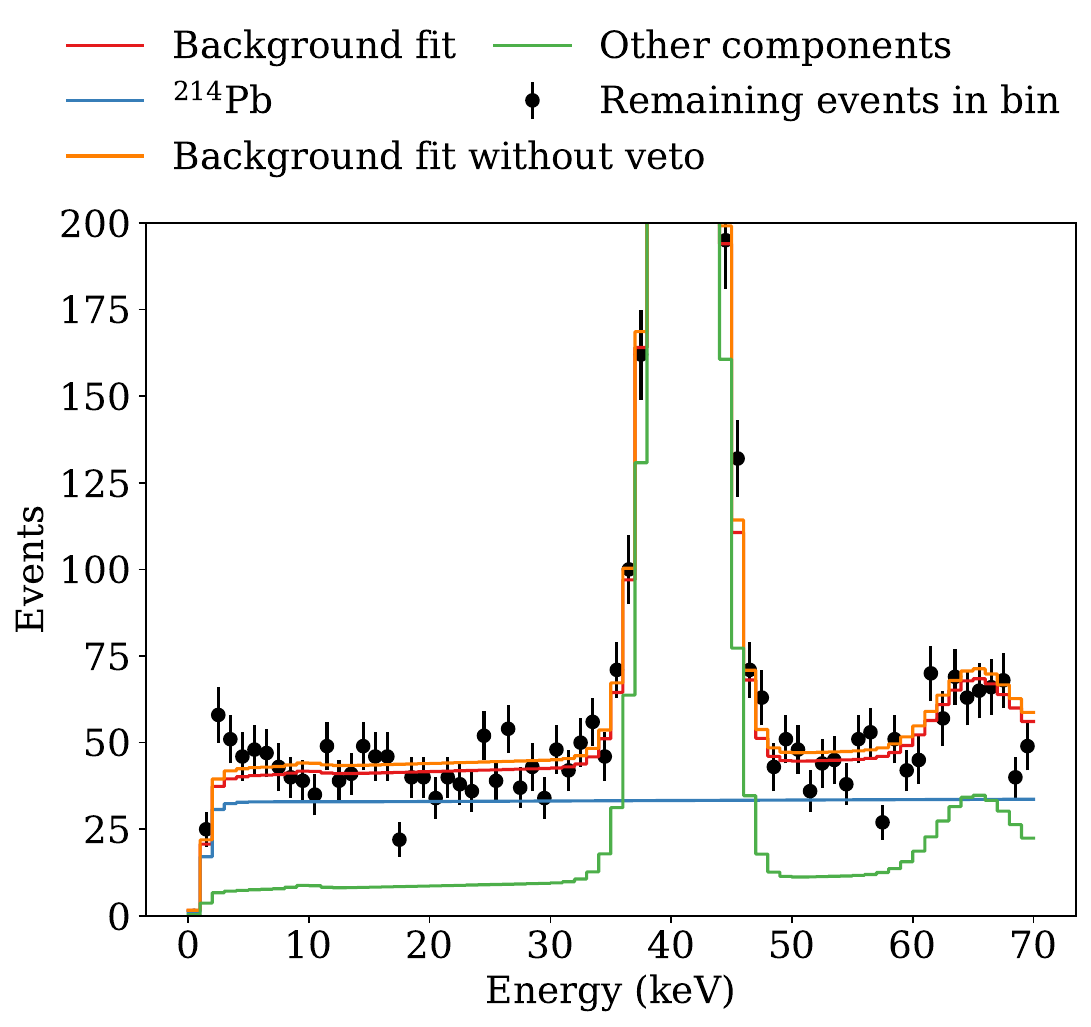}%
    \caption{Spectrum of events remaining after the software radon veto with 1-sigma poisson confidence intervals (black), compared with the expected background spectrum based on the signal-free spectral fit from~\cite{XENON:2020rca} and the inferred values of \(p_{\text{true}}\) and \(p_{\text{coinc}}\) (red). The \isotope{Pb}{214} component is shown in blue, and other background components are shown in green. The peaks at \(42\unit{keV}\) and \(64\) are due to \isotope{Kr}{83m} and \isotope{Xe}{124} decay, respectively~\cite{XENON:2020rca}. The combined fit with all components before applying the software radon veto is also shown in red to indicate the impact of this procedure.}\label{fig:lowER_spectrum}
\end{figure}

The expected background spectrum was then computed by multiplying the components of the background fit from~\cite{XENON:2020rca} that are not from \isotope{Pb}{214} with \(1-p_{\text{coinc}}\), multiplying the \isotope{Pb}{214} background with \(1-p_{\text{true}}\) and summing the two. This is shown in \autoref{fig:lowER_spectrum}. It can be seen that the red line is a good fit for the data.

% In XENON1T, an ER excess was observed with 285 events between \(1\unit{keV}\) and \(7\unit{keV}\)~\cite{XENON:2020rca}, a \(3.3 \sigma\) Poisson fluctuation over the expectation of 232 events. In the unvetoed data and scaled background fit, there are 275 events in the same energy range and an expectation of 220, corresponding to a \(3.6\sigma\) excess according to Poisson statistics, demonstrating a small improvement as expected from~\autoref{ssec:optimisation}. This also affirms that the excess is not related to the \isotope{Pb}{214} background, in agreement with the lack of an excess observed in the XENONnT low-energy electronic recoil search~\cite{XENON:2022ltv}.

\begin{figure}[htp]
 \centering
    \includegraphics[width=\columnwidth]{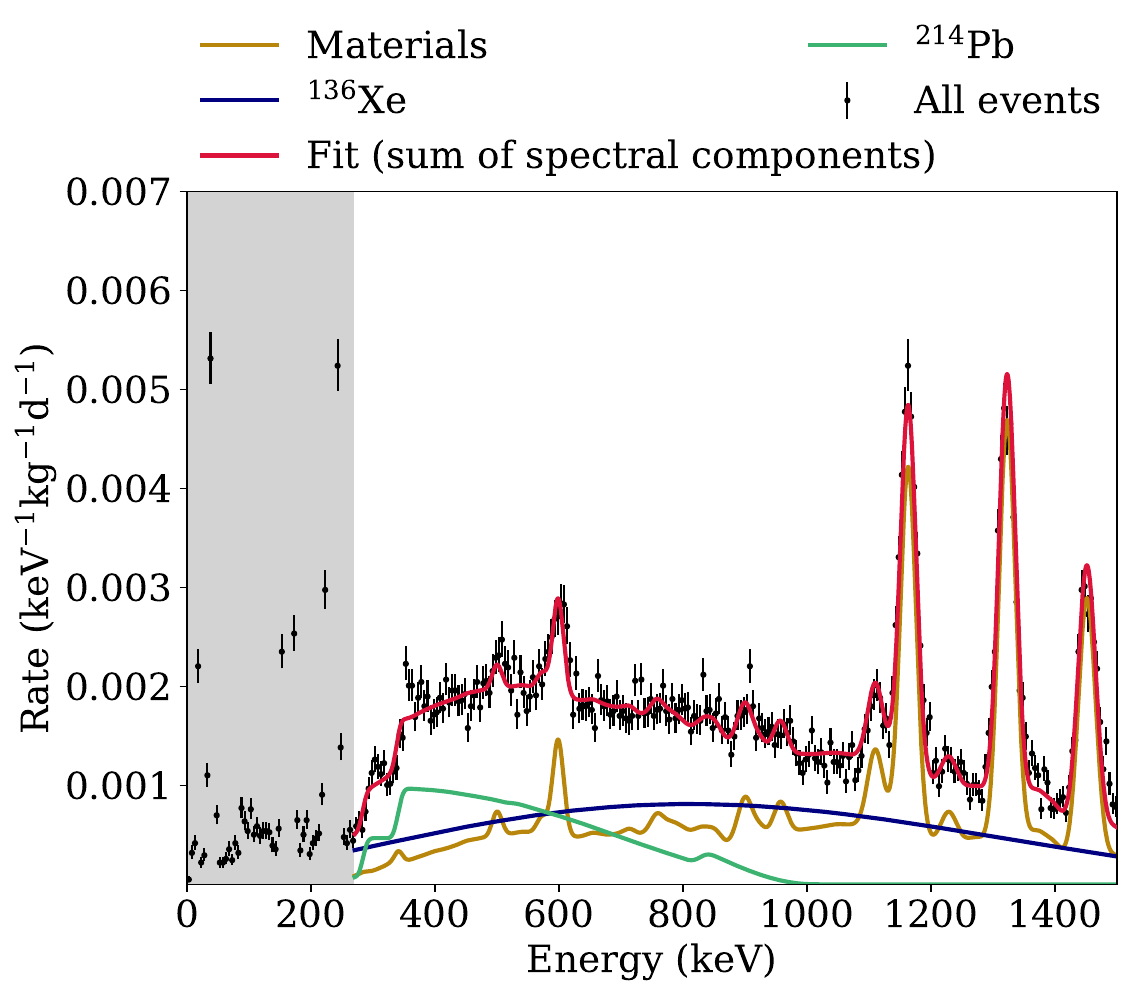}%
    \caption{Spectrum of events in the XENON1T double beta decay dataset~\cite{XENON:2022evz} corresponding to \(22.01 \unit{days}\) of exposure, before the application of the radon veto. Data points with \(5 \unit{keV}\) bins is shown in black. A spectral fit is shown in solid lines, with the summed fit in red. The grey shaded region indicates data that is not used for fitting.}\label{fig:DBD_spectrum}
\end{figure}

We can also demonstrate that this software radon veto indeed selects \isotope{Pb}{214} events by looking at the energy spectrum of vetoed events. To this end, a portion of the fiducialised data from the search for neutrinoless double-beta decays in XENON1T was used~\cite{XENON:2022evz}. This data corresponds to \(22.05\unit{days}\) of exposure with a fiducial mass of \(741 \pm 9 \unit{kg}\). A spectral fit between \(270 \unit{keV}\) and \(2000 \unit{keV}\) includes both spectral features due to \isotope{Bi}{214} excited states at \(295 \unit{keV}\) and \(352 \unit{keV}\), and the beta decay Q-value of \(1018 \unit{keV}\)~\cite{Zhu:2021qss}, but avoids low-energy features in the spectrum from \isotope{Kr}{83m} and \isotope{Xe}{131m}. The selected data with a spectral fit is shown in \autoref{fig:DBD_spectrum}.

The software radon veto was run on the dataset shown in \autoref{fig:DBD_spectrum} with thresholds of \(\xi_{Po}=9.7\) and \(\xi_{BiPo}=9.4\); these differ from those used in~\autoref{ssec:low_er_demo}. These thresholds were picked without an optimisation procedure, but did produce a cleaner sample of \isotope{Pb}{214} decays. 
% This change in the thresholds increases the enhancement factor, as defined by $p_{\text{true}}/p_{\text{coinc}}$, from $3.3^{+0.4}_{-0.5}$ to $6.4^{+0.9}_{-0.7}$. 
Following that, the same procedure used above for the ER analysis was used to fit \(p_{\text{true}}\) and \(p_{\text{coinc}}\); however, here events that are tagged as \isotope{Pb}{214} are examined instead. Thus, the components of the spectral fit that are not \isotope{Pb}{214} were multiplied with \(p_{\text{coinc}}\), and the \isotope{Pb}{214} component was multiplied by \(p_{\text{true}}\).

\begin{figure}[htp]
 \centering
    \includegraphics[width=0.94\columnwidth]{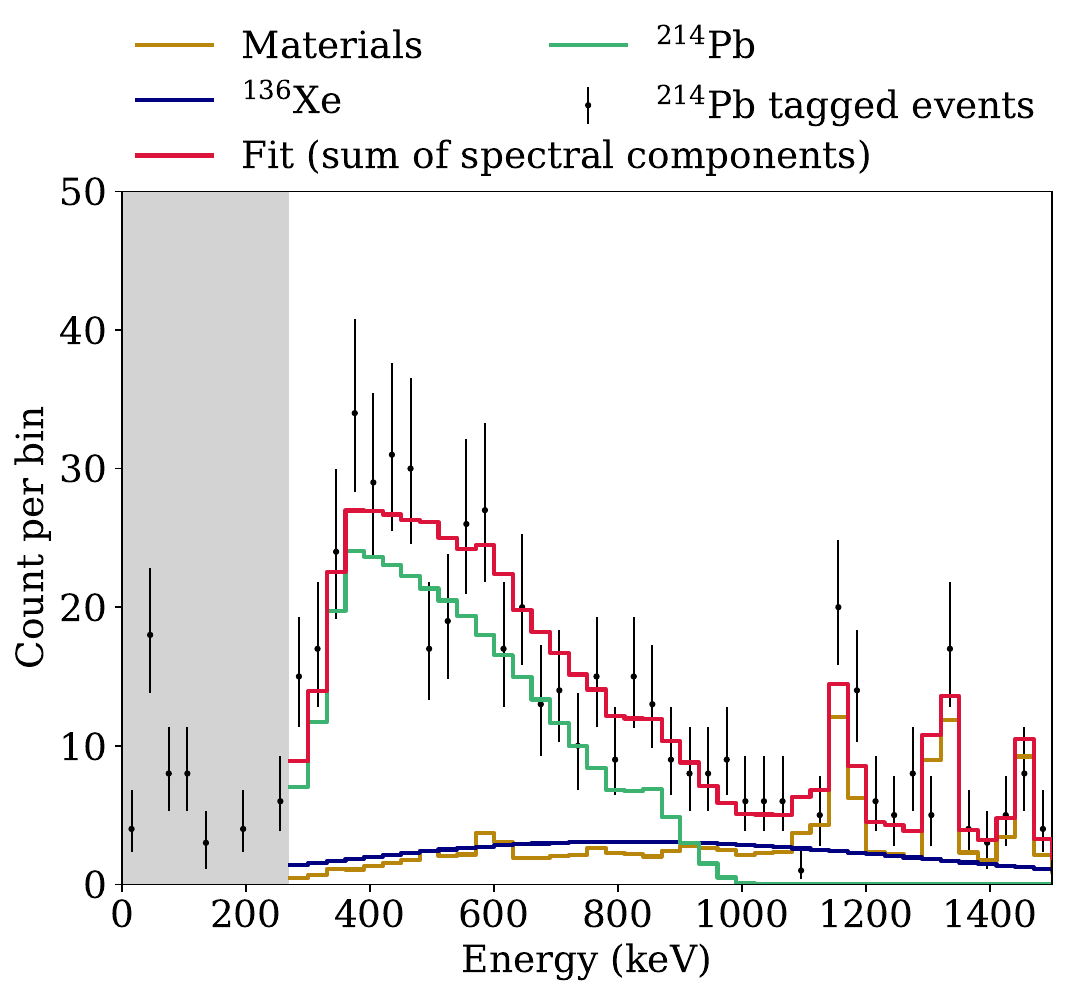}%
    \caption{The spectrum of events in the population of events tagged as \isotope{Pb}{214}. It can be seen both from the fit and from the shape of the spectrum that the \isotope{Pb}{214} fraction is greatly enhanced in the vetoed sample. The grey shaded region indicates data that is not used for fitting.}\label{fig:DBD_spectrum_vetoed}
\end{figure}

This result is shown in \autoref{fig:DBD_spectrum_vetoed}. It can be seen from the difference in spectral shape between \autoref{fig:DBD_spectrum} and \autoref{fig:DBD_spectrum_vetoed} that the tagged population is indeed dominated by the decay of \isotope{Pb}{214}. It is also possible to identify relevant spectral features at \(295 \unit{keV}\) and \(352 \unit{keV}\), as well as the Q-value of \(1018 \unit{keV}\), though there is insufficient statistics to resolve the two steps at \(295 \unit{keV}\) and \(352 \unit{keV}\) separately~\cite{Zhu:2021qss}. In particular, the \isotope{Pb}{214} decay endpoint can be clearly identified in the tagged population in \autoref{fig:DBD_spectrum_vetoed}, but not in the full dataset shown in \autoref{fig:DBD_spectrum}.

\subsection{Example of a recovered decay chain}

\begin{figure*}[htp]
 \centering
    \includegraphics[width=0.9\columnwidth]{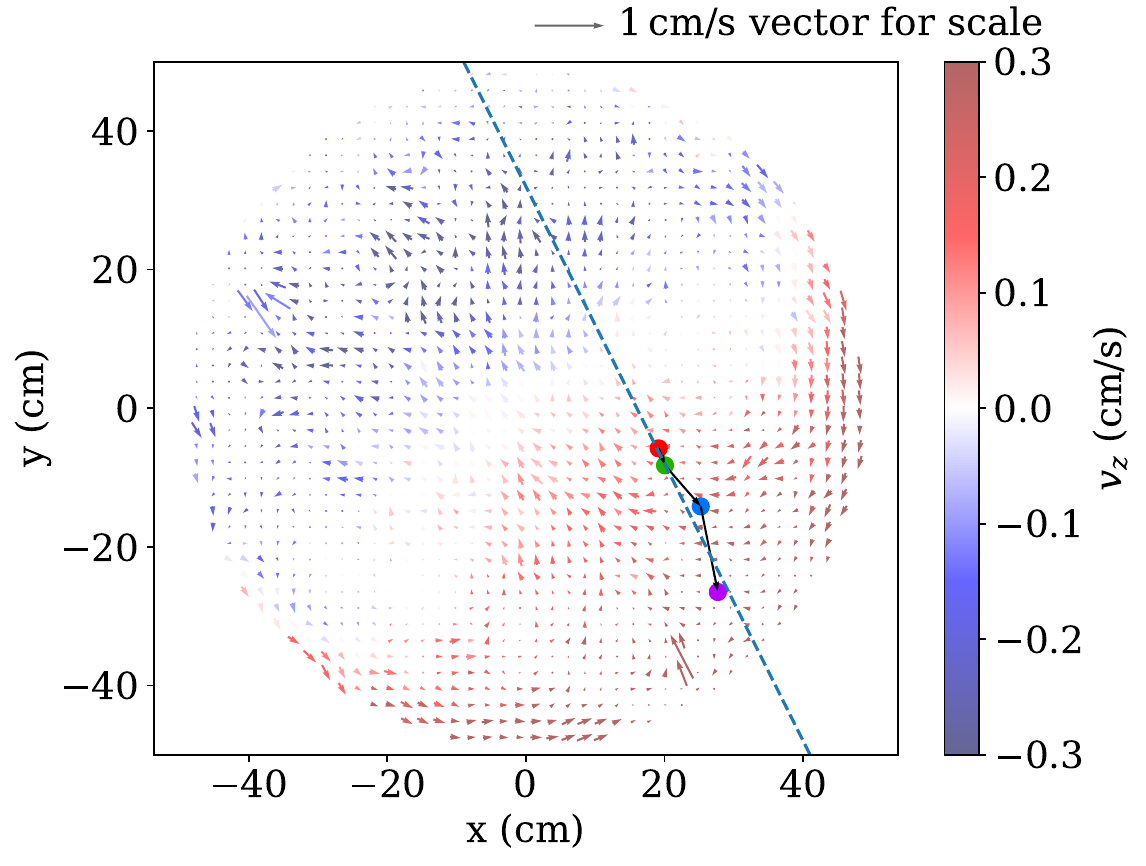}%
    \qquad
    \includegraphics[width=0.93\columnwidth]{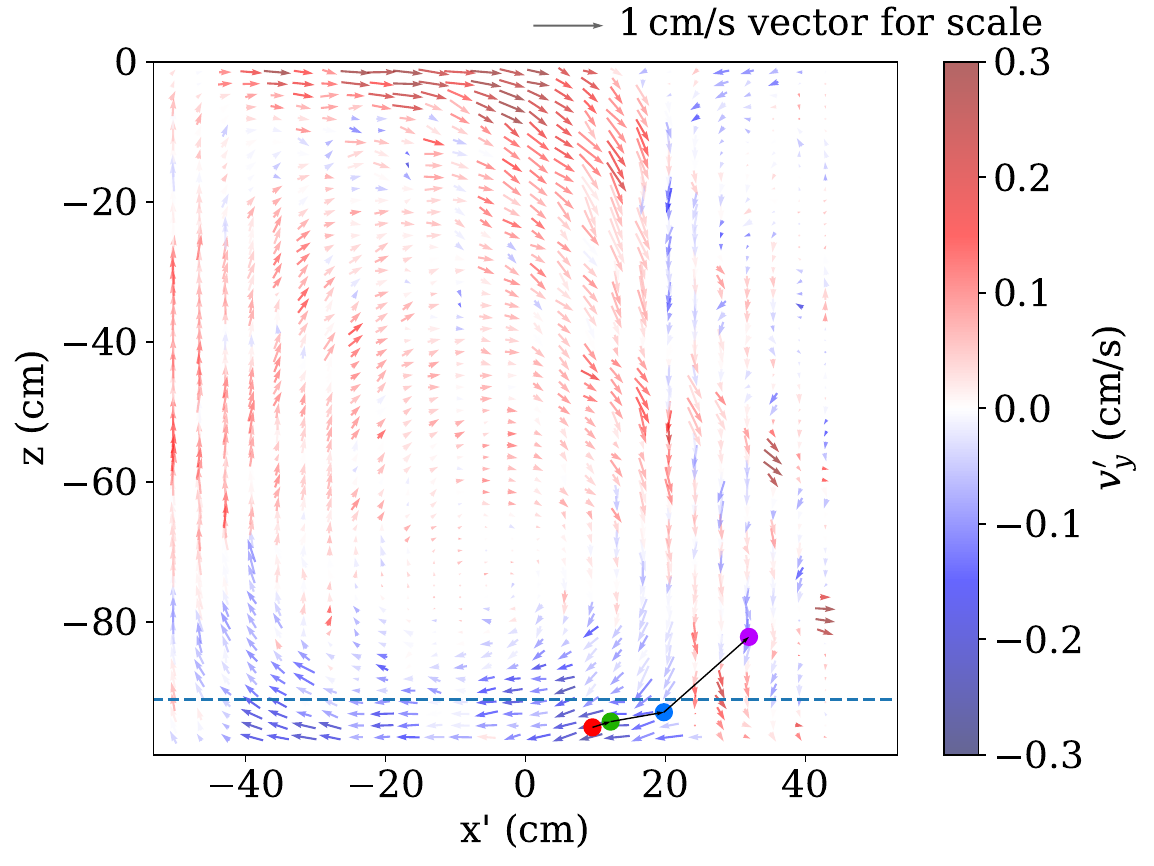}%
    \caption{The positions of a \isotope{Rn}{222} event (red), a \isotope{Po}{218} event (green), a \isotope{Pb}{214} event (blue), and a BiPo event (purple) are shown here, overlaid on top of the velocity field in the detector. The top view is shown on the left, and a side view is shown on the right. The blue dotted line on the top and side views are used to show the slices taken to create the side and top views, respectively.}\label{fig:full_chain_plots}
\end{figure*}

A reconstructed example of the portion of the decay chain that is used for the software radon veto is shown in this section (compare \autoref{fig:rn222_decay_chain}). The software radon veto was used to find the \isotope{Po}{218} and BiPo events from the \isotope{Pb}{214} event. Tagging a \isotope{Pb}{214} event only requires matching either a \isotope{Po}{218} or BiPo event, however, in the chosen example, both \isotope{Po}{218} and BiPo events were found. The \isotope{Rn}{222} event related to the \isotope{Po}{218} was then found via the matching procedure shown in \autoref{ssec:convection_measurement}. The four identified events can be seen in \autoref{fig:full_chain_plots}, laid over the same velocity field shown in \autoref{ssec:v_field_filtering}.
% , with waveforms shown in \autoref{fig:full_chain_waveforms}.
It can be seen that the events propagate along the velocity field. 

% All the waveforms have a single large S1 and a single large S2, except for the BiPo event. The BiPo event is made up of two decays, thus has at least two S1s and S2s. In this case, there are more than two S2s because the beta decay of \isotope{Bi}{214} has only a \(19.2 \%\) chance of decaying directly to the ground state; more typically decaying to short-lived excited states and emitting one or more gamma rays in addition to the beta particle~\cite{Zhu:2021qss}. The large S1s and S2s, corresponding to the relevant events, are also followed by single and few-electron signals largely due to photoionisation within the time window shown~\cite{XENONCollaborationSS:2021sgk}. These waveforms match our expectations for the respective event types.

% \begin{figure}[htp]
%  \centering
%     \includegraphics[width=\columnwidth]{plots/full_chain_waveforms.pdf}%
%     \caption{Waveforms of the events in \autoref{fig:full_chain_plots}. From top to bottom, the waveforms correspond to the identified \isotope{Rn}{222} (red), \isotope{Po}{218} (green), \isotope{Pb}{214} (blue), and BiPo (purple) events. Insets show the S1 waveforms with units identical to the main plots.}\label{fig:full_chain_waveforms}
% \end{figure}

\subsection{Projection of performance in XENONnT and future TPCs}\label{ssec:performance_xent}

To project the performance of this technique to XENONnT, a model of how the software radon veto performs under various conditions must be constructed. Such a model needs to be constructed because the analyses shown in earlier sections is data-driven and done using XENON1T data; repeating the full analysis on XENONnT data has not been done yet as of this publication. For each channel, the probability of incorrectly vetoing an event that is not \isotope{Pb}{214} is simply given by the size of the point cloud used to construct the veto volume, multiplied by the rate of \isotope{Po}{218} or BiPo events. When propagating particles along a 3-dimensional flow, chaotic mixing is expected to occur. This makes the point cloud size diverge exponentially with time~\cite{speetjens2021lagrangian}. As such, the growth of the point cloud volume can be modelled with a Lyapunov exponent -- the characteristic exponential divergence of two close trajectories~\cite{taylor2005classical}. The probability of incorrectly vetoing an event that is not \isotope{Pb}{214} as a function of the time the point cloud is propagated for is given by:
\begin{equation}\label{eq:p_coinc_model}
\begin{split}
    p_{\text{coinc}}(t) &= A \cdot C \int^{t}_0 e^{v \lambda \tau} d\tau\\
    &= A \cdot C \frac{e^{v \lambda t} - 1}{v \lambda} 
\end{split}
\end{equation}
where \(A\) is the activity of \isotope{Po}{218} or BiPo events, depending on the channel being modelled, \(v\) is the convection speed, \(t\) is the time the point cloud is being propagated, and \(C\) and \(\lambda\) are fitting constants. 

The probability of correctly vetoing an event that is \isotope{Pb}{214}, on the other hand, can be modelled with the exponential decay of the radioactive species, multiplied by the probability of there being a correctly reconstructed \isotope{Po}{218} alpha or BiPo event in the detector, \(p_{\text{branch}}\). As the efficiency of detecting alphas is high, the probability for the $^{218}\text{Po}$ channel is approximated as \(p_{\text{branch, Po}} = 1\). The probability for the BiPo channel has to account for the effect of plate-out onto surfaces in the detector~\cite{Bruenner:2020arp} and less efficient selections, and as such is taken to be the ratio of the BiPo rate as measured using fully-reconstructed BiPo events in the XENON1T detector and the rate of \(^{214}\text{Pb}\) events from the search of dark matter in the electronic recoil channel~\cite{XENON:2020rca} as \(p_{\text{branch, BiPo}} = 0.25\). As this includes selection efficiencies and plate-out, this number might change between detectors, but is kept constant here to estimate the XENONnT performance. The probability of correctly vetoing an event that is \isotope{Pb}{214} is thus given by:

\begin{equation}\label{eq:p_true_model}
\begin{split}
    p_{\text{true}}(t) &= p_{\text{branch}} \lambda_{\text{decay}} \int^{t}_0 e^{ - \lambda_{\text{decay}} \tau} d\tau\\
    &= p_{\text{branch}} \left(1 - e^{-\lambda_{\text{decay}} t}\right)
\end{split}
\end{equation}
where \(\lambda_{\text{decay}}\) is the decay constant of the radioactive species relevant to the specific channel, \(p_{\text{branch}}\) is the multiplicative factor stemming from selection efficiencies and plate out as detailed above, and \(t\) is the time the point cloud is being propagated. Equations~\eqref{eq:p_coinc_model} and~\eqref{eq:p_true_model} can then be combined to eliminate the time variable and produce
 \begin{equation}\label{eq:model_fit_eqn}
 p_{\text{coinc}} = \frac{A \cdot C}{v \lambda}\left(\left(1-\frac{p_{\text{true}}}{p_{\text{branch}}}\right)^{-\frac{v \lambda}{\lambda_{\text{decay}}}}-1\right).
 \end{equation}

\begin{figure}[htp]
 \centering
    \includegraphics[width=\columnwidth]{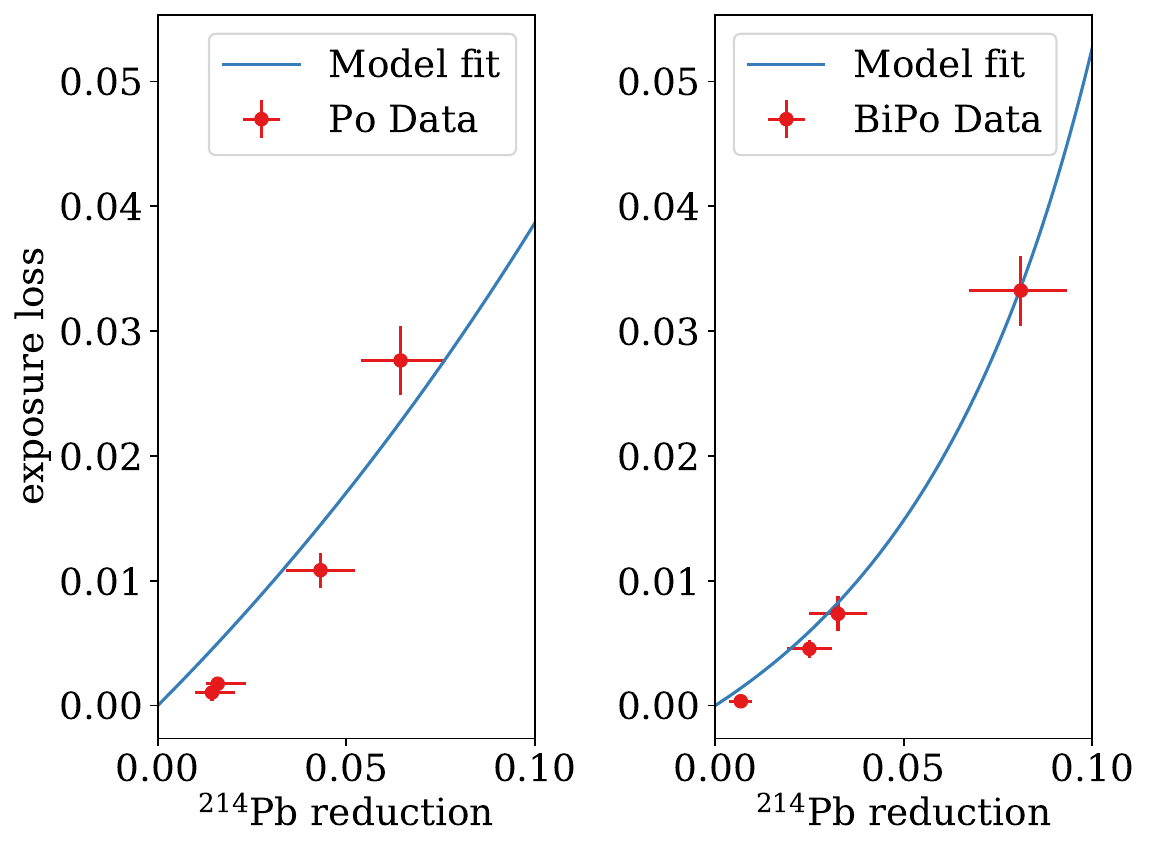}%
    \caption{The exposure loss versus the \isotope{Pb}{214} background reduction from the \isotope{Po}{218} (left) and BiPo (right) channels. The left and right plots correspond to a simultaneous fit on both datasets; the reason why the curve looks different in the two plots is due to the different half lives, and the different probability of there being a correctly reconstructed \isotope{Po}{218} alpha or BiPo event \((p_\text{branch})\).}\label{fig:performance_model_fit}
\end{figure}

With \autoref{eq:model_fit_eqn}, there are only two free parameters, \(C\) and \(\lambda\). These two free parameters can be fit by running the software radon veto on XENON1T data with different veto volumes using both \isotope{Po}{218} and BiPo channels. The resultant values from both the channels, $p_{\mathrm{true, Po}}$, $p_{\mathrm{true, BiPo}}$, $p_{\mathrm{coinc, Po}}$, and $p_{\mathrm{coinc, BiPo}}$, are used to fit the values of the fitting constants, as shown in \autoref{fig:performance_model_fit}. This is done in a single fit, thus the fit procedure only produces one value each of \(C\) and \(\lambda\). These data points differ from the exposure loss and \isotope{Pb}{214} reduction values in \autoref{fig:lowER_likelihood} as the values from the \isotope{Po}{218} and BiPo channels are presented separately.

The veto volume is parameterised by threshold parameters that attempt to find veto volumes with the highest probability content instead of using a simple time cut-off for how long to propagate the point cloud; that is, outlying points in a point cloud might be propagated for shorter amounts of time than points that are central to the point cloud. The extrapolation to XENONnT and future TPCs considers a constant integration time for each point cloud and is thus approximate. However, as can be seen in \autoref{fig:performance_model_fit}, it fits XENON1T data quite well when fit simultaneously on both the \isotope{Po}{218} and BiPo channels. 

\begin{table*}[htp]
    \centering
    \begin{tabular}{c c c c}
        \hline
        Convection speed (cm/s) \qquad & Sensitivity improvement  \qquad & \isotope{Pb}{214} background reduction \qquad & Exposure loss \qquad\\
        \(v_{\mathrm{convection}}\) & \(Z_{\mathrm{optim}}-1\) & \(1-b_{Pb}\) & \(1-s\)\\
        \hline
        0.8 & \(4.3\%\) & \(25\%\) & \(3.5\%\) \\
        0.4 & \(7.6\%\) & \(41\%\) & \(5.9\%\) \\
        0.2 & \(12\%\) & \(59\%\) & \(8.8\%\) \\
        0.1 & \(17\%\) & \(75\%\) & \(11\%\) \\
        \hline
    \end{tabular}
    \caption{Table showing the estimated optimal improvement in sensitivity \((Z_{\mathrm{optim}}-1)\), at various scenarios of convection speed \((v_{\mathrm{convection}})\) in XENONnT, together with the reduction in \isotope{Pb}{214} background \((1-b_{Pb})\) and the exposure loss \((1-s)\) at the stated optimal sensitivity improvement.}
    \label{tab:nT_projection}
\end{table*}

To project the performance of the software radon veto in XENONnT, the fit parameters from above are kept the same, but the activities are scaled down. The \isotope{Po}{218} activity in XENONnT is measured to be \(1.691 \pm 0.006_{\text{stat}} \pm 0.072_{\text{sys}} \unit{\mu Bq/kg}\), and the \isotope{Pb}{214} activity is measured to be \(1.31 \pm 0.17_{\text{stat}} \unit{\mu Bq/kg}\) in XENONnT~\cite{XENONCollaboration:2022kmb}. It should be noted that this is the XENONnT Science Run 0 radon level, and could be further lowered in future science runs depending on the mode of operation of the radon removal system~\cite{Murra:2022mlr}. The ratio between \isotope{Po}{218} and fully-reconstructed BiPo activities is kept the same from XENON1T. Due to the lower \isotope{Pb}{214} background, the fraction of the background attributed to \isotope{Pb}{214} is estimated to be \(\alpha=0.5\) here. The projected performance for various convection speeds, optimised for normalised sensitivity as defined in \autoref{eq:normalised_sensitivity}, is shown in \autoref{tab:nT_projection}.

\begin{figure}[htp]
 \centering
    \includegraphics[width=\columnwidth]{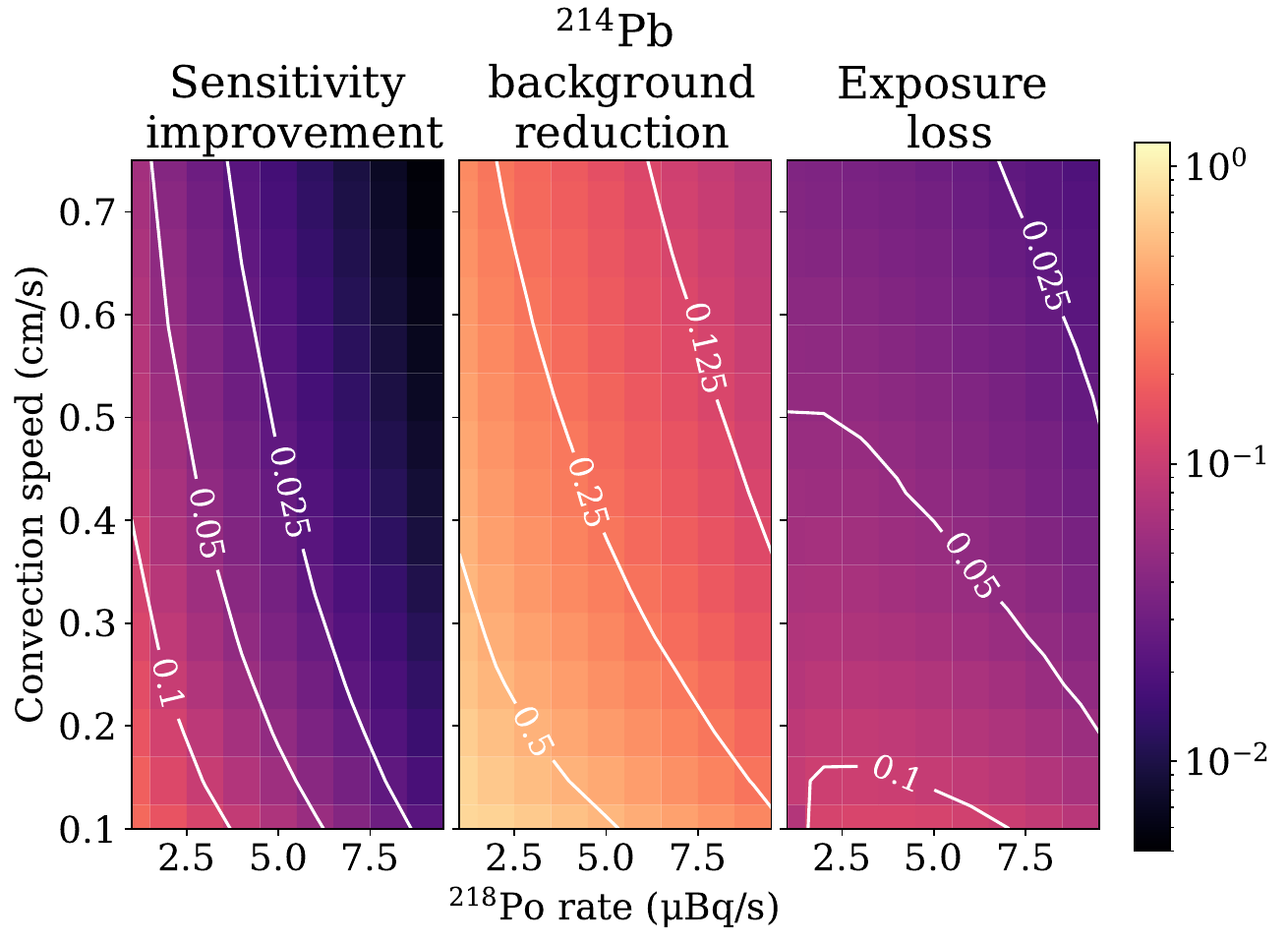}%
    \caption{Estimated optimal improvement in sensitivity \((Z_{\mathrm{optim}}-1)\) (left), reduction in \isotope{Pb}{214} background \((1-b_{Pb})\) (center), and exposure loss \((1-s)\), as a function of \isotope{Po}{218} activity and convection speed. The ratio between \isotope{Po}{218} and \isotope{Pb}{214} activities is kept at a constant $1.691/1.31$, based on~\cite{XENONCollaboration:2022kmb}. Contours for specific values of each panel are shown in white. It can be seen that we can expect significantly improved background reduction at lower activities and convection speeds.}\label{fig:3D_nT_proj}
\end{figure}

XENONnT is a larger detector than XENON1T; hence, due to considerations discussed in \autoref{ssec:convection_measurement} should be expected to have much lower convection speeds. However, we consider higher convection velocity conditions as well due to the introduction of liquid xenon recirculation, which may affect the convective flow in the TPC. As can be seen, due to the reduced radon level in XENONnT, the background reduction is improved greatly over XENON1T in all of the considered convection speed scenarios. These results can also be seen in~\autoref{fig:3D_nT_proj}.

% \subsection{Projection of performance in diffusion-limited regime}\label{ssec:projection_diffusion}

% As shown in \autoref{ssec:v_rms}, for large dual-phase TPCs, such as DARWIN/XLZD~\cite{DARWIN:2016hyl, Aalbers:2022dzr}, there might be insufficient heat flux to induce convection. 
% In such a situation, the movement of daughter nuclides after a radioactive decay becomes dominated by ion-drift~\cite{EXO-200:2015ura}. 
% It is noted in~\cite{EXO-200:2015ura} that the measured diffusion constant is significantly larger than what one might expect from true diffusion of ions in liquid xenon, and that the measured diffusion likely represents the effect of small-scale fluid flow. It is thus unclear if a true diffusion-limited regime can be achieved in any planned noble-liquid TPCs. In this section, the performance of software tagging of radon-chain backgrounds under both scenarios is discussed. 
Future large dual-phase TPCs might also not have a convective flow; in such a situation, the movement of daughter nuclides after a radioactive decay becomes dominated by ion-drift~\cite{EXO-200:2015ura}.
Here, we consider the performance of software tagging of radon-chain backgrounds in the limiting case of this regime, where any stochastic motion is entirely due to diffusion. For simplicity, and because of the unknown effects of plate-out and BiPo reconstruction in future detectors, only the \isotope{Po}{218} channel is considered here, resulting in a conservative estimate of the algorithm's performance.

A simple analytic model can be used to estimate the performance of software tagging in the true diffusion-limited regime. The probability density function of the displacement of a particle diffusing in one dimension is given by the 1D diffusion equation~\cite{einstein1956investigations}
\begin{equation}\label{eq:diffusion_equation}
    \frac{\partial \rho_x}{\partial t} = D \frac{\partial^2 \rho_x}{\partial x^2}.
\end{equation}
Using the 1D diffusion equation leads to no loss of generality because the distribution of displacement of a diffusing point is independent in different orthogonal axes. 

The solution to \autoref{eq:diffusion_equation} with an initial Dirac delta function, \(\delta(x)\), corresponding to the known position of the original particle, is a normal distribution with \(\mu=0\) and \(\sigma^2=2 D t\); in 3D, this corresponds to a spherical normal distribution with \(\sigma_x = \sigma_y = \sigma_z = \sqrt{2 D t}\). It can be noted here that these are also the Green's function of the 1D and 3D heat equations, respectively, as the isotropic diffusion equation is the heat equation~\cite{Polianin:alma99169543790201081}.

% However, the veto volume cannot be directly determined from this normal distribution. This is because this distribution does not take into account the position reconstruction uncertainty. This has to be introduced twice, for both the candidate event that is being tagged, and for the alphas. The position reconstruction uncertainty will depend on the performance of future detectors. However, as the sum of normally-distributed random variables simply involves summing the variances, the expected performance can be expressed in terms of the position reconstruction uncertainties.

% The veto volume \(V(t)\), after time \(t\), can then be given by the volume of an ellipsoid, where the semi-major axes are given by a multiple \(n\) of the standard deviation:
% \begin{equation}\label{eq:diffusion_veto_vol}
%     V(t) \approx \frac{4}{3} \pi n^3 \left(\sigma_{\mathrm{posrec, r}}^2 + 2 D t\right)\sqrt{\sigma_{\mathrm{posrec, z}}^2 + 2 D t}
% \end{equation}
% This multiple \(n\) is a free parameter that determines the size of the veto volume, and hence controls the tradeoff between background removal and exposure loss. A larger \(n\) corresponds to a bigger veto volume, which would remove more background but also result in a greater loss of exposure.

The true diffusion constant can be estimated using Einstein's relation~\cite{Richert:PhysRevLett.63.547} and the mobility of \(\mu = 0.219 \pm 0.004 \unit{cm^2/(kV \, s)}\) as measured by EXO-200~\cite{EXO-200:2015ura}:
\begin{equation}\label{eq:true_diffusion}
    \begin{split}
    D &= \frac{\mu k_b T}{q}\\
    &= \frac{k_b 0.219 \unit{cm^2/(kV \, s)} 170 \unit{K}}{q_e}\\
    &\approx 3.2 \times 10^{-6} \unit{cm^2/s}
    \end{split}
\end{equation}

With the diffusion constant shown in \autoref{eq:true_diffusion}, the daughter of a \isotope{Po}{218} decay would diffuse approximately \(\sqrt{3\times2\times D \times \left(5 \times 27.06 \unit{min}\right)} \approx 0.4 \unit{cm}\) in 5 half-lives. An activity of \(1.7 \unit{\mu Bq/kg}\) as achieved in XENONnT SR0~\cite{XENONCollaboration:2022kmb}, and a liquid xenon density of \(\sim 3\unit{g/cm^3}\)~\cite{Aprile:2009dv} corresponds to an activity per unit volume of \(4.6\times10^{-9}\unit{Bq/cm^3}\), or 0.4 decays per \(10 \unit{L}\) per day. As \(1 \unit{L} = (10\unit{cm})^3\), this implies that as long as the position reconstruction uncertainty remains significantly below \(10 \unit{cm}\), decays would be essentially spatially isolated without fluid flows, and hence one can reject radon-chain backgrounds with a tagging efficiency of near-unity.

\subsection{Application to \texorpdfstring{\isotope{Xe}{137}}{Xe-137}}
The decay of cosmogenic \isotope{Xe}{137} is expected to be a major background in the search for \(0\nu\beta\beta\) decay in \isotope{Xe}{136} in XENONnT~\cite{XENON:2022evz}, and next-generation liquid xenon TPCs~\cite{Aalbers:2022dzr, Baudis:2024jnk, DARWIN:2016hyl}. \isotope{Xe}{137} is produced due to the capture of muon-induced neutrons or radiogenic neutrons by \isotope{Xe}{136}~\cite{XENON:2022evz, LZ:2019qdm, Aalbers:2022dzr}, and subsequently undergoes beta-decay to \isotope{Cs}{137}, as shown in~\autoref{fig:xe_136_capture}. In this section, we focus on the \isotope{Xe}{137} background arising from muon-induced neutrons.
\tikzstyle{block} = [rectangle, draw, fill=white, 
    text width=2.4cm, text centered, rounded corners, minimum height=1cm]
\tikzstyle{arrow} = [thick,->,>=stealth]

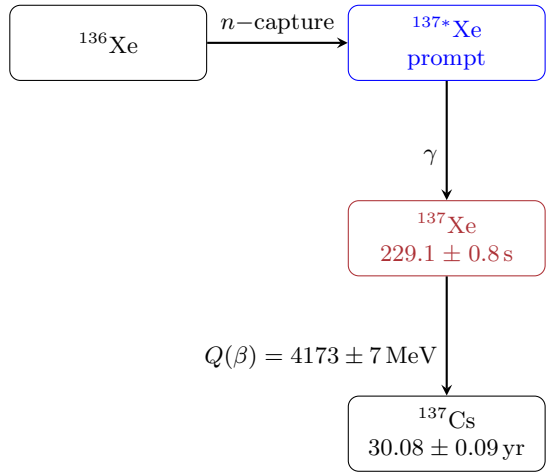
\begin{figure}[htp]
    \centering
    \begin{tikzpicture}[scale=1, node distance =1.5cm and 1.5cm, auto]
        \node at (0,0) [block] (Xe136) {\isotope{Xe}{136}};
        \node at (4.5,0) [block, text=blue, draw=blue] (Xe137s) {\isotope{Xe}{137*}\\prompt};
        \node at (4.5,-2.6cm) [block, text=Maroon, draw=Maroon] (Xe137) {\isotope{Xe}{137}\\\(229.1\pm0.8\unit{s}\)};
        \node at (4.5,-5.2cm) [block] (Cs137) {\isotope{Cs}{137}\\\(30.08\pm0.09\unit{yr}\)};

        \draw [arrow] (Xe136) -- node [align=right, anchor=south]{\(n-\)capture} (Xe137s);
        \draw [arrow] (Xe137s) -- node [align=right, anchor=north east]{\(\gamma\)} (Xe137);
        \draw [arrow] (Xe137) -- node [align=left, anchor=north east]{\(Q(\beta) = 4173\pm7 \unit{MeV}\)} (Cs137);
    \end{tikzpicture}
    \caption{Neutron capture of \isotope{Xe}{136} and subsequent decay of \isotope{Xe}{137}. Data regarding the decay of \isotope{Xe}{137} and \isotope{Cs}{137} retrieved using the NNDC ENSDF, with original data from Nuclear Data Sheets~\cite{Browne:2007wqt}. The isotope the decays to produce the relevant background, \isotope{Xe}{137}, is coloured red, whereas the excited state which produces the gamma events that are used for the tagging of the \isotope{Xe}{137} background are coloured blue.}
    \label{fig:xe_136_capture}
\end{figure}

A similar methodology to \autoref{ssec:performance_xent} can be used to estimate the performance of a \isotope{Xe}{137} veto; however, this estimate is more speculative. This is because the performance of such a veto would rely on the reconstruction of neutron-capture gammas and a detailed analysis to search of these neutron-capture events has not been done in this study. These neutron-captures gammas represent the progenitor events. Point clouds generated at the position of \isotope{Xe}{137} decay candidates are thus used to look for these neutron-capture events, which should appear as ER events that are coincident with muon veto triggers.

The relationship between \(p_\text{true}\) and \(p_\text{coinc}\) can be derived from \autoref{eq:model_fit_eqn}. However, the fit parameters from \autoref{ssec:performance_xent} have to be adapted for this study. The initial point cloud has to be much bigger, because the uncertainty on the true location of the neutron capture is not dominated by position reconstruction uncertainties, but by the mean-free-path of gammas. In the absence of a detailed analysis, the minimum attenuation between \(10^{-2} \unit{MeV}\) and \(10^{1} \unit{MeV}\) is conservatively applied. This is \(0.036 \unit{cm^2/g}\) according to the XCOM database~\cite{Berger:148746}, corresponding to a maximum mean-free-path of \(9.8 \unit{cm}\). Thus, to account for this, the fit parameter \(C\) in \autoref{eq:p_coinc_model} which should scale with the initial point cloud size, is divided by the position reconstruction uncertainty volume, and multiplied by the volume of a sphere with a radius of \(9.8 \unit{cm}\) in liquid xenon. Further, the half-life of \isotope{Xe}{137}, which is \(229.1 \pm 0.8 \unit{s}\)~\cite{Browne:2007wqt}, is applied.

\begin{figure}[htp]
 \centering
    \includegraphics[width=\columnwidth]{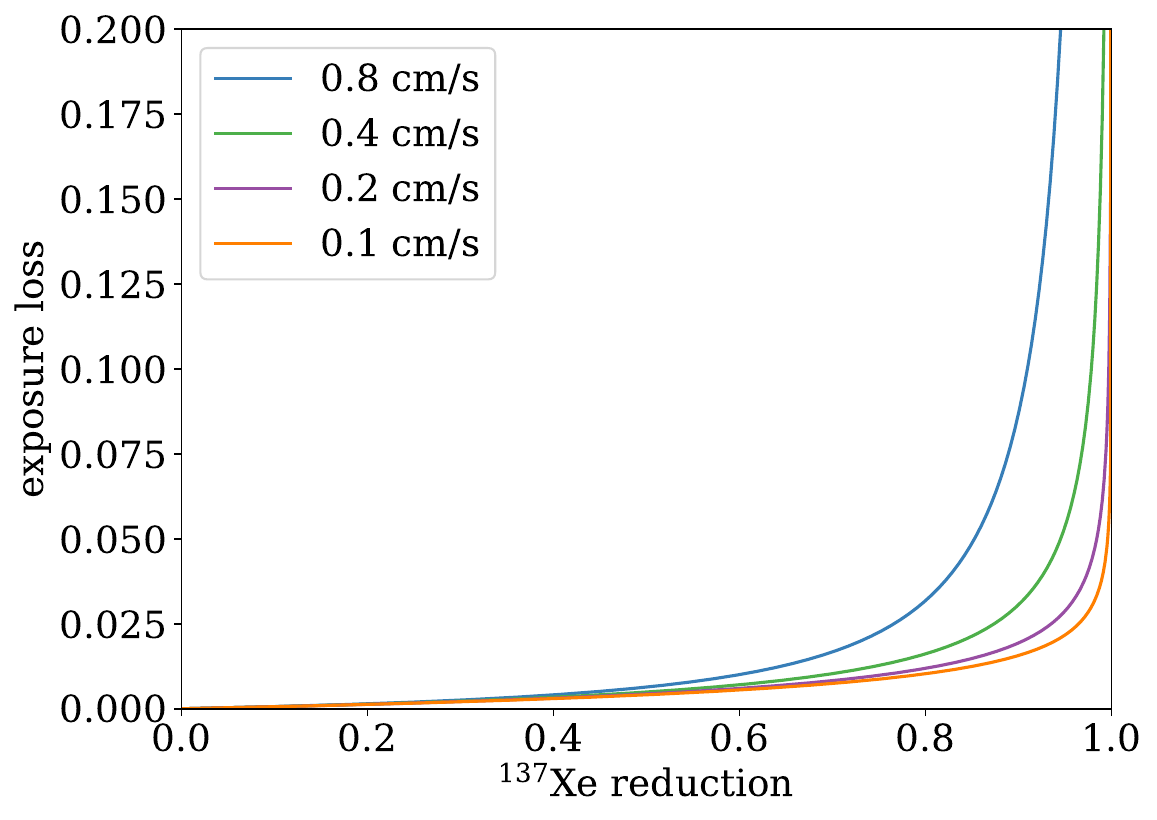}%
    \caption{Projected exposure loss versus \isotope{Xe}{137} background reduction when tagging \isotope{Xe}{137} backgrounds for different scenarios of convection velocity.}\label{fig:xe137_projection}
\end{figure}

The activity rate \(A\) is also different in this scenario. In XENONnT, the rate in the muon veto is observed to be \(\approx0.035 \unit{Hz}\). As the neutron capture time in liquid xenon is \(\sim 100 \unit{\mu s}\)~\cite{Amarasinghe:2022jgk}, a \(1 \unit{ms}\) window after each muon trigger to search for neutron captures can be considered, leading to a livetime fraction of \(3.5\times 10^{-5}\) within which neutron captures would be searched for. The emitted gammas are expected to be of energies \(\sim 1 \unit{MeV}\)~\cite{Amarasinghe:2022jgk} where \isotope{Xe}{136} decay is the dominant background. Thus, the background rate can be approximated using the fraction found above, multiplied by the rate of \isotope{Xe}{136} decays in natural xenon, \(\approx 4.2\unit{\mu Bq/kg}\)~\cite{EXO-200:2013xfn, XENON:2022evz}, resulting in \(A=1.5\times10^{-4}\unit{\mu Bq/kg}\). Using these values, which represent adaptations of the fit values used in \autoref{ssec:performance_xent}, the performance for different convection velocities is shown in \autoref{fig:xe137_projection}.

It can be seen that for all of the velocity scenarios, almost all of the cosmogenic \isotope{Xe}{137} background can be rejected. In particular, for convection velocities around or below \(0.2 \unit{cm/s}\), the reduction of the cosmogenic \isotope{Xe}{137} background approaches unity for a \(10 \%\) reduction in exposure. However, it should be noted that the reconstruction efficiency of neutron-capture gammas has not been measured, and will proportionally reduce \(p_\text{true}\). In addition, the projections presented here use the worst-case mean free path of $9.8 \unit{cm}$; in reality, the initial point cloud size could be potentially much smaller, depending on the spectrum of the gammas emitted after a neutron capture. 

\section{Conclusions} \label{sec:conclusion}

In this paper, the design and performance of an algorithm for tagging radon-chain backgrounds in liquid noble element TPCs were presented. The presented algorithm performs tagging of the \isotope{Pb}{214} background, which is part of the \isotope{Rn}{222} decay chain. This was demonstrated on XENON1T datasets used for the ER analysis and the search for neutrinoless double beta decay; the original analyses can be found in~\cite{XENON:2020rca, XENON:2022evz}. It was shown that for the ER analysis, an exposure loss of \(1.8\pm0.2\%\) and a \(6.2^{+0.4}_{-0.9}\%\) reduction in the \isotope{Pb}{214} background can be expected. The neutrinoless double beta decay data set is used to produce a high-purity sample of \isotope{Pb}{214} decay events, as can be seen from a spectral fit. This sample also displays relevant features in the spectrum such as the peak at \(352 \unit{keV}\) and falling off at the Q-value of approximately \(1 \unit{MeV}\).

While the demonstrated background reduction is small, the cost of such a software-based background-reduction technique can be minimal, making deployment cost effective. In addition, much higher performance can be expected in larger detectors with lower intrinsic radon levels, due to individual radon-chain events being further apart in the detector in both space and time. In XENONnT due to the lower radon level, performance is expected to be significantly higher than in XENON1T, with an optimal \isotope{Pb}{214} background reduction of between \(25\%\) and \(75\%\), depending on the convection speed in the detector, with a corresponding exposure loss of between \(3.5\%\) and \(11\%\). 
% In future detectors where there is no significant convective motion, it was conservatively estimated that a \(75\%\) reduction in \isotope{Pb}{214} background should be possible with a \(20\%\) exposure loss if there is small-scale fluid flow as measured in EXO-200~\cite{EXO-200:2015ura}. 
If the motion is dominated by diffusion, near-perfect tagging of radon chain backgrounds can be expected.

The fact that the performance of a software veto for \isotope{Pb}{214} backgrounds improves with larger detectors and lower intrinsic radon levels makes it complementary to hardware-based approaches such as the cryogenic distillation system used by XENONnT~\cite{Murra:2022mlr}, or a charcoal trap ~\cite{Pushkin:2018wdl}. 
This is because as detector size increases, these hardware-based approaches require increasing mass flow rates to retain the same performance, whereas algorithmic approaches do not suffer from this scaling. In addition, software-based approaches perform better if the radon level is already low due to radiopurity controls or hardware-based radon removal methods; in the limiting case where there is on average much less than one \isotope{Po}{218} in the TPC at any given time, there can simply be a veto on all data within a few half-lives of a \isotope{Po}{218} alpha decay to remove almost all of the \isotope{Pb}{214} background. It should be noted that there are also hardware approaches that do require increasing mass flow rates to retain performance, such as material selection and screening~\cite{XENON:2017fdb, LZ:2020fty, XENON:2021mrg}, detector design~\cite{Dierle:2022zzh}, and material coating~\cite{Jorg:2022spz}.

The methods outlined in this paper can also be used to suppress radon chain backgrounds in liquid argon TPCs, where hardware-based approaches for the mitigation of radon-chain backgrounds are similarly being pursued~\cite{Avasthi:2022tjr}.

In addition, the performance of a similar approach applied to reduce the cosmogenic \isotope{Xe}{137} background was also estimated. This background is expected to be a major background in the search for \(0\nu\beta\beta\) decay in \isotope{Xe}{136} in XENONnT~\cite{XENON:2022evz}, LZ~\cite{LZ:2019qdm}, and next-generation liquid xenon TPCs~\cite{Aalbers:2022dzr}. 
% As such, a software veto for the \isotope{Xe}{137} background can be invaluable for the search for \(0\nu\beta\beta\) decay. 
It was found that if neutron capture gammas can be selected with high efficiency, then the \isotope{Xe}{137} background can be tagged in XENONnT with an efficiency of \(>90\%\), resulting in a \(<9 \%\) background reduction, depending on the convection speed.

\begin{acknowledgments}
We gratefully acknowledge support from the National Science
Foundation, Swiss National Science Foundation, German Ministry for
Education and Research, Max Planck Gesellschaft, Deutsche
Forschungsgemeinschaft, Helmholtz Association, Dutch Research Council
(NWO), Weizmann Institute of Science, Israeli Science Foundation,
Binational Science Foundation, R\'egion des Pays de la Loire, Knut and
Alice Wallenberg Foundation, Kavli Foundation, JSPS Kakenhi and JST
FOREST Program in Japan, Tsinghua University Initiative Scientific
Research Program and Istituto Nazionale di Fisica Nucleare. This
project has received funding/support from the European Union’s Horizon
2020 research and innovation programme under the Marie
Sk\l{}odowska-Curie grant agreement No 860881-HIDDeN. Data processing
is performed using infrastructures from the Open Science Grid, the
European Grid Initiative and the Dutch national e-infrastructure with
the support of SURF Cooperative. We are grateful to Laboratori
Nazionali del Gran Sasso for hosting and supporting the XENON project.

\end{acknowledgments}

% \nocite{*}

\bibliography{main}% Produces the bibliography via BibTeX.

\begin{thebibliography}{63}%
\makeatletter
\providecommand \@ifxundefined [1]{%
 \@ifx{#1\undefined}
}%
\providecommand \@ifnum [1]{%
 \ifnum #1\expandafter \@firstoftwo
 \else \expandafter \@secondoftwo
 \fi
}%
\providecommand \@ifx [1]{%
 \ifx #1\expandafter \@firstoftwo
 \else \expandafter \@secondoftwo
 \fi
}%
\providecommand \natexlab [1]{#1}%
\providecommand \enquote  [1]{``#1''}%
\providecommand \bibnamefont  [1]{#1}%
\providecommand \bibfnamefont [1]{#1}%
\providecommand \citenamefont [1]{#1}%
\providecommand \href@noop [0]{\@secondoftwo}%
\providecommand \href [0]{\begingroup \@sanitize@url \@href}%
\providecommand \@href[1]{\@@startlink{#1}\@@href}%
\providecommand \@@href[1]{\endgroup#1\@@endlink}%
\providecommand \@sanitize@url [0]{\catcode `\\12\catcode `\$12\catcode `\&12\catcode `\#12\catcode `\^12\catcode `\_12\catcode `\%12\relax}%
\providecommand \@@startlink[1]{}%
\providecommand \@@endlink[0]{}%
\providecommand \url  [0]{\begingroup\@sanitize@url \@url }%
\providecommand \@url [1]{\endgroup\@href {#1}{\urlprefix }}%
\providecommand \urlprefix  [0]{URL }%
\providecommand \Eprint [0]{\href }%
\providecommand \doibase [0]{http://dx.doi.org/}%
\providecommand \selectlanguage [0]{\@gobble}%
\providecommand \bibinfo  [0]{\@secondoftwo}%
\providecommand \bibfield  [0]{\@secondoftwo}%
\providecommand \translation [1]{[#1]}%
\providecommand \BibitemOpen [0]{}%
\providecommand \bibitemStop [0]{}%
\providecommand \bibitemNoStop [0]{.\EOS\space}%
\providecommand \EOS [0]{\spacefactor3000\relax}%
\providecommand \BibitemShut  [1]{\csname bibitem#1\endcsname}%
\let\auto@bib@innerbib\@empty
%</preamble>
\bibitem [{\citenamefont {Aprile}\ \emph {et~al.}(2017{\natexlab{a}})\citenamefont {Aprile} \emph {et~al.}}]{XENON:2017lvq}%
  \BibitemOpen
  \bibfield  {author} {\bibinfo {author} {\bibfnamefont {E.}~\bibnamefont {Aprile}} \emph {et~al.} (\bibinfo {collaboration} {XENON}),\ }\href {\doibase 10.1140/epjc/s10052-017-5326-3} {\bibfield  {journal} {\bibinfo  {journal} {Eur. Phys. J. C}\ }\textbf {\bibinfo {volume} {77}},\ \bibinfo {pages} {881} (\bibinfo {year} {2017}{\natexlab{a}})},\ \Eprint {http://arxiv.org/abs/1708.07051} {arXiv:1708.07051 [astro-ph.IM]} \BibitemShut {NoStop}%
\bibitem [{\citenamefont {Aprile}\ \emph {et~al.}(2020{\natexlab{a}})\citenamefont {Aprile} \emph {et~al.}}]{XENON:2020kmp}%
  \BibitemOpen
  \bibfield  {author} {\bibinfo {author} {\bibfnamefont {E.}~\bibnamefont {Aprile}} \emph {et~al.} (\bibinfo {collaboration} {XENON}),\ }\href {\doibase 10.1088/1475-7516/2020/11/031} {\bibfield  {journal} {\bibinfo  {journal} {JCAP}\ }\textbf {\bibinfo {volume} {11}},\ \bibinfo {pages} {031} (\bibinfo {year} {2020}{\natexlab{a}})},\ \Eprint {http://arxiv.org/abs/2007.08796} {arXiv:2007.08796 [physics.ins-det]} \BibitemShut {NoStop}%
\bibitem [{\citenamefont {Akerib}\ \emph {et~al.}(2020{\natexlab{a}})\citenamefont {Akerib} \emph {et~al.}}]{LZ:2019sgr}%
  \BibitemOpen
  \bibfield  {author} {\bibinfo {author} {\bibfnamefont {D.~S.}\ \bibnamefont {Akerib}} \emph {et~al.} (\bibinfo {collaboration} {LZ}),\ }\href {\doibase 10.1016/j.nima.2019.163047} {\bibfield  {journal} {\bibinfo  {journal} {Nucl. Instrum. Meth. A}\ }\textbf {\bibinfo {volume} {953}},\ \bibinfo {pages} {163047} (\bibinfo {year} {2020}{\natexlab{a}})},\ \Eprint {http://arxiv.org/abs/1910.09124} {arXiv:1910.09124 [physics.ins-det]} \BibitemShut {NoStop}%
\bibitem [{\citenamefont {Aprile}\ \emph {et~al.}(2018)\citenamefont {Aprile} \emph {et~al.}}]{XENON:2018voc}%
  \BibitemOpen
  \bibfield  {author} {\bibinfo {author} {\bibfnamefont {E.}~\bibnamefont {Aprile}} \emph {et~al.} (\bibinfo {collaboration} {XENON}),\ }\href {\doibase 10.1103/PhysRevLett.121.111302} {\bibfield  {journal} {\bibinfo  {journal} {Phys. Rev. Lett.}\ }\textbf {\bibinfo {volume} {121}},\ \bibinfo {pages} {111302} (\bibinfo {year} {2018})},\ \Eprint {http://arxiv.org/abs/1805.12562} {arXiv:1805.12562 [astro-ph.CO]} \BibitemShut {NoStop}%
\bibitem [{\citenamefont {Aprile}\ \emph {et~al.}(2022{\natexlab{a}})\citenamefont {Aprile} \emph {et~al.}}]{XENON:2022evz}%
  \BibitemOpen
  \bibfield  {author} {\bibinfo {author} {\bibfnamefont {E.}~\bibnamefont {Aprile}} \emph {et~al.} (\bibinfo {collaboration} {XENON}),\ }\href {\doibase 10.1103/PhysRevC.106.024328} {\bibfield  {journal} {\bibinfo  {journal} {Phys. Rev. C}\ }\textbf {\bibinfo {volume} {106}},\ \bibinfo {pages} {024328} (\bibinfo {year} {2022}{\natexlab{a}})},\ \Eprint {http://arxiv.org/abs/2205.04158} {arXiv:2205.04158 [hep-ex]} \BibitemShut {NoStop}%
\bibitem [{\citenamefont {Anton}\ \emph {et~al.}(2019)\citenamefont {Anton} \emph {et~al.}}]{EXO-200:2019rkq}%
  \BibitemOpen
  \bibfield  {author} {\bibinfo {author} {\bibfnamefont {G.}~\bibnamefont {Anton}} \emph {et~al.} (\bibinfo {collaboration} {EXO-200}),\ }\href {\doibase 10.1103/PhysRevLett.123.161802} {\bibfield  {journal} {\bibinfo  {journal} {Phys. Rev. Lett.}\ }\textbf {\bibinfo {volume} {123}},\ \bibinfo {pages} {161802} (\bibinfo {year} {2019})},\ \Eprint {http://arxiv.org/abs/1906.02723} {arXiv:1906.02723 [hep-ex]} \BibitemShut {NoStop}%
\bibitem [{\citenamefont {Akerib}\ \emph {et~al.}(2021)\citenamefont {Akerib} \emph {et~al.}}]{LZ:2021blo}%
  \BibitemOpen
  \bibfield  {author} {\bibinfo {author} {\bibfnamefont {D.~S.}\ \bibnamefont {Akerib}} \emph {et~al.} (\bibinfo {collaboration} {LZ}),\ }\href {\doibase 10.1103/PhysRevC.104.065501} {\bibfield  {journal} {\bibinfo  {journal} {Phys. Rev. C}\ }\textbf {\bibinfo {volume} {104}},\ \bibinfo {pages} {065501} (\bibinfo {year} {2021})},\ \Eprint {http://arxiv.org/abs/2104.13374} {arXiv:2104.13374 [physics.ins-det]} \BibitemShut {NoStop}%
\bibitem [{\citenamefont {Aprile}\ \emph {et~al.}(2019{\natexlab{a}})\citenamefont {Aprile} \emph {et~al.}}]{XENON:2019dti}%
  \BibitemOpen
  \bibfield  {author} {\bibinfo {author} {\bibfnamefont {E.}~\bibnamefont {Aprile}} \emph {et~al.} (\bibinfo {collaboration} {XENON}),\ }\href {\doibase 10.1038/s41586-019-1124-4} {\bibfield  {journal} {\bibinfo  {journal} {Nature}\ }\textbf {\bibinfo {volume} {568}},\ \bibinfo {pages} {532} (\bibinfo {year} {2019}{\natexlab{a}})},\ \Eprint {http://arxiv.org/abs/1904.11002} {arXiv:1904.11002 [nucl-ex]} \BibitemShut {NoStop}%
\bibitem [{\citenamefont {Aprile}\ \emph {et~al.}(2020{\natexlab{b}})\citenamefont {Aprile} \emph {et~al.}}]{XENON:2020rca}%
  \BibitemOpen
  \bibfield  {author} {\bibinfo {author} {\bibfnamefont {E.}~\bibnamefont {Aprile}} \emph {et~al.} (\bibinfo {collaboration} {XENON}),\ }\href {\doibase 10.1103/PhysRevD.102.072004} {\bibfield  {journal} {\bibinfo  {journal} {Phys. Rev. D}\ }\textbf {\bibinfo {volume} {102}},\ \bibinfo {pages} {072004} (\bibinfo {year} {2020}{\natexlab{b}})},\ \Eprint {http://arxiv.org/abs/2006.09721} {arXiv:2006.09721 [hep-ex]} \BibitemShut {NoStop}%
\bibitem [{\citenamefont {Aprile}\ \emph {et~al.}(2022{\natexlab{b}})\citenamefont {Aprile} \emph {et~al.}}]{XENON:2022ltv}%
  \BibitemOpen
  \bibfield  {author} {\bibinfo {author} {\bibfnamefont {E.}~\bibnamefont {Aprile}} \emph {et~al.} (\bibinfo {collaboration} {XENON}),\ }\href {\doibase 10.1103/PhysRevLett.129.161805} {\bibfield  {journal} {\bibinfo  {journal} {Phys. Rev. Lett.}\ }\textbf {\bibinfo {volume} {129}},\ \bibinfo {pages} {161805} (\bibinfo {year} {2022}{\natexlab{b}})},\ \Eprint {http://arxiv.org/abs/2207.11330} {arXiv:2207.11330 [hep-ex]} \BibitemShut {NoStop}%
\bibitem [{\citenamefont {Aalbers}\ \emph {et~al.}(2016)\citenamefont {Aalbers} \emph {et~al.}}]{DARWIN:2016hyl}%
  \BibitemOpen
  \bibfield  {author} {\bibinfo {author} {\bibfnamefont {J.}~\bibnamefont {Aalbers}} \emph {et~al.} (\bibinfo {collaboration} {DARWIN}),\ }\href {\doibase 10.1088/1475-7516/2016/11/017} {\bibfield  {journal} {\bibinfo  {journal} {JCAP}\ }\textbf {\bibinfo {volume} {11}},\ \bibinfo {pages} {017} (\bibinfo {year} {2016})},\ \Eprint {http://arxiv.org/abs/1606.07001} {arXiv:1606.07001 [astro-ph.IM]} \BibitemShut {NoStop}%
\bibitem [{\citenamefont {Zhang}\ \emph {et~al.}(2019)\citenamefont {Zhang} \emph {et~al.}}]{PandaX:2018wtu}%
  \BibitemOpen
  \bibfield  {author} {\bibinfo {author} {\bibfnamefont {H.}~\bibnamefont {Zhang}} \emph {et~al.} (\bibinfo {collaboration} {PandaX}),\ }\href {\doibase 10.1007/s11433-018-9259-0} {\bibfield  {journal} {\bibinfo  {journal} {Sci. China Phys. Mech. Astron.}\ }\textbf {\bibinfo {volume} {62}},\ \bibinfo {pages} {31011} (\bibinfo {year} {2019})},\ \Eprint {http://arxiv.org/abs/1806.02229} {arXiv:1806.02229 [physics.ins-det]} \BibitemShut {NoStop}%
\bibitem [{\citenamefont {Aprile}\ \emph {et~al.}(2022{\natexlab{c}})\citenamefont {Aprile} \emph {et~al.}}]{XENON:2021mrg}%
  \BibitemOpen
  \bibfield  {author} {\bibinfo {author} {\bibfnamefont {E.}~\bibnamefont {Aprile}} \emph {et~al.} (\bibinfo {collaboration} {XENON}),\ }\href {\doibase 10.1140/epjc/s10052-022-10345-6} {\bibfield  {journal} {\bibinfo  {journal} {Eur. Phys. J. C}\ }\textbf {\bibinfo {volume} {82}},\ \bibinfo {pages} {599} (\bibinfo {year} {2022}{\natexlab{c}})},\ \Eprint {http://arxiv.org/abs/2112.05629} {arXiv:2112.05629 [physics.ins-det]} \BibitemShut {NoStop}%
\bibitem [{\citenamefont {Singh}\ \emph {et~al.}(2019)\citenamefont {Singh} \emph {et~al.}}]{Singh:2019qyr}%
  \BibitemOpen
  \bibfield  {author} {\bibinfo {author} {\bibfnamefont {B.}~\bibnamefont {Singh}} \emph {et~al.},\ }\href {\doibase 10.1016/j.nds.2019.100524} {\bibfield  {journal} {\bibinfo  {journal} {Nucl. Data Sheets}\ }\textbf {\bibinfo {volume} {160}},\ \bibinfo {pages} {405} (\bibinfo {year} {2019})}\BibitemShut {NoStop}%
\bibitem [{\citenamefont {Abe}\ \emph {et~al.}(2012)\citenamefont {Abe}, \citenamefont {Hieda}, \citenamefont {Hiraide}, \citenamefont {Hirano}, \citenamefont {Kishimoto}, \citenamefont {Kobayashi}, \citenamefont {Koshio}, \citenamefont {Liu}, \citenamefont {Martens}, \citenamefont {Moriyama}, \citenamefont {Nakahata}, \citenamefont {Nishiie}, \citenamefont {Ogawa}, \citenamefont {Sekiya}, \citenamefont {Shinozaki}, \citenamefont {Suzuki}, \citenamefont {Takachio}, \citenamefont {Takeda}, \citenamefont {Ueshima}, \citenamefont {Umemoto}, \citenamefont {Yamashita}, \citenamefont {Hosokawa}, \citenamefont {Murata}, \citenamefont {Otsuka}, \citenamefont {Takeuchi}, \citenamefont {Kusaba}, \citenamefont {Motoki}, \citenamefont {Nishijima}, \citenamefont {Tasaka}, \citenamefont {Fujii}, \citenamefont {Murayama}, \citenamefont {Nakamura}, \citenamefont {Fukuda}, \citenamefont {Itow}, \citenamefont {Masuda}, \citenamefont {Nishitani}, \citenamefont {Takiya}, \citenamefont {Uchida}, \citenamefont {Kim}, \citenamefont
  {Kim}, \citenamefont {Lee}, \citenamefont {Lee},\ and\ \citenamefont {Lee}}]{Abe2012RadonRF}%
  \BibitemOpen
  \bibfield  {author} {\bibinfo {author} {\bibfnamefont {K.}~\bibnamefont {Abe}}, \bibinfo {author} {\bibfnamefont {K.}~\bibnamefont {Hieda}}, \bibinfo {author} {\bibfnamefont {K.}~\bibnamefont {Hiraide}}, \bibinfo {author} {\bibfnamefont {S.}~\bibnamefont {Hirano}}, \bibinfo {author} {\bibfnamefont {Y.}~\bibnamefont {Kishimoto}}, \bibinfo {author} {\bibfnamefont {K.}~\bibnamefont {Kobayashi}}, \bibinfo {author} {\bibfnamefont {Y.}~\bibnamefont {Koshio}}, \bibinfo {author} {\bibfnamefont {J.}~\bibnamefont {Liu}}, \bibinfo {author} {\bibfnamefont {K.}~\bibnamefont {Martens}}, \bibinfo {author} {\bibfnamefont {S.}~\bibnamefont {Moriyama}}, \bibinfo {author} {\bibfnamefont {M.}~\bibnamefont {Nakahata}}, \bibinfo {author} {\bibfnamefont {H.}~\bibnamefont {Nishiie}}, \bibinfo {author} {\bibfnamefont {H.}~\bibnamefont {Ogawa}}, \bibinfo {author} {\bibfnamefont {H.}~\bibnamefont {Sekiya}}, \bibinfo {author} {\bibfnamefont {A.}~\bibnamefont {Shinozaki}}, \bibinfo {author} {\bibfnamefont {Y.}~\bibnamefont {Suzuki}},
  \bibinfo {author} {\bibfnamefont {O.}~\bibnamefont {Takachio}}, \bibinfo {author} {\bibfnamefont {A.}~\bibnamefont {Takeda}}, \bibinfo {author} {\bibfnamefont {K.}~\bibnamefont {Ueshima}}, \bibinfo {author} {\bibfnamefont {D.}~\bibnamefont {Umemoto}}, \bibinfo {author} {\bibfnamefont {M.}~\bibnamefont {Yamashita}}, \bibinfo {author} {\bibfnamefont {K.}~\bibnamefont {Hosokawa}}, \bibinfo {author} {\bibfnamefont {A.}~\bibnamefont {Murata}}, \bibinfo {author} {\bibfnamefont {K.}~\bibnamefont {Otsuka}}, \bibinfo {author} {\bibfnamefont {Y.}~\bibnamefont {Takeuchi}}, \bibinfo {author} {\bibfnamefont {F.}~\bibnamefont {Kusaba}}, \bibinfo {author} {\bibfnamefont {D.}~\bibnamefont {Motoki}}, \bibinfo {author} {\bibfnamefont {K.}~\bibnamefont {Nishijima}}, \bibinfo {author} {\bibfnamefont {S.}~\bibnamefont {Tasaka}}, \bibinfo {author} {\bibfnamefont {K.}~\bibnamefont {Fujii}}, \bibinfo {author} {\bibfnamefont {I.}~\bibnamefont {Murayama}}, \bibinfo {author} {\bibfnamefont {S.}~\bibnamefont {Nakamura}}, \bibinfo
  {author} {\bibfnamefont {Y.}~\bibnamefont {Fukuda}}, \bibinfo {author} {\bibfnamefont {Y.}~\bibnamefont {Itow}}, \bibinfo {author} {\bibfnamefont {K.}~\bibnamefont {Masuda}}, \bibinfo {author} {\bibfnamefont {Y.}~\bibnamefont {Nishitani}}, \bibinfo {author} {\bibfnamefont {H.}~\bibnamefont {Takiya}}, \bibinfo {author} {\bibfnamefont {H.}~\bibnamefont {Uchida}}, \bibinfo {author} {\bibfnamefont {Y.~D.}\ \bibnamefont {Kim}}, \bibinfo {author} {\bibfnamefont {Y.~H.}\ \bibnamefont {Kim}}, \bibinfo {author} {\bibfnamefont {K.~B.}\ \bibnamefont {Lee}}, \bibinfo {author} {\bibfnamefont {M.~K.}\ \bibnamefont {Lee}}, \ and\ \bibinfo {author} {\bibfnamefont {J.-M.}\ \bibnamefont {Lee}},\ }\href {https://api.semanticscholar.org/CorpusID:54791532} {\bibfield  {journal} {\bibinfo  {journal} {Nuclear Instruments \& Methods in Physics Research Section A-accelerators Spectrometers Detectors and Associated Equipment}\ }\textbf {\bibinfo {volume} {661}},\ \bibinfo {pages} {50} (\bibinfo {year} {2012})}\BibitemShut {NoStop}%
\bibitem [{\citenamefont {Arthurs}\ \emph {et~al.}(2020)\citenamefont {Arthurs}, \citenamefont {Huang}, \citenamefont {Amarasinghe}, \citenamefont {Miller},\ and\ \citenamefont {Lorenzon}}]{Arthurs:2020ans}%
  \BibitemOpen
  \bibfield  {author} {\bibinfo {author} {\bibfnamefont {M.}~\bibnamefont {Arthurs}}, \bibinfo {author} {\bibfnamefont {D.~Q.}\ \bibnamefont {Huang}}, \bibinfo {author} {\bibfnamefont {C.}~\bibnamefont {Amarasinghe}}, \bibinfo {author} {\bibfnamefont {E.}~\bibnamefont {Miller}}, \ and\ \bibinfo {author} {\bibfnamefont {W.}~\bibnamefont {Lorenzon}},\ }\href@noop {} {\  (\bibinfo {year} {2020})},\ \Eprint {http://arxiv.org/abs/2009.06069} {arXiv:2009.06069 [physics.ins-det]} \BibitemShut {NoStop}%
\bibitem [{\citenamefont {Chen}\ \emph {et~al.}(2024)\citenamefont {Chen}, \citenamefont {Gibbons}, \citenamefont {Haselschwardt}, \citenamefont {Kravitz}, \citenamefont {Xia},\ and\ \citenamefont {Sorensen}}]{Chen:2023llu}%
  \BibitemOpen
  \bibfield  {author} {\bibinfo {author} {\bibfnamefont {H.}~\bibnamefont {Chen}}, \bibinfo {author} {\bibfnamefont {R.}~\bibnamefont {Gibbons}}, \bibinfo {author} {\bibfnamefont {S.~J.}\ \bibnamefont {Haselschwardt}}, \bibinfo {author} {\bibfnamefont {S.}~\bibnamefont {Kravitz}}, \bibinfo {author} {\bibfnamefont {Q.}~\bibnamefont {Xia}}, \ and\ \bibinfo {author} {\bibfnamefont {P.}~\bibnamefont {Sorensen}},\ }\href {\doibase 10.1103/PhysRevD.109.L071102} {\bibfield  {journal} {\bibinfo  {journal} {Phys. Rev. D}\ }\textbf {\bibinfo {volume} {109}},\ \bibinfo {pages} {L071102} (\bibinfo {year} {2024})},\ \Eprint {http://arxiv.org/abs/2312.15082} {arXiv:2312.15082 [hep-ex]} \BibitemShut {NoStop}%
\bibitem [{\citenamefont {Akerib}\ \emph {et~al.}(2018)\citenamefont {Akerib} \emph {et~al.}}]{LUX:2016wel}%
  \BibitemOpen
  \bibfield  {author} {\bibinfo {author} {\bibfnamefont {D.~S.}\ \bibnamefont {Akerib}} \emph {et~al.} (\bibinfo {collaboration} {LUX}),\ }\href {\doibase 10.1016/j.astropartphys.2017.10.014} {\bibfield  {journal} {\bibinfo  {journal} {Astropart. Phys.}\ }\textbf {\bibinfo {volume} {97}},\ \bibinfo {pages} {80} (\bibinfo {year} {2018})},\ \Eprint {http://arxiv.org/abs/1605.03844} {arXiv:1605.03844 [physics.ins-det]} \BibitemShut {NoStop}%
\bibitem [{\citenamefont {Akerib}\ \emph {et~al.}(2020{\natexlab{b}})\citenamefont {Akerib} \emph {et~al.}}]{LZ:2020fty}%
  \BibitemOpen
  \bibfield  {author} {\bibinfo {author} {\bibfnamefont {D.~S.}\ \bibnamefont {Akerib}} \emph {et~al.} (\bibinfo {collaboration} {LZ}),\ }\href {\doibase 10.1140/epjc/s10052-020-8420-x} {\bibfield  {journal} {\bibinfo  {journal} {Eur. Phys. J. C}\ }\textbf {\bibinfo {volume} {80}},\ \bibinfo {pages} {1044} (\bibinfo {year} {2020}{\natexlab{b}})},\ \Eprint {http://arxiv.org/abs/2006.02506} {arXiv:2006.02506 [physics.ins-det]} \BibitemShut {NoStop}%
\bibitem [{\citenamefont {Murra}\ \emph {et~al.}(2022)\citenamefont {Murra}, \citenamefont {Schulte}, \citenamefont {Huhmann},\ and\ \citenamefont {Weinheimer}}]{Murra:2022mlr}%
  \BibitemOpen
  \bibfield  {author} {\bibinfo {author} {\bibfnamefont {M.}~\bibnamefont {Murra}}, \bibinfo {author} {\bibfnamefont {D.}~\bibnamefont {Schulte}}, \bibinfo {author} {\bibfnamefont {C.}~\bibnamefont {Huhmann}}, \ and\ \bibinfo {author} {\bibfnamefont {C.}~\bibnamefont {Weinheimer}},\ }\href {\doibase 10.1140/epjc/s10052-022-11001-9} {\bibfield  {journal} {\bibinfo  {journal} {Eur. Phys. J. C}\ }\textbf {\bibinfo {volume} {82}},\ \bibinfo {pages} {1104} (\bibinfo {year} {2022})},\ \Eprint {http://arxiv.org/abs/2205.11492} {arXiv:2205.11492 [physics.ins-det]} \BibitemShut {NoStop}%
\bibitem [{\citenamefont {Pushkin}\ \emph {et~al.}(2018)\citenamefont {Pushkin} \emph {et~al.}}]{Pushkin:2018wdl}%
  \BibitemOpen
  \bibfield  {author} {\bibinfo {author} {\bibfnamefont {K.}~\bibnamefont {Pushkin}} \emph {et~al.},\ }\href {\doibase 10.1016/j.nima.2018.06.076} {\bibfield  {journal} {\bibinfo  {journal} {Nucl. Instrum. Meth. A}\ }\textbf {\bibinfo {volume} {903}},\ \bibinfo {pages} {267} (\bibinfo {year} {2018})},\ \Eprint {http://arxiv.org/abs/1805.11306} {arXiv:1805.11306 [physics.ins-det]} \BibitemShut {NoStop}%
\bibitem [{\citenamefont {Aprile}\ \emph {et~al.}(2017{\natexlab{b}})\citenamefont {Aprile} \emph {et~al.}}]{XENON:2017fdb}%
  \BibitemOpen
  \bibfield  {author} {\bibinfo {author} {\bibfnamefont {E.}~\bibnamefont {Aprile}} \emph {et~al.} (\bibinfo {collaboration} {XENON}),\ }\href {\doibase 10.1140/epjc/s10052-017-5329-0} {\bibfield  {journal} {\bibinfo  {journal} {Eur. Phys. J. C}\ }\textbf {\bibinfo {volume} {77}},\ \bibinfo {pages} {890} (\bibinfo {year} {2017}{\natexlab{b}})},\ \Eprint {http://arxiv.org/abs/1705.01828} {arXiv:1705.01828 [physics.ins-det]} \BibitemShut {NoStop}%
\bibitem [{\citenamefont {Avasthi}\ \emph {et~al.}(2022)\citenamefont {Avasthi} \emph {et~al.}}]{Avasthi:2022tjr}%
  \BibitemOpen
  \bibfield  {author} {\bibinfo {author} {\bibfnamefont {A.}~\bibnamefont {Avasthi}} \emph {et~al.},\ }in\ \href@noop {} {\emph {\bibinfo {booktitle} {{Snowmass 2021}}}}\ (\bibinfo {year} {2022})\ \Eprint {http://arxiv.org/abs/2203.08821} {arXiv:2203.08821 [physics.ins-det]} \BibitemShut {NoStop}%
\bibitem [{\citenamefont {Fujii}\ \emph {et~al.}(2015)\citenamefont {Fujii}, \citenamefont {Endo}, \citenamefont {Torigoe}, \citenamefont {Nakamura}, \citenamefont {Haruyama}, \citenamefont {Kasami}, \citenamefont {Mihara}, \citenamefont {Saito}, \citenamefont {Sasaki},\ and\ \citenamefont {Tawara}}]{FUJII2015293}%
  \BibitemOpen
  \bibfield  {author} {\bibinfo {author} {\bibfnamefont {K.}~\bibnamefont {Fujii}}, \bibinfo {author} {\bibfnamefont {Y.}~\bibnamefont {Endo}}, \bibinfo {author} {\bibfnamefont {Y.}~\bibnamefont {Torigoe}}, \bibinfo {author} {\bibfnamefont {S.}~\bibnamefont {Nakamura}}, \bibinfo {author} {\bibfnamefont {T.}~\bibnamefont {Haruyama}}, \bibinfo {author} {\bibfnamefont {K.}~\bibnamefont {Kasami}}, \bibinfo {author} {\bibfnamefont {S.}~\bibnamefont {Mihara}}, \bibinfo {author} {\bibfnamefont {K.}~\bibnamefont {Saito}}, \bibinfo {author} {\bibfnamefont {S.}~\bibnamefont {Sasaki}}, \ and\ \bibinfo {author} {\bibfnamefont {H.}~\bibnamefont {Tawara}},\ }\href {\doibase https://doi.org/10.1016/j.nima.2015.05.065} {\bibfield  {journal} {\bibinfo  {journal} {Nuclear Instruments and Methods in Physics Research Section A: Accelerators, Spectrometers, Detectors and Associated Equipment}\ }\textbf {\bibinfo {volume} {795}},\ \bibinfo {pages} {293} (\bibinfo {year} {2015})}\BibitemShut {NoStop}%
\bibitem [{\citenamefont {Szydagis}\ \emph {et~al.}(2022)\citenamefont {Szydagis} \emph {et~al.}}]{Szydagis:2022ikv}%
  \BibitemOpen
  \bibfield  {author} {\bibinfo {author} {\bibfnamefont {M.}~\bibnamefont {Szydagis}} \emph {et~al.},\ }\href@noop {} {\  (\bibinfo {year} {2022})},\ \Eprint {http://arxiv.org/abs/2211.10726} {arXiv:2211.10726 [hep-ex]} \BibitemShut {NoStop}%
\bibitem [{\citenamefont {Aprile}\ \emph {et~al.}(2019{\natexlab{b}})\citenamefont {Aprile} \emph {et~al.}}]{XENON:2019ykp}%
  \BibitemOpen
  \bibfield  {author} {\bibinfo {author} {\bibfnamefont {E.}~\bibnamefont {Aprile}} \emph {et~al.} (\bibinfo {collaboration} {XENON}),\ }\href {\doibase 10.1103/PhysRevD.100.052014} {\bibfield  {journal} {\bibinfo  {journal} {Phys. Rev. D}\ }\textbf {\bibinfo {volume} {100}},\ \bibinfo {pages} {052014} (\bibinfo {year} {2019}{\natexlab{b}})},\ \Eprint {http://arxiv.org/abs/1906.04717} {arXiv:1906.04717 [physics.ins-det]} \BibitemShut {NoStop}%
\bibitem [{\citenamefont {Aprile}\ \emph {et~al.}(2006)\citenamefont {Aprile}, \citenamefont {Dahl}, \citenamefont {DeViveiros}, \citenamefont {Gaitskell}, \citenamefont {Giboni}, \citenamefont {Kwong}, \citenamefont {Majewski}, \citenamefont {Ni}, \citenamefont {Shutt},\ and\ \citenamefont {Yamashita}}]{Aprile:2006kx}%
  \BibitemOpen
  \bibfield  {author} {\bibinfo {author} {\bibfnamefont {E.}~\bibnamefont {Aprile}}, \bibinfo {author} {\bibfnamefont {C.~E.}\ \bibnamefont {Dahl}}, \bibinfo {author} {\bibfnamefont {L.}~\bibnamefont {DeViveiros}}, \bibinfo {author} {\bibfnamefont {R.}~\bibnamefont {Gaitskell}}, \bibinfo {author} {\bibfnamefont {K.~L.}\ \bibnamefont {Giboni}}, \bibinfo {author} {\bibfnamefont {J.}~\bibnamefont {Kwong}}, \bibinfo {author} {\bibfnamefont {P.}~\bibnamefont {Majewski}}, \bibinfo {author} {\bibfnamefont {K.}~\bibnamefont {Ni}}, \bibinfo {author} {\bibfnamefont {T.}~\bibnamefont {Shutt}}, \ and\ \bibinfo {author} {\bibfnamefont {M.}~\bibnamefont {Yamashita}},\ }\href {\doibase 10.1103/PhysRevLett.97.081302} {\bibfield  {journal} {\bibinfo  {journal} {Phys. Rev. Lett.}\ }\textbf {\bibinfo {volume} {97}},\ \bibinfo {pages} {081302} (\bibinfo {year} {2006})},\ \Eprint {http://arxiv.org/abs/astro-ph/0601552} {arXiv:astro-ph/0601552} \BibitemShut {NoStop}%
\bibitem [{\citenamefont {J\"org}\ \emph {et~al.}(2022)\citenamefont {J\"org}, \citenamefont {Cichon}, \citenamefont {Eurin}, \citenamefont {H\"otzsch}, \citenamefont {Undagoitia~Marrod\'an},\ and\ \citenamefont {Rupp}}]{Jorg:2021hzu}%
  \BibitemOpen
  \bibfield  {author} {\bibinfo {author} {\bibfnamefont {F.}~\bibnamefont {J\"org}}, \bibinfo {author} {\bibfnamefont {D.}~\bibnamefont {Cichon}}, \bibinfo {author} {\bibfnamefont {G.}~\bibnamefont {Eurin}}, \bibinfo {author} {\bibfnamefont {L.}~\bibnamefont {H\"otzsch}}, \bibinfo {author} {\bibfnamefont {T.}~\bibnamefont {Undagoitia~Marrod\'an}}, \ and\ \bibinfo {author} {\bibfnamefont {N.}~\bibnamefont {Rupp}},\ }\href {\doibase 10.1140/epjc/s10052-022-10259-3} {\bibfield  {journal} {\bibinfo  {journal} {Eur. Phys. J. C}\ }\textbf {\bibinfo {volume} {82}},\ \bibinfo {pages} {361} (\bibinfo {year} {2022})},\ \Eprint {http://arxiv.org/abs/2109.13735} {arXiv:2109.13735 [physics.ins-det]} \BibitemShut {NoStop}%
\bibitem [{\citenamefont {Singh}\ \emph {et~al.}(2011)\citenamefont {Singh}, \citenamefont {Jain},\ and\ \citenamefont {Tuli}}]{Singh:2011yau}%
  \BibitemOpen
  \bibfield  {author} {\bibinfo {author} {\bibfnamefont {S.}~\bibnamefont {Singh}}, \bibinfo {author} {\bibfnamefont {A.~K.}\ \bibnamefont {Jain}}, \ and\ \bibinfo {author} {\bibfnamefont {J.~K.}\ \bibnamefont {Tuli}},\ }\href {\doibase 10.1016/j.nds.2011.10.002} {\bibfield  {journal} {\bibinfo  {journal} {Nucl. Data Sheets}\ }\textbf {\bibinfo {volume} {112}},\ \bibinfo {pages} {2851} (\bibinfo {year} {2011})}\BibitemShut {NoStop}%
\bibitem [{\citenamefont {Zhu}\ and\ \citenamefont {McCutchan}(2021)}]{Zhu:2021qss}%
  \BibitemOpen
  \bibfield  {author} {\bibinfo {author} {\bibfnamefont {S.}~\bibnamefont {Zhu}}\ and\ \bibinfo {author} {\bibfnamefont {E.~A.}\ \bibnamefont {McCutchan}},\ }\href {\doibase 10.1016/j.nds.2021.06.001} {\bibfield  {journal} {\bibinfo  {journal} {Nucl. Data Sheets}\ }\textbf {\bibinfo {volume} {175}},\ \bibinfo {pages} {1} (\bibinfo {year} {2021})}\BibitemShut {NoStop}%
\bibitem [{\citenamefont {Shamsuzzoha~Basunia}(2014)}]{ShamsuzzohaBasunia:2014yyr}%
  \BibitemOpen
  \bibfield  {author} {\bibinfo {author} {\bibfnamefont {M.}~\bibnamefont {Shamsuzzoha~Basunia}},\ }\href {\doibase 10.1016/j.nds.2014.09.004} {\bibfield  {journal} {\bibinfo  {journal} {Nucl. Data Sheets}\ }\textbf {\bibinfo {volume} {121}},\ \bibinfo {pages} {561} (\bibinfo {year} {2014})}\BibitemShut {NoStop}%
\bibitem [{\citenamefont {Kondev}(2008)}]{Kondev:2008roc}%
  \BibitemOpen
  \bibfield  {author} {\bibinfo {author} {\bibfnamefont {F.~G.}\ \bibnamefont {Kondev}},\ }\href {\doibase 10.1016/j.nds.2008.05.002} {\bibfield  {journal} {\bibinfo  {journal} {Nucl. Data Sheets}\ }\textbf {\bibinfo {volume} {109}},\ \bibinfo {pages} {1527} (\bibinfo {year} {2008})}\BibitemShut {NoStop}%
\bibitem [{\citenamefont {Aprile}\ \emph {et~al.}(2017{\natexlab{c}})\citenamefont {Aprile} \emph {et~al.}}]{XENON:2016rze}%
  \BibitemOpen
  \bibfield  {author} {\bibinfo {author} {\bibfnamefont {E.}~\bibnamefont {Aprile}} \emph {et~al.} (\bibinfo {collaboration} {XENON}),\ }\href {\doibase 10.1103/PhysRevD.95.072008} {\bibfield  {journal} {\bibinfo  {journal} {Phys. Rev. D}\ }\textbf {\bibinfo {volume} {95}},\ \bibinfo {pages} {072008} (\bibinfo {year} {2017}{\natexlab{c}})},\ \Eprint {http://arxiv.org/abs/1611.03585} {arXiv:1611.03585 [physics.ins-det]} \BibitemShut {NoStop}%
\bibitem [{\citenamefont {Malling}(2014)}]{Malling:2014oxk}%
  \BibitemOpen
  \bibfield  {author} {\bibinfo {author} {\bibfnamefont {D.~C.}\ \bibnamefont {Malling}},\ }\emph {\bibinfo {title} {{Measurement and Analysis of WIMP Detection Backgrounds, and Characteri- zation and Performance of the Large Underground Xenon Dark Matter Search Experiment}}},\ \href {\doibase 10.7301/Z0057D9F} {Ph.D. thesis},\ \bibinfo  {school} {Brown U.} (\bibinfo {year} {2014})\BibitemShut {NoStop}%
\bibitem [{\citenamefont {Pedregosa}\ \emph {et~al.}(2011)\citenamefont {Pedregosa}, \citenamefont {Varoquaux}, \citenamefont {Gramfort}, \citenamefont {Michel}, \citenamefont {Thirion}, \citenamefont {Grisel}, \citenamefont {Blondel}, \citenamefont {Prettenhofer}, \citenamefont {Weiss}, \citenamefont {Dubourg}, \citenamefont {Vanderplas}, \citenamefont {Passos}, \citenamefont {Cournapeau}, \citenamefont {Brucher}, \citenamefont {Perrot},\ and\ \citenamefont {{{\'E}}douard Duchesnay}}]{pedregosa:JMLR:v12:pedregosa11a}%
  \BibitemOpen
  \bibfield  {author} {\bibinfo {author} {\bibfnamefont {F.}~\bibnamefont {Pedregosa}}, \bibinfo {author} {\bibfnamefont {G.}~\bibnamefont {Varoquaux}}, \bibinfo {author} {\bibfnamefont {A.}~\bibnamefont {Gramfort}}, \bibinfo {author} {\bibfnamefont {V.}~\bibnamefont {Michel}}, \bibinfo {author} {\bibfnamefont {B.}~\bibnamefont {Thirion}}, \bibinfo {author} {\bibfnamefont {O.}~\bibnamefont {Grisel}}, \bibinfo {author} {\bibfnamefont {M.}~\bibnamefont {Blondel}}, \bibinfo {author} {\bibfnamefont {P.}~\bibnamefont {Prettenhofer}}, \bibinfo {author} {\bibfnamefont {R.}~\bibnamefont {Weiss}}, \bibinfo {author} {\bibfnamefont {V.}~\bibnamefont {Dubourg}}, \bibinfo {author} {\bibfnamefont {J.}~\bibnamefont {Vanderplas}}, \bibinfo {author} {\bibfnamefont {A.}~\bibnamefont {Passos}}, \bibinfo {author} {\bibfnamefont {D.}~\bibnamefont {Cournapeau}}, \bibinfo {author} {\bibfnamefont {M.}~\bibnamefont {Brucher}}, \bibinfo {author} {\bibfnamefont {M.}~\bibnamefont {Perrot}}, \ and\ \bibinfo {author} {\bibnamefont
  {{{\'E}}douard Duchesnay}},\ }\href {http://jmlr.org/papers/v12/pedregosa11a.html} {\bibfield  {journal} {\bibinfo  {journal} {Journal of Machine Learning Research}\ }\textbf {\bibinfo {volume} {12}},\ \bibinfo {pages} {2825} (\bibinfo {year} {2011})}\BibitemShut {NoStop}%
\bibitem [{\citenamefont {Masson}(2018)}]{Masson:2018pte}%
  \BibitemOpen
  \bibfield  {author} {\bibinfo {author} {\bibfnamefont {D.}~\bibnamefont {Masson}},\ }\emph {\bibinfo {title} {{Novel Ideas and Techniques for Large Dark Matter Detectors}}},\ \href@noop {} {Ph.D. thesis},\ \bibinfo  {school} {Purdue U.} (\bibinfo {year} {2018})\BibitemShut {NoStop}%
\bibitem [{\citenamefont {Aprile}\ \emph {et~al.}(2021)\citenamefont {Aprile} \emph {et~al.}}]{XENON:2020fbs}%
  \BibitemOpen
  \bibfield  {author} {\bibinfo {author} {\bibfnamefont {E.}~\bibnamefont {Aprile}} \emph {et~al.} (\bibinfo {collaboration} {XENON}),\ }\href {\doibase 10.1140/epjc/s10052-020-08777-z} {\bibfield  {journal} {\bibinfo  {journal} {Eur. Phys. J. C}\ }\textbf {\bibinfo {volume} {81}},\ \bibinfo {pages} {337} (\bibinfo {year} {2021})},\ \Eprint {http://arxiv.org/abs/2009.13981} {arXiv:2009.13981 [physics.ins-det]} \BibitemShut {NoStop}%
\bibitem [{\citenamefont {Ahlers}\ \emph {et~al.}(2022)\citenamefont {Ahlers}, \citenamefont {Bodenschatz}, \citenamefont {Hartmann}, \citenamefont {He}, \citenamefont {Lohse}, \citenamefont {Reiter}, \citenamefont {Stevens}, \citenamefont {Verzicco}, \citenamefont {Wedi}, \citenamefont {Weiss}, \citenamefont {Zhang}, \citenamefont {Zwirner},\ and\ \citenamefont {Shishkina}}]{Guenter:PhysRevLett.128.084501}%
  \BibitemOpen
  \bibfield  {author} {\bibinfo {author} {\bibfnamefont {G.}~\bibnamefont {Ahlers}}, \bibinfo {author} {\bibfnamefont {E.}~\bibnamefont {Bodenschatz}}, \bibinfo {author} {\bibfnamefont {R.}~\bibnamefont {Hartmann}}, \bibinfo {author} {\bibfnamefont {X.}~\bibnamefont {He}}, \bibinfo {author} {\bibfnamefont {D.}~\bibnamefont {Lohse}}, \bibinfo {author} {\bibfnamefont {P.}~\bibnamefont {Reiter}}, \bibinfo {author} {\bibfnamefont {R.~J. A.~M.}\ \bibnamefont {Stevens}}, \bibinfo {author} {\bibfnamefont {R.}~\bibnamefont {Verzicco}}, \bibinfo {author} {\bibfnamefont {M.}~\bibnamefont {Wedi}}, \bibinfo {author} {\bibfnamefont {S.}~\bibnamefont {Weiss}}, \bibinfo {author} {\bibfnamefont {X.}~\bibnamefont {Zhang}}, \bibinfo {author} {\bibfnamefont {L.}~\bibnamefont {Zwirner}}, \ and\ \bibinfo {author} {\bibfnamefont {O.}~\bibnamefont {Shishkina}},\ }\href {\doibase 10.1103/PhysRevLett.128.084501} {\bibfield  {journal} {\bibinfo  {journal} {Phys. Rev. Lett.}\ }\textbf {\bibinfo {volume} {128}},\ \bibinfo {pages}
  {084501} (\bibinfo {year} {2022})}\BibitemShut {NoStop}%
\bibitem [{\citenamefont {{Ye}}(2020)}]{Ye:2020PhDT.........9Y}%
  \BibitemOpen
  \bibfield  {author} {\bibinfo {author} {\bibfnamefont {J.}~\bibnamefont {{Ye}}},\ }\emph {\bibinfo {title} {{Searches for WIMPs and axions with the XENON1T experiment}}},\ \href@noop {} {Ph.D. thesis},\ \bibinfo  {school} {University of California, San Diego} (\bibinfo {year} {2020})\BibitemShut {NoStop}%
\bibitem [{\citenamefont {HALDANE}(1948)}]{HALDANE:10.1093/biomet/35.3-4.414}%
  \BibitemOpen
  \bibfield  {author} {\bibinfo {author} {\bibfnamefont {J.~B.~S.}\ \bibnamefont {HALDANE}},\ }\href {\doibase 10.1093/biomet/35.3-4.414} {\bibfield  {journal} {\bibinfo  {journal} {Biometrika}\ }\textbf {\bibinfo {volume} {35}},\ \bibinfo {pages} {414} (\bibinfo {year} {1948})},\ \Eprint {http://arxiv.org/abs/https://academic.oup.com/biomet/article-pdf/35/3-4/414/785927/35-3-4-414.pdf} {https://academic.oup.com/biomet/article-pdf/35/3-4/414/785927/35-3-4-414.pdf} \BibitemShut {NoStop}%
\bibitem [{\citenamefont {Lopuhaa}\ and\ \citenamefont {Rousseeuw}(1991)}]{Lopuhaa:10.1214/aos/1176347978}%
  \BibitemOpen
  \bibfield  {author} {\bibinfo {author} {\bibfnamefont {H.~P.}\ \bibnamefont {Lopuhaa}}\ and\ \bibinfo {author} {\bibfnamefont {P.~J.}\ \bibnamefont {Rousseeuw}},\ }\href {\doibase 10.1214/aos/1176347978} {\bibfield  {journal} {\bibinfo  {journal} {The Annals of Statistics}\ }\textbf {\bibinfo {volume} {19}},\ \bibinfo {pages} {229 } (\bibinfo {year} {1991})}\BibitemShut {NoStop}%
\bibitem [{\citenamefont {Albert}\ \emph {et~al.}(2015)\citenamefont {Albert} \emph {et~al.}}]{EXO-200:2015ura}%
  \BibitemOpen
  \bibfield  {author} {\bibinfo {author} {\bibfnamefont {J.~B.}\ \bibnamefont {Albert}} \emph {et~al.} (\bibinfo {collaboration} {EXO-200}),\ }\href {\doibase 10.1103/PhysRevC.92.045504} {\bibfield  {journal} {\bibinfo  {journal} {Phys. Rev. C}\ }\textbf {\bibinfo {volume} {92}},\ \bibinfo {pages} {045504} (\bibinfo {year} {2015})},\ \Eprint {http://arxiv.org/abs/1506.00317} {arXiv:1506.00317 [nucl-ex]} \BibitemShut {NoStop}%
\bibitem [{\citenamefont {Aalbers}\ \emph {et~al.}(2022{\natexlab{a}})\citenamefont {Aalbers} \emph {et~al.}}]{LZ:2022oxb}%
  \BibitemOpen
  \bibfield  {author} {\bibinfo {author} {\bibfnamefont {J.}~\bibnamefont {Aalbers}} \emph {et~al.} (\bibinfo {collaboration} {LZ}),\ }\href@noop {} {\  (\bibinfo {year} {2022}{\natexlab{a}})},\ \Eprint {http://arxiv.org/abs/2211.17120} {arXiv:2211.17120 [hep-ex]} \BibitemShut {NoStop}%
\bibitem [{\citenamefont {Bridson}\ \emph {et~al.}(2007)\citenamefont {Bridson}, \citenamefont {Houriham},\ and\ \citenamefont {Nordenstam}}]{Bridson:10.1145/1276377.1276435}%
  \BibitemOpen
  \bibfield  {author} {\bibinfo {author} {\bibfnamefont {R.}~\bibnamefont {Bridson}}, \bibinfo {author} {\bibfnamefont {J.}~\bibnamefont {Houriham}}, \ and\ \bibinfo {author} {\bibfnamefont {M.}~\bibnamefont {Nordenstam}},\ }\href {\doibase 10.1145/1276377.1276435} {\bibfield  {journal} {\bibinfo  {journal} {ACM Trans. Graph.}\ }\textbf {\bibinfo {volume} {26}},\ \bibinfo {pages} {46–es} (\bibinfo {year} {2007})}\BibitemShut {NoStop}%
\bibitem [{\citenamefont {Ester}\ \emph {et~al.}(1996)\citenamefont {Ester}, \citenamefont {Kriegel}, \citenamefont {Sander},\ and\ \citenamefont {Xu}}]{Ester:10.5555/3001460.3001507}%
  \BibitemOpen
  \bibfield  {author} {\bibinfo {author} {\bibfnamefont {M.}~\bibnamefont {Ester}}, \bibinfo {author} {\bibfnamefont {H.-P.}\ \bibnamefont {Kriegel}}, \bibinfo {author} {\bibfnamefont {J.}~\bibnamefont {Sander}}, \ and\ \bibinfo {author} {\bibfnamefont {X.}~\bibnamefont {Xu}},\ }in\ \href@noop {} {\emph {\bibinfo {booktitle} {Proceedings of the Second International Conference on Knowledge Discovery and Data Mining}}},\ \bibinfo {series and number} {KDD'96}\ (\bibinfo  {publisher} {AAAI Press},\ \bibinfo {year} {1996})\ p.\ \bibinfo {pages} {226–231}\BibitemShut {NoStop}%
\bibitem [{\citenamefont {Cowan}\ \emph {et~al.}(2011)\citenamefont {Cowan}, \citenamefont {Cranmer}, \citenamefont {Gross},\ and\ \citenamefont {Vitells}}]{Cowan:2010js}%
  \BibitemOpen
  \bibfield  {author} {\bibinfo {author} {\bibfnamefont {G.}~\bibnamefont {Cowan}}, \bibinfo {author} {\bibfnamefont {K.}~\bibnamefont {Cranmer}}, \bibinfo {author} {\bibfnamefont {E.}~\bibnamefont {Gross}}, \ and\ \bibinfo {author} {\bibfnamefont {O.}~\bibnamefont {Vitells}},\ }\href {\doibase 10.1140/epjc/s10052-011-1554-0} {\bibfield  {journal} {\bibinfo  {journal} {Eur. Phys. J. C}\ }\textbf {\bibinfo {volume} {71}},\ \bibinfo {pages} {1554} (\bibinfo {year} {2011})},\ \bibinfo {note} {[Erratum: Eur.Phys.J.C 73, 2501 (2013)]},\ \Eprint {http://arxiv.org/abs/1007.1727} {arXiv:1007.1727 [physics.data-an]} \BibitemShut {NoStop}%
\bibitem [{\citenamefont {Speetjens}\ \emph {et~al.}(2021)\citenamefont {Speetjens}, \citenamefont {Metcalfe},\ and\ \citenamefont {Rudman}}]{speetjens2021lagrangian}%
  \BibitemOpen
  \bibfield  {author} {\bibinfo {author} {\bibfnamefont {M.}~\bibnamefont {Speetjens}}, \bibinfo {author} {\bibfnamefont {G.}~\bibnamefont {Metcalfe}}, \ and\ \bibinfo {author} {\bibfnamefont {M.}~\bibnamefont {Rudman}},\ }\href@noop {} {\bibfield  {journal} {\bibinfo  {journal} {Applied Mechanics Reviews}\ }\textbf {\bibinfo {volume} {73}} (\bibinfo {year} {2021})}\BibitemShut {NoStop}%
\bibitem [{\citenamefont {Taylor}(2005)}]{taylor2005classical}%
  \BibitemOpen
  \bibfield  {author} {\bibinfo {author} {\bibfnamefont {J.}~\bibnamefont {Taylor}},\ }\enquote {\bibinfo {title} {Classical mechanics},}\ \ (\bibinfo  {publisher} {University Science Books},\ \bibinfo {year} {2005})\ Chap.~\bibinfo {chapter} {12}\BibitemShut {NoStop}%
\bibitem [{\citenamefont {Bruenner}\ \emph {et~al.}(2021)\citenamefont {Bruenner}, \citenamefont {Cichon}, \citenamefont {Eurin}, \citenamefont {Herrero~G\'omez}, \citenamefont {J\"org}, \citenamefont {Marrod\'an~Undagoitia}, \citenamefont {Simgen},\ and\ \citenamefont {Rupp}}]{Bruenner:2020arp}%
  \BibitemOpen
  \bibfield  {author} {\bibinfo {author} {\bibfnamefont {S.}~\bibnamefont {Bruenner}}, \bibinfo {author} {\bibfnamefont {D.}~\bibnamefont {Cichon}}, \bibinfo {author} {\bibfnamefont {G.}~\bibnamefont {Eurin}}, \bibinfo {author} {\bibfnamefont {P.}~\bibnamefont {Herrero~G\'omez}}, \bibinfo {author} {\bibfnamefont {F.}~\bibnamefont {J\"org}}, \bibinfo {author} {\bibfnamefont {T.}~\bibnamefont {Marrod\'an~Undagoitia}}, \bibinfo {author} {\bibfnamefont {H.}~\bibnamefont {Simgen}}, \ and\ \bibinfo {author} {\bibfnamefont {N.}~\bibnamefont {Rupp}},\ }\href {\doibase 10.1140/epjc/s10052-021-09047-2} {\bibfield  {journal} {\bibinfo  {journal} {Eur. Phys. J. C}\ }\textbf {\bibinfo {volume} {81}},\ \bibinfo {pages} {343} (\bibinfo {year} {2021})},\ \Eprint {http://arxiv.org/abs/2009.08828} {arXiv:2009.08828 [physics.ins-det]} \BibitemShut {NoStop}%
\bibitem [{\citenamefont {Aprile}\ \emph {et~al.}(2022{\natexlab{d}})\citenamefont {Aprile} \emph {et~al.}}]{XENONCollaboration:2022kmb}%
  \BibitemOpen
  \bibfield  {author} {\bibinfo {author} {\bibfnamefont {E.}~\bibnamefont {Aprile}} \emph {et~al.} (\bibinfo {collaboration} {(XENON Collaboration)\textdagger{}\textdagger{}, XENON}),\ }\href {\doibase 10.1103/PhysRevLett.129.161805} {\bibfield  {journal} {\bibinfo  {journal} {Phys. Rev. Lett.}\ }\textbf {\bibinfo {volume} {129}},\ \bibinfo {pages} {161805} (\bibinfo {year} {2022}{\natexlab{d}})},\ \Eprint {http://arxiv.org/abs/2207.11330} {arXiv:2207.11330 [hep-ex]} \BibitemShut {NoStop}%
\bibitem [{\citenamefont {Einstein}(1956)}]{einstein1956investigations}%
  \BibitemOpen
  \bibfield  {author} {\bibinfo {author} {\bibfnamefont {A.}~\bibnamefont {Einstein}},\ }\href@noop {} {\emph {\bibinfo {title} {Investigations on the theory of the Brownian movement}}},\ Dover Books on Physics\ (\bibinfo  {publisher} {Dover Publications},\ \bibinfo {address} {Mineola, NY},\ \bibinfo {year} {1956})\BibitemShut {NoStop}%
\bibitem [{\citenamefont {Polianin}(2002)}]{Polianin:alma99169543790201081}%
  \BibitemOpen
  \bibfield  {author} {\bibinfo {author} {\bibfnamefont {A.~D. A.~D.}\ \bibnamefont {Polianin}},\ }\href@noop {} {\emph {\bibinfo {title} {Handbook of linear partial differential equations for engineers and scientists}}},\ \bibinfo {edition} {second edition.}\ ed.\ (\bibinfo  {publisher} {Chapman \& Hall/CRC},\ \bibinfo {address} {Boca Raton},\ \bibinfo {year} {2002})\BibitemShut {NoStop}%
\bibitem [{\citenamefont {Richert}\ \emph {et~al.}(1989)\citenamefont {Richert}, \citenamefont {Pautmeier},\ and\ \citenamefont {B\"assler}}]{Richert:PhysRevLett.63.547}%
  \BibitemOpen
  \bibfield  {author} {\bibinfo {author} {\bibfnamefont {R.}~\bibnamefont {Richert}}, \bibinfo {author} {\bibfnamefont {L.}~\bibnamefont {Pautmeier}}, \ and\ \bibinfo {author} {\bibfnamefont {H.}~\bibnamefont {B\"assler}},\ }\href {\doibase 10.1103/PhysRevLett.63.547} {\bibfield  {journal} {\bibinfo  {journal} {Phys. Rev. Lett.}\ }\textbf {\bibinfo {volume} {63}},\ \bibinfo {pages} {547} (\bibinfo {year} {1989})}\BibitemShut {NoStop}%
\bibitem [{\citenamefont {Aprile}\ and\ \citenamefont {Doke}(2010)}]{Aprile:2009dv}%
  \BibitemOpen
  \bibfield  {author} {\bibinfo {author} {\bibfnamefont {E.}~\bibnamefont {Aprile}}\ and\ \bibinfo {author} {\bibfnamefont {T.}~\bibnamefont {Doke}},\ }\href {\doibase 10.1103/RevModPhys.82.2053} {\bibfield  {journal} {\bibinfo  {journal} {Rev. Mod. Phys.}\ }\textbf {\bibinfo {volume} {82}},\ \bibinfo {pages} {2053} (\bibinfo {year} {2010})},\ \Eprint {http://arxiv.org/abs/0910.4956} {arXiv:0910.4956 [physics.ins-det]} \BibitemShut {NoStop}%
\bibitem [{\citenamefont {Aalbers}\ \emph {et~al.}(2022{\natexlab{b}})\citenamefont {Aalbers} \emph {et~al.}}]{Aalbers:2022dzr}%
  \BibitemOpen
  \bibfield  {author} {\bibinfo {author} {\bibfnamefont {J.}~\bibnamefont {Aalbers}} \emph {et~al.},\ }\href@noop {} {\  (\bibinfo {year} {2022}{\natexlab{b}})},\ \Eprint {http://arxiv.org/abs/2203.02309} {arXiv:2203.02309 [physics.ins-det]} \BibitemShut {NoStop}%
\bibitem [{\citenamefont {Baudis}(2024)}]{Baudis:2024jnk}%
  \BibitemOpen
  \bibfield  {author} {\bibinfo {author} {\bibfnamefont {L.}~\bibnamefont {Baudis}},\ }\href {\doibase 10.1016/j.nuclphysb.2024.116473} {\bibfield  {journal} {\bibinfo  {journal} {Nucl. Phys. B}\ }\textbf {\bibinfo {volume} {1003}},\ \bibinfo {pages} {116473} (\bibinfo {year} {2024})},\ \Eprint {http://arxiv.org/abs/2404.19524} {arXiv:2404.19524 [astro-ph.IM]} \BibitemShut {NoStop}%
\bibitem [{\citenamefont {Akerib}\ \emph {et~al.}(2020{\natexlab{c}})\citenamefont {Akerib} \emph {et~al.}}]{LZ:2019qdm}%
  \BibitemOpen
  \bibfield  {author} {\bibinfo {author} {\bibfnamefont {D.~S.}\ \bibnamefont {Akerib}} \emph {et~al.} (\bibinfo {collaboration} {LZ}),\ }\href {\doibase 10.1103/PhysRevC.102.014602} {\bibfield  {journal} {\bibinfo  {journal} {Phys. Rev. C}\ }\textbf {\bibinfo {volume} {102}},\ \bibinfo {pages} {014602} (\bibinfo {year} {2020}{\natexlab{c}})},\ \Eprint {http://arxiv.org/abs/1912.04248} {arXiv:1912.04248 [nucl-ex]} \BibitemShut {NoStop}%
\bibitem [{\citenamefont {Browne}\ and\ \citenamefont {Tuli}(2007)}]{Browne:2007wqt}%
  \BibitemOpen
  \bibfield  {author} {\bibinfo {author} {\bibfnamefont {E.}~\bibnamefont {Browne}}\ and\ \bibinfo {author} {\bibfnamefont {J.~K.}\ \bibnamefont {Tuli}},\ }\href {\doibase 10.1016/j.nds.2007.09.002} {\bibfield  {journal} {\bibinfo  {journal} {Nucl. Data Sheets}\ }\textbf {\bibinfo {volume} {108}},\ \bibinfo {pages} {2173} (\bibinfo {year} {2007})}\BibitemShut {NoStop}%
\bibitem [{\citenamefont {Berger}\ \emph {et~al.}(2010)\citenamefont {Berger}, \citenamefont {Hubbell}, \citenamefont {Seltzer}, \citenamefont {Chang}, \citenamefont {Coursey}, \citenamefont {Sukumar}, \citenamefont {Zucker},\ and\ \citenamefont {Olsen}}]{Berger:148746}%
  \BibitemOpen
  \bibfield  {author} {\bibinfo {author} {\bibfnamefont {M.}~\bibnamefont {Berger}}, \bibinfo {author} {\bibfnamefont {J.}~\bibnamefont {Hubbell}}, \bibinfo {author} {\bibfnamefont {S.}~\bibnamefont {Seltzer}}, \bibinfo {author} {\bibfnamefont {J.}~\bibnamefont {Chang}}, \bibinfo {author} {\bibfnamefont {J.}~\bibnamefont {Coursey}}, \bibinfo {author} {\bibfnamefont {R.}~\bibnamefont {Sukumar}}, \bibinfo {author} {\bibfnamefont {D.}~\bibnamefont {Zucker}}, \ and\ \bibinfo {author} {\bibfnamefont {K.}~\bibnamefont {Olsen}},\ }\href {http://physics.nist.gov/xcom} {\enquote {\bibinfo {title} {Xcom: Photon cross section database (version 1.5)},}\ } (\bibinfo {year} {2010})\BibitemShut {NoStop}%
\bibitem [{\citenamefont {Amarasinghe}\ \emph {et~al.}(2022)\citenamefont {Amarasinghe}, \citenamefont {Coronel}, \citenamefont {Huang}, \citenamefont {Liu}, \citenamefont {Arthurs}, \citenamefont {Steinfeld}, \citenamefont {Gaitskell},\ and\ \citenamefont {Lorenzon}}]{Amarasinghe:2022jgk}%
  \BibitemOpen
  \bibfield  {author} {\bibinfo {author} {\bibfnamefont {C.~S.}\ \bibnamefont {Amarasinghe}}, \bibinfo {author} {\bibfnamefont {R.}~\bibnamefont {Coronel}}, \bibinfo {author} {\bibfnamefont {D.~Q.}\ \bibnamefont {Huang}}, \bibinfo {author} {\bibfnamefont {Y.}~\bibnamefont {Liu}}, \bibinfo {author} {\bibfnamefont {M.}~\bibnamefont {Arthurs}}, \bibinfo {author} {\bibfnamefont {S.}~\bibnamefont {Steinfeld}}, \bibinfo {author} {\bibfnamefont {R.}~\bibnamefont {Gaitskell}}, \ and\ \bibinfo {author} {\bibfnamefont {W.}~\bibnamefont {Lorenzon}},\ }\href {\doibase 10.1103/PhysRevD.106.032007} {\bibfield  {journal} {\bibinfo  {journal} {Phys. Rev. D}\ }\textbf {\bibinfo {volume} {106}},\ \bibinfo {pages} {032007} (\bibinfo {year} {2022})},\ \Eprint {http://arxiv.org/abs/2204.03109} {arXiv:2204.03109 [physics.ins-det]} \BibitemShut {NoStop}%
\bibitem [{\citenamefont {Albert}\ \emph {et~al.}(2014)\citenamefont {Albert} \emph {et~al.}}]{EXO-200:2013xfn}%
  \BibitemOpen
  \bibfield  {author} {\bibinfo {author} {\bibfnamefont {J.~B.}\ \bibnamefont {Albert}} \emph {et~al.} (\bibinfo {collaboration} {EXO-200}),\ }\href {\doibase 10.1103/PhysRevC.89.015502} {\bibfield  {journal} {\bibinfo  {journal} {Phys. Rev. C}\ }\textbf {\bibinfo {volume} {89}},\ \bibinfo {pages} {015502} (\bibinfo {year} {2014})},\ \Eprint {http://arxiv.org/abs/1306.6106} {arXiv:1306.6106 [nucl-ex]} \BibitemShut {NoStop}%
\bibitem [{\citenamefont {Dierle}\ \emph {et~al.}(2023)\citenamefont {Dierle}, \citenamefont {Brown}, \citenamefont {Fischer}, \citenamefont {Glade-Beucke}, \citenamefont {Grigat}, \citenamefont {Kuger}, \citenamefont {Lindemann}, \citenamefont {Silva},\ and\ \citenamefont {Schumann}}]{Dierle:2022zzh}%
  \BibitemOpen
  \bibfield  {author} {\bibinfo {author} {\bibfnamefont {J.}~\bibnamefont {Dierle}}, \bibinfo {author} {\bibfnamefont {A.}~\bibnamefont {Brown}}, \bibinfo {author} {\bibfnamefont {H.}~\bibnamefont {Fischer}}, \bibinfo {author} {\bibfnamefont {R.}~\bibnamefont {Glade-Beucke}}, \bibinfo {author} {\bibfnamefont {J.}~\bibnamefont {Grigat}}, \bibinfo {author} {\bibfnamefont {F.}~\bibnamefont {Kuger}}, \bibinfo {author} {\bibfnamefont {S.}~\bibnamefont {Lindemann}}, \bibinfo {author} {\bibfnamefont {M.~R.}\ \bibnamefont {Silva}}, \ and\ \bibinfo {author} {\bibfnamefont {M.}~\bibnamefont {Schumann}},\ }\href {\doibase 10.1140/epjc/s10052-022-11151-w} {\bibfield  {journal} {\bibinfo  {journal} {Eur. Phys. J. C}\ }\textbf {\bibinfo {volume} {83}},\ \bibinfo {pages} {9} (\bibinfo {year} {2023})},\ \Eprint {http://arxiv.org/abs/2209.00362} {arXiv:2209.00362 [physics.ins-det]} \BibitemShut {NoStop}%
\bibitem [{\citenamefont {J\"org}(2022)}]{Jorg:2022spz}%
  \BibitemOpen
  \bibfield  {author} {\bibinfo {author} {\bibfnamefont {F.}~\bibnamefont {J\"org}},\ }\emph {\bibinfo {title} {{From 222Rn measurements in XENONnT and HeXe to radon mitigation in future liquid xenon experiments}}},\ \href {\doibase 10.11588/heidok.00031915} {Ph.D. thesis},\ \bibinfo  {school} {Heidelberg U.} (\bibinfo {year} {2022})\BibitemShut {NoStop}%
\end{thebibliography}%

\end{document}